\documentclass[12pts,oneside]{amsart}
\usepackage{color,pstricks,caption,cite}
\usepackage{hyperref}

\newtheorem{theorem}{Theorem}
\newtheorem{note}[theorem]{Note}
\newtheorem{corollary}[theorem]{Corollary}
\newtheorem{lemma}[theorem]{Lemma}

\newcommand{\mystrut}{\rule[-1.0ex]{0pt}{3.5ex}}

\newcommand{\Proof}{\medskip\noindent\emph{Proof: }}
\newcommand{\cqfd}{\hfill $\Box$\medskip}

\newcommand{\B}{\mathcal B}
\newcommand{\D}{\mathcal D}
\newcommand{\Tran}{{\mathcal C}}

\newcommand{\boldm}{\mbox{\boldmath$m$}}
\newcommand{\boldn}{\mbox{\boldmath$n$}}

\newcommand{\boldQ}{\mbox{\boldmath$Q$}}

\newcommand{\boldR}{\mbox{\boldmath$R$}}

\newcommand{\boldC}{\mbox{\boldmath$C$}}

\newcommand{\sboldm}{\mbox{\boldmath$\scriptstyle m$}}
\newcommand{\sboldn}{\mbox{\boldmath$\scriptstyle n$}}

\newcommand{\sboldC}{\mbox{\boldmath$\scriptstyle C$}}

\newcommand{\boldmp}{\mbox{\boldmath$m$}^{\natural}}

\newcommand{\tboldn}{\mbox{\boldmath$\tilde n$}}
\newcommand{\tboldQ}{\mbox{\boldmath$\tilde Q$}}

\newcommand{\tsboldn}{\mbox{\boldmath$\scriptstyle\tilde n$}}

\newcommand{\boldN}{\boldsymbol{N}}

\newcommand{\qbinom}[2]{{\genfrac{[}{]}{0pt}{}{#1}{#2}}_q}
\newcommand{\qbinomq}[3]{{\genfrac{[}{]}{0pt}{}{#1}{#2}}_{#3}}
\newcommand{\mytri}[3]{U_{#1}(#3,#2)}

\newcommand{\mytriq}[4]{U_{#1}(#3,#2;#4)}

\newcommand{\habf}{h}
\newcommand{\hbone}{\habf^{(0)}}
\newcommand{\hbnnn}{\habf^{(n)}}
\newcommand{\unwt}{w^{\circ}}
\newcommand{\Lbone}{L'}

\newcommand{\ABFGFe}[3]{{\hat A}^{#3;#1}_{#2}} % args: "p", "a,b", "e,f"
\newcommand{\ABFsetef}[3]{{\mathcal A}^{#3;#1}_{#2}}
\newcommand{\ABFpabLef}{\ABFsetef{p}{a,b}{e,f}(L)}
\newcommand{\ABFpabLMef}{\ABFsetef{p}{a,b}{e,f}(L,m)}

\newcommand{\RABFGFe}[3]{{\hat R}^{#3;#1}_{#2}} % args: "p", "a,b", "e,f"
\newcommand{\RABFsetef}[3]{{\mathcal R}^{#3;#1}_{#2}}

\newcommand{\evenval}{\xi}
\newcommand{\mnos}{{m}}
\newcommand{\vword}{{v}}

\newcommand{\hh}{h}
\newcommand{\ha}{a}
\newcommand{\hb}{b}

\newcommand{\hhunwt}{{{\hat{w}}^{\circ}}}
\newcommand{\HLGFe}[3]{{\hat H}^{#3;#1}_{#2}} % args: "p", "a,b", "e,f"
\newcommand{\HLsetef}[3]{{\mathcal H}^{#3;#1}_{#2}} % args: "t", "a,b"
\newcommand{\HLtabLef}{\HLsetef{t}{a,b}{e,f}(L)}

\newcommand{\HLGFinf}[2]{{\hat H}^{#1}_{#2}} % args: "p", "a,b"
\newcommand{\HLtabinf}{{\mathcal H}^{t}_{a,b}}

\newcommand{\oHLGFe}[3]{{\overline H}^{#3;#1}_{#2}}

\newcommand{\QL}{T^{{\mathtt L}}}
\newcommand{\QR}{T^{{\mathtt R}}}
\newcommand{\boldQL}{\mbox{\boldmath$T$}^{{\mathtt L}}}
\newcommand{\boldQR}{\mbox{\boldmath$T$}^{{\mathtt R}}}

\newcommand{\sboldQR}{\mbox{\boldmath$\scriptstyle T$}^{{\mathtt R}}}

\newcommand{\tsboldQR}{\mbox{\boldmath$\scriptstyle\tilde T$}{}^{{\mathtt R}}}
\newcommand{\pabs}[1]{|#1|^{+}}

\def\ZZ{{\mathbb{Z}}}

\def\HZZ{{\tfrac12\mathbb{Z}}}
\def\HZZp{{\tfrac12\mathbb{Z}_{\ge0}}}
\def\ZZph{{\mathbb{Z}+\tfrac12}}

\newcommand{\elrm}[1]{#1}  % verbatim.

\begin{document}
\title[$M(p,2p\pm1)$ characters via half-lattice paths]{
           A quartet of fermionic expressions for $M(k,2k\pm1)$ Virasoro
           characters via half-lattice paths}
%\thanks{\today}
%\subjclass[2000]{
%        Primary 82B23; Secondary 05A15, 05A19, 17B68, 81T40.}
%\thanks{Research supported by the Research Council (RC)}
%\dedicatory{Version: \today.}

\author{Olivier Blondeau-Fournier}
\address{D\'epartement de physique, de g\'enie physique et
d'optique, Universit\'e Laval,  Qu\'ebec, Canada, G1K 7P4.}
\email{olivier.b-fournier.1@ulaval.ca}

\author{Pierre Mathieu}
\address{D\'epartement de physique, de g\'enie physique et
d'optique, Universit\'e Laval,  Qu\'ebec, Canada, G1K 7P4.}
\email{pmathieu@phy.ulaval.ca}

\author{Trevor A Welsh}
\address{D\'epartement de physique, de g\'enie physique et
d'optique, Universit\'e Laval,  Qu\'ebec, Canada, G1K 7P4.}
\address{
Institute of Mathematics,
King's College,
University of Aberdeen,
Aberdeen,
United Kingdom,
AB24 3UE.}
\email{trevor.welsh@utoronro.ca}

\date{}

\begin{abstract}
We derive new fermionic expressions for the characters of the
Virasoro minimal models $M(k,2k\pm1)$  by analysing the
recently introduced half-lattice paths.
These fermionic expressions display a quasiparticle formulation
characteristic of the
$\phi_{2,1}$ and $\phi_{1,5}$ integrable perturbations.
We find that they arise by imposing
a simple restriction on the RSOS quasiparticle states
of the unitary models $M(p,p+1)$.  
In fact, four fermionic expressions are obtained
for each generating function of half-lattice paths of
finite length $L$, and these lead to four distinct expressions
for most characters $\chi^{k,2k\pm1}_{r,s}$.
These are direct analogues of Melzer's expressions
for $M(p,p+1)$, and their proof entails
revisiting, reworking and refining a proof of Melzer's identities
which used combinatorial transforms on lattice paths.

We also derive a bosonic version of the generating functions
of length $L$ half-lattice paths, this expression being notable in
that it involves $q$-trinomial coefficients.
Taking the $L\to\infty$ limit shows that the generating
functions for infinite length half-lattice paths are indeed
the Virasoro characters $\chi^{k,2k\pm1}_{r,s}$.
\end{abstract}

% \begin{keyword}
% %% keywords here, in the form: keyword \sep keyword
% %% PACS codes here, in the form: \PACS code \sep code
% %% MSC codes here, in the form: \MSC code \sep code
% %% or \MSC[2008] code \sep code (2000 is the default)
% Virasoro algebra\sep
% minimal model character\sep
% half-lattice path\sep
% fermionic expression
% \MSC[2010]
% Primary 05E10\sep
% Secondary 17B68\sep
% 82B23
% \end{keyword}

\maketitle

\section{Introduction}

\subsection{Fermionic expressions for Virasoro characters}

The royal road for generating and proving fermionic expressions
for Virasoro characters in minimal models is the path description
of the states induced by an underlying statistical model
\cite{Melzer1994,Warnaar1996a,Warnaar1996b,FLPW2000,%
FodaWelsh1999,FodaWelsh2000,Welsh2005,JacobMathieu2009}.
The statistical model generally considered is the Forrester-Baxter
RSOS model \cite{FB1985}, which is defined for any pair of
relatively coprime integers $p$ and $p'$, and referred to here
as the RSOS$(p,p')$ model.
The RSOS$(p,p+1)$ model is then the Andrews-Baxter-Forrester (ABF) model
\cite{ABF1984}.
The combinatorial techniques for computing the generating functions
of the paths lead to expressions that are manifestly positive,
these expressions being termed fermionic.
On the other hand, an inclusion-exclusion calculation shows that
these generating functions are also given by
\begin{equation}
\label{Eq:Rocha}
\chi^{p, p'}_{r, s}(q)=
{\frac{1}{(q)_\infty}}\sum_{\lambda=-\infty}^\infty
(q^{\lambda^2pp'+\lambda(p'r-ps)}-q^{(\lambda p+r)(\lambda p'+s)}),
\end{equation}
for $1\le r<p$ and $1\le s<p'$,
where $(q)_\infty=\prod_{i=1}^\infty(1-q^i)$.
This bosonic type expression is the well-known Rocha-Caridi
\cite{Rocha-Caridi1985}
expression for the (normalised) Virasoro character $\chi^{p,p'}_{r,s}$
from the minimal model $M(p,p')$.

A fermionic expression encodes a construction of the Hilbert space
by a filling process in which each species
of quasiparticle is subject to a generalized exclusion principle.
This is expected to provide a formulation of the minimal model under
study that is attuned to a particular integrable perturbation
\cite{KKMM1993b,BMc1998},
in the sense that when the specified perturbation is switched on,
the quasiparticles become genuine
particles off-criticality.

Most fermionic characters derived so far can be linked to the
RSOS$(p,p')$ model in a regime that corresponds to a $\phi_{1,3}$
perturbation \cite{Huse1984}.
The corresponding RSOS$(p,p')$ paths are defined on an integer lattice,
with path segments orientated either in the NE or SE direction.

\begin{figure}[ht]
\caption{{%\footnotesize
A typical half-lattice path truncated at length $L=25$.
Observe that all valleys are at integer heights.
For those paths related to Virasoro states, the starting point
is integer.
Infinite length paths that have finite weight necessarily end with
an infinite tail of NE-SE edges oscillating between
$b$ and $b+\frac12$ where $b$ is integer.
In this case, $b=3$.}}
\vskip-0.2cm
\label{TypicalDualBij}
\begin{center}
\psset{yunit=0.32cm,xunit=0.28cm}
\begin{pspicture}(0,-0.5)(50,7.5)
%dotted grid
\psset{linewidth=0.25pt,linestyle=dotted, dotsep=1.0pt,linecolor=gray}
\psline{-}(0,2)(50,2) \psline{-}(0,4)(50,4)
%dashed grid
\psset{linewidth=0.25pt,linestyle=dashed, dash=2.5pt 1.5pt,linecolor=gray}
\psline{-}(0,3)(50,3) \psline{-}(0,5)(50,5)
\psline{-}(1,1)(1,6) \psline{-}(2,1)(2,6) \psline{-}(3,1)(3,6)
\psline{-}(4,1)(4,6) \psline{-}(5,1)(5,6) \psline{-}(6,1)(6,6)
\psline{-}(7,1)(7,6) \psline{-}(8,1)(8,6) \psline{-}(9,1)(9,6)
\psline{-}(10,1)(10,6) \psline{-}(11,1)(11,6) \psline{-}(12,1)(12,6)
\psline{-}(13,1)(13,6) \psline{-}(14,1)(14,6) \psline{-}(15,1)(15,6)
\psline{-}(16,1)(16,6) \psline{-}(17,1)(17,6) \psline{-}(18,1)(18,6)
\psline{-}(19,1)(19,6) \psline{-}(20,1)(20,6) \psline{-}(21,1)(21,6)
\psline{-}(22,1)(22,6) \psline{-}(23,1)(23,6) \psline{-}(24,1)(24,6)
\psline{-}(25,1)(25,6) \psline{-}(26,1)(26,6) \psline{-}(27,1)(27,6)
\psline{-}(28,1)(28,6) \psline{-}(29,1)(29,6) \psline{-}(30,1)(30,6)
\psline{-}(31,1)(31,6) \psline{-}(32,1)(32,6) \psline{-}(33,1)(33,6)
\psline{-}(34,1)(34,6) \psline{-}(35,1)(35,6) \psline{-}(36,1)(36,6)
\psline{-}(37,1)(37,6) \psline{-}(38,1)(38,6) \psline{-}(39,1)(39,6)
\psline{-}(40,1)(40,6) \psline{-}(41,1)(41,6) \psline{-}(42,1)(42,6)
\psline{-}(43,1)(43,6) \psline{-}(44,1)(44,6) \psline{-}(45,1)(45,6)
\psline{-}(46,1)(46,6) \psline{-}(47,1)(47,6) \psline{-}(48,1)(48,6)
\psline{-}(49,1)(49,6)
%axes
\psset{linewidth=0.25pt,fillstyle=none,linestyle=solid,linecolor=black}
%\psline{->}(0,1)(46.5,1)
\psline{-}(0,1)(50,1)
\psline{-}(0,6)(50,6)
\psline{-}(0,1)(0,6)
\psline{-}(50,1)(50,6)
%numeros
\rput(-0.5,1){\scriptsize $1$}\rput(-0.5,3){\scriptsize $2$}
\rput(-0.5,5){\scriptsize $3$}
\rput(0,0.5){\scriptsize $0$} \rput(2,0.5){\scriptsize $1$}
\rput(4,0.5){\scriptsize $2$} \rput(6,0.5){\scriptsize $3$}
\rput(8,0.5){\scriptsize $4$} \rput(10,0.5){\scriptsize $5$}
\rput(12,0.5){\scriptsize $6$} \rput(14,0.5){\scriptsize $7$}
\rput(16,0.5){\scriptsize $8$} \rput(18,0.5){\scriptsize $9$}
\rput(20,0.5){\scriptsize $10$} \rput(22,0.5){\scriptsize $11$}
\rput(24,0.5){\scriptsize $12$} \rput(26,0.5){\scriptsize $13$}
\rput(28,0.5){\scriptsize $14$} \rput(30,0.5){\scriptsize $15$}
\rput(32,0.5){\scriptsize $16$} \rput(34,0.5){\scriptsize $17$}
\rput(36,0.5){\scriptsize $18$} \rput(38,0.5){\scriptsize $19$}
\rput(40,0.5){\scriptsize $20$} \rput(42,0.5){\scriptsize $21$}
\rput(44,0.5){\scriptsize $22$} \rput(46,0.5){\scriptsize $23$}
\rput(48,0.5){\scriptsize $24$} \rput(50,0.5){\scriptsize $25$}
%path
\psset{linewidth=0.7pt,fillstyle=none,linestyle=solid,linecolor=black}
\psline(0,1)(1,2)(2,1)(3,2)(4,3)(5,4)(6,3)(7,2)(8,1)(9,2)(10,3)(11,4)
 (12,3)(13,4)(14,5)(15,4)(16,3)(17,2)(18,1)(19,2)(20,3)(21,4)(22,3)
 (23,2)(24,1)(25,2)(26,1)(27,2)(28,3)(29,2)(30,1)(31,2)(32,3)(33,4)
 (34,5)(35,6)(36,5)(37,6)(38,5)(39,4)(40,3)(41,4)(42,5)(43,6)(44,5)
 (45,6)(46,5)(47,6)(48,5)(49,6)(50,5)
\end{pspicture}
\end{center}
\end{figure}
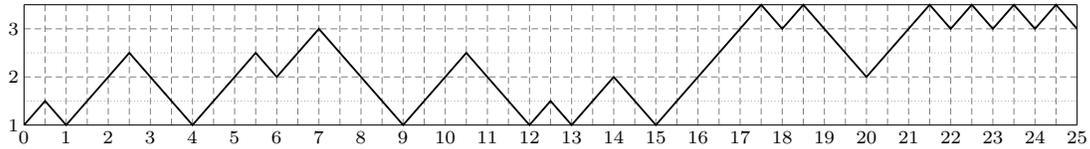

%The aim of this work is to derive fermionic characters using a
%different type of paths, which, as we argue below,
%is associated with the $\phi_{2,1}/\phi_{1,5}$ perturbation. 
%These so-called half-lattice paths
%are refined versions of paths defined on an integer lattice.
The aim of this work is to highlight the use of a different type
of paths for deriving fermionic characters.
These so-called half-lattice paths are,
as we argue below,
associated with the $\phi_{2,1}/\phi_{1,5}$ perturbation. 
The half-lattice paths
are refined versions of paths defined on an integer lattice.
Informally, the half-lattice paths lie on a half-integer grid and
also have path segments orientated in either the NE or SE direction.
However, they are not simply rescaled versions of integer lattice
paths since they are defined with an extra constraint on the position
of the valleys:
{each valley must lie at an integer height.}
An example is given in Figure \ref{TypicalDualBij}.

It will turn out that the generating functions for
the half-lattice paths are the $M(k,2k\pm1)$ instances of \eqref{Eq:Rocha}.
Therefore, the fermionic expressions for the same generating
functions are the fermionic expressions that we seek for
the characters $\chi^{k,2k\pm1}_{r,s}$.
It will also turn out that these expressions
are intimately related to those for the unitary $M(p,p+1)$ models
described by the ABF paths.
From the path perspective, the reason for this is clear:
the weight functions of the two types of paths are similar in that,
in each case, the only contributing vertices are those that lie
between two NE or two SE edges, with each such vertex contributing
half its horizontal position.
In fact, for the unitary models $M(p,p+1)$, there are
four fermionic expressions for each character $\chi^{p,p+1}_{r,s}$
\cite{KKMM1993b,Melzer1994}.
The identities arising from equating these with the
corresponding instances of \eqref{Eq:Rocha} are
often referred to as Melzer's identities.
Our main result, {Theorem \ref{Thrm:HLferms} below, gives}
$M(k,2k\pm1)$ analogues of all these four sets of identities.

It is first convenient to calculate
the generating functions of paths truncated to a fixed length $L$.
These polynomial generating functions are called finitized characters
because they yield $\chi^{p,p'}_{r,s}$ in the limit $L\to\infty$.
This is seen by obtaining a bosonic type expression for the
generating functions which yields \eqref{Eq:Rocha} as $L\to\infty$.
In the case of the RSOS$(p,p')$ models, the
bosonic expressions for the finitized characters are given
in terms of $q$-binomials.
The bosonic expressions that we derive for the finitized
$M(k,2k\pm1)$ characters are notable in that they are given
instead in terms of $q$-trinomials.

Each of the four fermionic expressions for the unitary
characters $\chi^{p,p+1}_{r,s}$ has a finitized version
which yields the former in the limit $L\to\infty$.
On equating these expressions with the bosonic
expressions for the finitized characters,
bosonic-fermionic polynomial identities are obtained.
These were also originally due to Melzer \cite{Melzer1994}.
For the finitized $M(k,2k\pm1)$ characters, we also
obtain four fermionic expressions for the finitized
characters as generating functions of length $L$ half-lattice paths,
and each of these also implies a novel bosonic-fermionic polynomial identity.

With this work, we thus initiate a systematic analysis of
fermionic forms pertaining to the $\phi_{2,1}/\phi_{1,5}$ perturbation. 
Earlier work in this direction,
which did not involve paths,
was carried out in \cite{BMcP1998,BMc2000}.

\subsection{Melzer identities in the unitary case}

Given that these are the objects we intend to generalize,
let us recall the four different fermionic expressions for
the unitary minimal model irreducible characters $\chi^{p,p+1}_{r,s}$,
where $1\le r<p$ and $1\le s\le p$.
These are written compactly using the following notation:
\begin{equation}
\tboldn=(n_2,n_3,\ldots,n_{p-1}),\qquad
\boldm=(m_1,m_2,\ldots,m_{p-2}),\qquad
\text{and}\qquad
\pabs{i}=\begin{cases}i&\text{if $i>0$,}\\
0&\text{if $i\leq 0$}\end{cases}
\end{equation}
(the indices of the components of $\tboldn$ start at 2 here
to make simpler the correspondence with expressions that appear later).
Each of the four expressions is a multiple summation over all
non-negative integer vectors $\tboldn$:
\begin{equation}\label{Eq:ABFferms}
\begin{split}
\chi^{p,p+1}_{r,s}
&=
q^{-\frac14(s-r)(s-r-1)}
\hskip-1mm
\sum_{\tsboldn\in\ZZ^{p-2}_{\ge0}}
\hskip-1mm
q^{\frac14\sboldm\sboldC{\sboldm}^T-\frac12 m_{\ell}}
\frac1{(q)_{m_1}}
\prod_{i=2}^{p-2}
\qbinom{n_i+m_i}{n_i}\!,\\
&\text{where }\quad m_i=2\sum_{i<k<p}(k-i)n_k -\Delta_i
              \quad\text{ for $1\le i<p$,}
\end{split}
\end{equation}
but they differ in the linear term in the exponent, indexed by $\ell$,
and the value of $\Delta_i$ used to define $m_i$; these are tabulated
in Table~\ref{Tab:ABF}.
\begin{table}[ht]
\caption{Parameters for the four cases of the expressions \eqref{Eq:ABFferms}.} 
\label{Tab:ABF}
\begin{center}
\begin{tabular}{|l|c|c|c|c|}\hline
\mystrut & $r$ & $s$ & $\ell$ &  $ \Delta_i$\\
\hline\hline
\mystrut{\rm (a)}& $\ne 1$ & $\ne 1$ & $s-1$
                 & $\pabs{s-1-i}+\pabs{r-i}$\\
\hline
\mystrut{\rm (b)}& $\ne 1$ & $\ne p$ & $p-s$
                 & $\pabs{p-s-i}+\pabs{r-i}+p-1-i$\\
\hline
\mystrut{\rm (c)}& $\ne p-1$ & $\ne p$ & $p-s$
                 & $\pabs{p-s-i}+\pabs{p-r-i}$\\
\hline
\mystrut{\rm (d)}& $\ne p-1$ & $\ne 1$ & $s-1$
                 & $\pabs{s-1-i}+\pabs{p-r-i}+p-1-i$\\
\hline
\end{tabular}
\end{center}
\end{table}

The $q$-binomial
appearing in \eqref{Eq:ABFferms} is defined by
\begin{equation}\label{Eq:qBinomialDef}
\qbinom{n+m}{n}=
\begin{cases}
\genfrac{}{}{}{0}{(q)_{n+m}}{(q)_n(q)_m}&\text{if $n,m\ge0$,}\\
  0&\text{otherwise,}
\end{cases}
\end{equation}
where $(q)_n=\prod_{i=1}^n (1-q^i)$ with $(q)_0=1$.

Note that $m_{p-1}=0$ in each case.
The matrix $\boldC=\boldC^{(p-2)}$,
where $\boldC^{(n)}$ is the
$n\times n$ tri-diagonal matrix with entries $C_{ij}$ for
$1\le i,j\le n$ given by, when the indices are in this range,
$C_{i,i}=2$ and $C_{i,i\pm1}=-1$.
Thus, $\boldC^{(n)}$ is the Cartan matrix
of the finite-dimensional simple Lie algebra $A_n$.

In Table \ref{Tab:ABF}, it is seen that certain cases of
$r$ and $s$ are excluded.
This is solely to prevent repetition of (essentially)
identical expressions in the extreme cases where
$r=1$, $r=p-1$, $s=1$ and $s=p$.
For these values, the equivalence between the various cases
of \eqref{Eq:ABFferms} is readily seen after, perhaps,
shifting the value of $n_{p-1}$.
For \eqref{Eq:ABFferms} to be correct in the excluded cases
where $\ell=0$, it would also be necessary to set $m_{0}=0$.
In this paper, we obtain various expressions of a form similar to
\eqref{Eq:ABFferms}, and corresponding comments apply
(in these later expressions, however, care would need to be taken
in the excluded cases for which $\ell=0$, because $m_{0}$ might already
refer to a non-zero value).

Expressions (\ref{Eq:ABFferms}a) and (\ref{Eq:ABFferms}c)
(meaning (\ref{Eq:ABFferms}) with entries (a) and (c)
of Table \ref{Tab:ABF} respectively) were first
conjectured in \cite{KKMM1993b}, one being obtained
from the other using the equivalence
$\chi^{p,p+1}_{r,s}=\chi^{p,p+1}_{p-r,p+1-s}$.
The expressions (\ref{Eq:ABFferms}b) and (\ref{Eq:ABFferms}d) were
conjectured in \cite{Melzer1994}, with again one being obtained
from the other using the above equivalence.
In addition, \cite{Melzer1994} gave proofs for all four expressions
in the cases of $p=3$ and $p=4$.
A proof of (\ref{Eq:ABFferms}a) and (\ref{Eq:ABFferms}c) in the special
case where $s=1$ was given in \cite{Berkovich1994} using the
technique of \lq\lq telescopic expansion\rq\rq.
A similar technique was used later in \cite{Schilling1996a} to give
a proof of all four expressions for each $\chi^{p,p+1}_{r,s}$.
Independently, a
complete proof of (\ref{Eq:ABFferms}a) and (\ref{Eq:ABFferms}c)
was given in \cite{Warnaar1996a,Warnaar1996b}
using a lattice path construction.
A different lattice path proof of all four expressions
was given in \cite{FodaWelsh1999}.
We rework and refine this latter proof here.

\subsection{Non-unitary Melzer-type identities}

To give our fermionic expressions for $M(k,2k\pm1)$ in a uniform manner,
we extend the definition \eqref{Eq:Rocha} for $\chi^{p,p'}_{r,s}$
so that it applies for $p\in\HZZ$.
The symmetry relation
\begin{equation}\label{Eq:RochaSwitch1}
\chi^{p,p'}_{r,s}
=
\chi^{p'/2,2p}_{s/2,2r},
\end{equation}
which is a direct consequence of \eqref{Eq:Rocha},
then transforms the cases where $p\in\ZZph$ and $s$ and $p'$ are even,
to genuine minimal-model characters.

The expressions make use of a modified version of the $q$-binomial
defined by (compare with \eqref{Eq:qBinomialDef})
\begin{equation}\label{Eq:qModBinomialDef}
\qbinom{n+m}{n}'=
\begin{cases}
\genfrac{}{}{}{0}{(q)_{n+m}}{(q)_n(q)_m}&\text{if $n,m\ge0$,}\\
  1&\text{if $n=0$ and $m=-1$,}\\
  0&\text{otherwise.}
\end{cases}
\end{equation}
Also, they are expressed in terms of  the \emph{parity}
vectors $\boldQ^{(c,j)}$ and $\boldR^{(c,j)}$,
defined for $1\le c\le j+1$ by setting
\begin{equation}\label{Eq:QparDef}
\boldQ^{(c,j)}=(Q^{(c,j)}_1,Q^{(c,j)}_2,\ldots,Q^{(c,j)}_j)
\quad\text{with}\quad
Q^{(c,j)}_i=
\pabs{c-1-i}\bmod2;
\end{equation}
and setting
\begin{equation}\label{Eq:RparDef}
\boldR^{(c,j)}=(R^{(c,j)}_1,R^{(c,j)}_2,\ldots,R^{(c,j)}_j)
\quad\text{with}\quad
R^{(c,j)}_i=\bigl(\pabs{i+c-j-1}+c+1\bigr)\bmod2.
\end{equation}
For example,
\begin{equation}
\boldQ^{(8,12)}=(0,1,0,1,0,1,0,0,0,0,0,0),\;
\boldR^{(7,12)}=(0,0,0,0,0,0,1,0,1,0,1,0),\;
\boldR^{(6,12)}=(1,1,1,1,1,1,1,0,1,0,1,0).
\end{equation}
In each of these cases, a tilde indicates that the first
component should be dropped.
Therefore, for example,
$\tboldQ{}^{(8,12)}=(1,0,1,0,1,0,0,0,0,0,0)$.

\begin{theorem}\label{Thrm:HLferms}
Let $t\in\HZZ$ and $a,r\in\ZZ$ with $1\le a\le t$ and $1\le r<t$,
and set $\boldC=\boldC^{(2t-3)}$.
Then, we have four expressions for $\chi^{t,2t+1}_{r,2a}$,
distinguished by the
values of $\ell$, $\Delta_i$,
$\boldQL=(\QL_1,\ldots,\QL_{2t-2})$ and $\boldQR=(\QR_1,\ldots,\QR_{2t-2})$,
and the type of $q$-binomial {$[\,]_q^*$},
as given in Table~\ref{tabHL}.
These expressions are then:
\begin{equation}\label{Eq:HLferms}
\begin{split}
\chi^{t,2t+1}_{r,2a}
&=
q^{-\frac12(a-r)(a-r-\frac12)}
\hskip-1.5mm
\sum_{\tsboldn\in\ZZ^{2t-3}_{\ge0}}
\hskip-1mm
q^{\frac18\sboldm\sboldC\sboldm^T
   -\frac14 m_{\ell}
   +\frac12\tsboldn\cdot\tsboldQR}
\frac1{(q)_{\hat m_1}}
\prod_{i=2}^{2t-3}
\qbinom{n_i+\hat m_i}{n_i}^*
,\\
\text{where}\quad
 m_i=2&\sum_{i<k<2t-1}(k-i)n_k -\Delta_i
\quad\text{and}\quad
 \hat m_i=\frac12(m_i-\QL_i-\QR_i)
\quad\text{for}\quad
 1\le i\le 2t-2.
\end{split}
\end{equation}
Here, each sum is over all non-negative integer vectors
$\tboldn=(n_2,n_3,\ldots,n_{2t-2})$, with
$\boldm=(m_1,m_2,\ldots,m_{2t-3})$ obtained from $\tboldn$ as indicated.
Note that $m_{2t-2}=0$ in each case, and for each $i<2t-2$,
$\hat m_i$ is an integer equal to either
$\frac12m_i$, $\frac12(m_i-1)$ or $\frac12m_i-1$.

\begin{table}[ht]
\caption{Parameters for the four cases of the expressions \eqref{Eq:HLferms}.} 
\vskip.2cm
\label{tabHL}
\begin{center}
\begin{tabular}{|l|c|c|c|c|c|c|c|}\hline
\mystrut
 & $r$ & $a$ & $\ell$ & $\Delta_i$ & $\boldQL$ & $\boldQR$ &$[\,]_q^*$\\
\hline\hline
\mystrut {\rm (a)} & $\ne1$ & $\ne 1$ & $2a-2$
   & $\pabs{2a-2-i}+\pabs{2r-1-i}$
   & $\boldQ^{(2a-1,2t-2)}$ & $\boldQ^{(2r,2t-2)}$&$[\,]_q$\\
\hline
\mystrut {\rm (b)} & $\ne1$ & $\ne t$ & $2t-2a$
   & $\pabs{2t-2a-i}+\pabs{2r-1-i}+2t-2-i$
   & $\boldR^{(2a-1,2t-2)}$ & $\boldQ^{(2r,2t-2)}$&$[\,]_q'$\\
\hline
\mystrut {\rm (c)} & $\ne t-\frac12$ & $\ne t$ & $2t-2a$
   & $ \pabs{2t-2a-i}+\pabs{2t-2r-i}$
   & $\boldR^{(2a-1,2t-2)}$ & $\boldR^{(2r-1,2t-2)}$&$[\,]_q'$\\
\hline
\mystrut {\rm (d)} & $\ne t-\frac12$ & $\ne 1$ & $2a-2$
   & $\pabs{2a-2-i}+\pabs{2t-2r-i}+2t-2-i$
   & $\boldQ^{(2a-1,2t-2)}$ & $\boldR^{(2r-1,2t-2)}$&$[\,]_q'$\\
\hline
\end{tabular}
\end{center}
\end{table}
\end{theorem}

In the above theorem, the $t\in\ZZ$ cases encompass all the characters
of the minimal models $M(k,2k+1)$, also giving four expressions for
each character
(except for the $a=1,$ $a=t$ and $r=1$ cases,
for which there are one or two expressions).
For the characters $\chi^{k,2k+1}_{r,s}$ with $s$ even, this is
immediate. Because $\chi^{k,2k+1}_{r,s}=\chi^{k,2k+1}_{k-r,2k+1-s}$,
the cases of odd $s$ are also covered.
The $t\in\ZZph$ cases, in fact, encompass all the characters of the
minimal models $M(k,2k-1)$, giving four expressions for each character
(except for the $a=1$, $a=t$, $r=1$ and $r=t-\frac12$ cases).
This follows because, from \eqref{Eq:RochaSwitch1},
\begin{equation}\label{Eq:RochaSwitch}
\chi^{t,2t+1}_{r,2a}
=
\chi^{t+\frac12,2t}_{a,2r},
\end{equation}
and thus if $k=t+\frac12$, we obtain an $M(k,2k-1)$ character.
Again, all characters are covered because 
$\chi^{k,2k-1}_{a,2r}=\chi^{k,2k-1}_{k-a,2k-1-2r}$.

The cases $a=r=1$ of (\ref{Eq:HLferms}c) 
are implicit in
\cite[eqn.~(9.4)]{BMcP1998}
(see \cite[IVD]{Warnaar1999} for a proof).
The expressions obtained in \cite{JacobMathieu2007} for the characters
of $M(k,2k+1)$ are equivalent to the expressions (\ref{Eq:HLferms}d)
(however, the proof in \cite{JacobMathieu2007} is incomplete).
Expressions for the $M(k,2k-1)$ characters, also equivalent to
(\ref{Eq:HLferms}d), were obtained similarly in \cite{Blondeau-Fournier2011}.

\subsection{Remarks on the structure of these non-unitary
                 Melzer-type identities}
\label{Sec:Intro-Rem}

A number of aspects of the four expressions \eqref{Eq:HLferms} for
$M(k,2k\pm1)$ characters are particularly noteworthy
when compared with previously obtained fermionic expressions for
minimal model Virasoro characters.

(1).
Our first observation is that each expression of \eqref{Eq:HLferms} for
$M(k,2k+1)$, when considered in terms of the excitations of
quasiparticles, involves a restriction of the corresponding
expression of \eqref{Eq:ABFferms} for the unitary model
$M(2k-1,2k)$, with the quasiparticles permitted only to inhabit alternate
excited states.
Likewise, each expression of \eqref{Eq:HLferms} for
$M(k,2k-1)$ may be considered a similar restriction of the
corresponding expression of \eqref{Eq:ABFferms} for
the unitary model $M(2k-2,2k-1)$.
Also, because each $\hat m_i$ is an integer equal to either
$\frac12m_i$, $\frac12(m_i-1)$ or $\frac12m_i-1$,
the number of moves that the $i$th quasiparticle can make is then
about 1/2 that for the corresponding unitary case.

(2).
These expressions \eqref{Eq:HLferms} differ from the fermionic expressions
for the same $M(k,2k\pm1)$ characters obtained
\cite{BMc1996lmp,BMcS1998,FodaWelsh2000,JacobMathieu2009}
via the RSOS statistical models.
In particular, they encode the excitations of a different number of
species of quasiparticles: the $M(k,2k+1)$ and $M(k,2k-1)$
expressions given in Theorem \ref{Thrm:HLferms} involve
$2k-3$ and $2k-4$ species of quasiparticles respectively
(the number of species being counted by the number of entries
of the vector $\tboldn$), 
whereas the expressions derived from the corresponding RSOS models 
involve $k-1$ and $k-2$ species of quasiparticles respectively.

(3).
The expressions \eqref{Eq:ABFferms} give four distinct expressions
for each unitary character $\chi^{p,p+1}_{r,s}$
(except for the exceptional cases where $r=1$, $r=p-1$,
$s=1$ or $s=p$, where there are fewer than four expressions).
These expressions are not trivially equivalent.
However, the knowledge that
$\chi^{p,p'}_{r,s}=\chi^{p,p'}_{p-r,p'-s}$
shows that the expressions (\ref{Eq:ABFferms}a) and (\ref{Eq:ABFferms}c)
are indeed equivalent.
The expressions (\ref{Eq:ABFferms}b) and (\ref{Eq:ABFferms}d) are also
equivalent for the same reason.
In the unitary case, the equivalence
$\chi^{p,p+1}_{r,s}=\chi^{p,p+1}_{p-r,p+1-s}$
can be seen to arise from the up-down symmetry of the ABF paths.

In the case of the half-lattice paths, however,
the up-down symmetry is lost because
the valley restriction is not preserved under up-down reflection.
Thus, the equivalence of expressions
(\ref{Eq:HLferms}a) and (\ref{Eq:HLferms}c)
cannot be attributed to the up-down symmetry
that accounts for the corresponding equivalence in the unitary case.
A similar statement is true for the expressions
(\ref{Eq:HLferms}b) and (\ref{Eq:HLferms}d).

(4).
The modified $q$-binomials need to be employed in some cases.
Note \ref{Note:Mod} gives specific information, and indicates,
in particular, why the usual $q$-binomial is sufficient
in  (\ref{Eq:HLferms}a).

(5).
In addition to the four fermionic expressions for $\chi^{t,2t+1}_{r,2a}$
given in Theorem \ref{Thrm:HLferms},
a further two fermionic expressions arise naturally from our analysis.
These, however, are for the sum
$\chi^{t,2t+1}_{r,2a}+q^{a-r}\chi^{t,2t+1}_{r-1,2a}$
of two characters.

(6).
The key tool in the derivation of the fermionic formulae is
a combinatorial transform that increases by one half-integer unit
the vertical size of the path grid, thereby relating the character
of a model with parameter $t$ to the one with $t+\tfrac12$.
It thus interpolates between the two classes of models
$M(k,2k-1)$ and $M(k,2k+1)$.
Specifically, it provides the following ``combinatorial flow''
between minimal models
\begin{equation}
M(2,5)\xrightarrow{\phantom{(2,1)}}
M(3,5)\xrightarrow{\phantom{(1,5)}}
M(3,7)\xrightarrow{\phantom{(2,1)}}
M(4,7)\xrightarrow{\phantom{(1,5)}}
M(4,9)\xrightarrow{\phantom{(2,1)}}
M(5,9)\xrightarrow{\phantom{(1,5)}}\cdots.
\end{equation}
This is discussed further in the following subsection.

\subsection{Relation with integrable perturbations}
\label{Sec:Intro-Rel}

Remark (2) in Section \ref{Sec:Intro-Rem}, concerning the number of
quasiparticles, is a clear indication that these fermionic forms are
associated with an integrable perturbation \cite{Zamolodchikov1989}
other than $\phi_{1,3}$.
The model $M(p,p')$ perturbed by the $\phi_{1,3}$ field is described
by the restricted sine-Gordon model
\cite{LeClair1989,Smirnov1989,Smirnov1990},
whose lattice regularization is the RSOS$(p,p')$ model.
The underlying structure is that of the affine Lie algebra $A_1^{(1)}$,
which manifests itself in the path description through
there being two possible orientations for the edges,
two being the dimension of the fundamental representation of $A_1$.

The other integrable perturbations, $\phi_{1,2}$ and the pair
$\phi_{2,1}/\phi_{1,5}$ (depending upon which one is relevant),
are associated with the affine Lie algebra $A_2^{(2)}$.
The corresponding field theory is the restricted
Bullough-Dodd-Zhiber-Mikhailov-Shabat model
(with imaginary coupling constant) \cite{Smirnov1991,Efthimiou1993},
a Toda-type model which comes in two versions due to the
asymmetry between the two simple roots of $A_2^{(2)}$.

\begin{figure}[ht]
\caption{Allowed edge orientations between two integer heights ($c$ is integer)}
\label{Fig:Ori}
\begin{center}
\psset{yunit=0.32cm,xunit=0.28cm}
\begin{pspicture}(0,1.0)(40,6.5)
\rput[bl](0,0){
%dashed grid
\psset{linewidth=0.25pt,linestyle=dashed, dash=2.5pt 1.5pt,linecolor=gray}
\psline{-}(0,1)(5,1) \psline{-}(0,2)(5,2) \psline{-}(0,3)(5,3)
 \psline{-}(0,4)(5,4) \psline{-}(0,5)(5,5)
\psline{-}(1,1)(1,5) \psline{-}(2,1)(2,5) \psline{-}(3,1)(3,5)
\psline{-}(4,1)(4,5) %\psline{-}(5,1)(5,5)
%path
\psset{linewidth=0.7pt,fillstyle=none,linestyle=solid,linecolor=black}
\psline(1,3)(3,5)%(4,2)(5,1)
%annotate
\rput(-5,3){\scriptsize Allowed:}
\rput(-.5,3){\scriptsize $c$}
\rput(-.9,5){\scriptsize $c\!+\!1$}
\rput(-.9,1){\scriptsize $c\!-\!1$}
}
\rput[bl](10,0){
%dashed grid
\psset{linewidth=0.25pt,linestyle=dashed, dash=2.5pt 1.5pt,linecolor=gray}
\psline{-}(0,1)(5,1) \psline{-}(0,2)(5,2) \psline{-}(0,3)(5,3)
 \psline{-}(0,4)(5,4) \psline{-}(0,5)(5,5)
\psline{-}(1,1)(1,5) \psline{-}(2,1)(2,5) \psline{-}(3,1)(3,5)
\psline{-}(4,1)(4,5) %\psline{-}(5,1)(5,5)
%path
\psset{linewidth=0.7pt,fillstyle=none,linestyle=solid,linecolor=black}
\psline(1,3)(3,1)%(5,1)
}
\rput[bl](20,0){
%dashed grid
\psset{linewidth=0.25pt,linestyle=dashed, dash=2.5pt 1.5pt,linecolor=gray}
\psline{-}(0,1)(5,1) \psline{-}(0,2)(5,2) \psline{-}(0,3)(5,3)
 \psline{-}(0,4)(5,4) \psline{-}(0,5)(5,5)
\psline{-}(1,1)(1,5) \psline{-}(2,1)(2,5) \psline{-}(3,1)(3,5)
\psline{-}(4,1)(4,5)
%path
\psset{linewidth=0.7pt,fillstyle=none,linestyle=solid,linecolor=black}
\psline(1,3)(2,4)(3,3)
}

\rput[bl](36,0){
%dashed grid
\psset{linewidth=0.25pt,linestyle=dashed, dash=2.5pt 1.5pt,linecolor=gray}
\psline{-}(0,1)(5,1) \psline{-}(0,2)(5,2) \psline{-}(0,3)(5,3)
 \psline{-}(0,4)(5,4) \psline{-}(0,5)(5,5)
\psline{-}(1,1)(1,5) \psline{-}(2,1)(2,5) \psline{-}(3,1)(3,5)
\psline{-}(4,1)(4,5)
%path
\psset{linewidth=0.7pt,fillstyle=none,linestyle=solid,linecolor=black}
\psline(1,3)(2,2)(3,3)
%annotate
\rput(-4,3){\scriptsize Forbidden:}
}
\end{pspicture}
\end{center}
\end{figure}
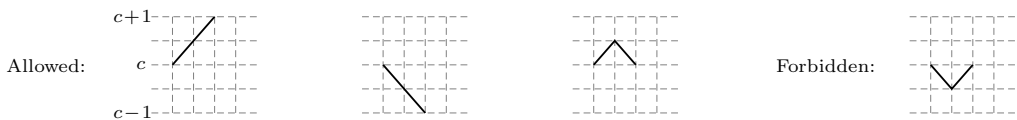

Statistical models associated with $A_2^{(2)}$ are described in
\cite{IzerginKorepin1981,Kuniba1991,WNS1993},
with \cite{KKMMNN1992} dealing more generally with the case
of an arbitrary affine Lie algebra.
For the $A_2^{(2)}$ cases, application of
Baxter's corner transfer matrix method \cite{Baxter1982}
to these models gives rise to paths that have three possible
directions at each step.
In \cite{KKMMNN1992}, their
paths are shown to provide a realisation of the crystal graph of
the level 1 representation of $A^{(2)}_2$,
with the three directions corresponding to the three nodes of the
relevant perfect crystal
(and corresponding to the dimension of the defining representation of $A_2$).
The half-lattice paths may be obtained from these paths
by splitting each of the segments in half:
each NE (resp.~SE) segment is split into two NE (resp.~SE) segments of
half the length, while each E segment is split into a half-length
NE segment followed by a half-length SE segment.
This yields the three possible shapes on the half-lattice
depicted on the left of Figure \ref{Fig:Ori}.
Note that the forbidden fourth shape there, a half-length SE segment followed
by a half-length NE segment, does not arise:
this corresponds to the restriction on valleys in our
definition of the half-lattice paths.%
\footnote{
In this paper, we are interested in half-lattice paths
whose heights are restricted: when recast back in terms
of lattice paths with segments orientated in the NE, SE or E directions,
such paths are sometimes referred to as Motzkin or Riordan paths
(these correspond to whether, in Section \ref{Sec:HL},
$t\in\ZZph$ or $t\in\ZZ$ respectively).
}
Although the redefinition seems somewhat arbitrary
from this point of view,
the weighting for the half-lattice paths is much simpler,
being similar to that for the original ABF paths
(compare \eqref{Eq:ABFwtsDef} and \eqref{Eq:WtsDef}).
For this reason, the half-lattice paths can be analysed using
combinatorial techniques similar to those developed for the ABF paths.

We also note that the `three edge-orientations' pattern is
manifest in the bosonic form for the finitized characters,
which is expressed in terms of $q$-trinomial coefficients
(see Section \ref{Sec:BosonicGF}).

In order to pinpoint exactly the perturbation associated with
these new paths, we rely on our last remark in Section \ref{Sec:Intro-Rem}.
It displays a combinatorial flow that is precisely the reverse
of the $\phi_{2,1}/\phi_{1,5}$ renormalization-group flow observed in
\cite{Martins1992,Martins1993} (see also \cite{RST1996,DDT2000}):
\begin{equation}
\cdots \xrightarrow{(1,5)}
M(5,9)\xrightarrow{(2,1)}
M(4,9)\xrightarrow{(1,5)}
M(4,7)\xrightarrow{(2,1)}
M(3,7)\xrightarrow{(1,5)}
M(3,5)\xrightarrow{(2,1)}
M(2,5),
\end{equation}
where the label above each arrow indicates the perturbating field.
The correspondence between the two types of flows selects thus
the $\phi_{2,1}/\phi_{1,5}$ perturbation.
That such a correspondence is reliable is supported by the fact that
for ABF paths, the analogous combinatorial transform
(called the $\Tran$-transform and defined
in Section \ref{Sec:CombTran} below)
relates the paths of the RSOS$(p-1,p)$ model to those of RSOS$(p,p+1)$,
which is the reverse of the $\phi_{1,3}$ renormalization-group flow
between unitary models \cite{Zamolodchikov1987,LudwigCardy1987}.
Moreover, the latter combinatorial transform is decomposed
in \cite{FLPW2000} into a sequence `$\D\B\D$' of three transformations
and also applied to non-unitary models.
Their combined effect is to transform paths between the models
in the following sequence:
\begin{equation}\label{Eq:Cflow}
\text{RSOS}(p,p')\xrightarrow{\hskip3mm\D\hskip3mm}
  \text{RSOS}(p'-p,p')\xrightarrow{\hskip3mm\B\hskip3mm}
  \text{RSOS}(p'-p,2p'-p)\xrightarrow{\hskip3mm\D\hskip3mm}
  \text{RSOS}(p',2p'-p).
\end{equation}
Correspondingly, the renormalization-group flow along the
$\phi_{1,3}$ perturbation relates precisely the
minimal models $M(p',2p'-p)$ and $M(p,p')$ \cite{Lassig1992,Ahn1992}.

\subsection{Relation to previous works}

Half-lattice paths were originally introduced in \cite{JacobMathieu2007}
as a description of the $M(p,2p+1)$ minimal models.
However, no precise \emph{a priori} argument pointed toward the
correspondence between these particular paths and the states of
irreducible modules for these specific non-unitary models.
We remedied this in \cite{BfMW2010} by providing a bijection between
these half-integer paths and the paths for the RSOS$(p,2p+1)$ model.

In this original formulation of the half-integer paths
\cite{JacobMathieu2007,BfMW2010},
the peaks were forced to have integer heights.
In the conclusion of \cite{BfMW2010}, we pointed out that
on modifying the restriction so that peaks are forced to have
half-integer heights,
one observes a correspondence with the states in the
irreducible modules of the $M(p+1,2p+1)$ minimal models.
This observation was put on firm foundations in \cite{BfMW2012} by
providing a bijection between these paths and the RSOS$(p+1,2p+1)$ paths.

In \cite{BfMW2012}, it was pointed out that half-integer paths pertaining
to these two classes of models can be treated uniformly by redefining
the paths to be the up-down reflections of the original ones.
In this way, the restriction becomes identical for the two cases:
the valleys must lie at integer heights.
This is the description used in the present work.

\subsection{Organization of the article}

Section \ref{Sec:CombTran} is devoted to reformulating the
combinatorial transforms used in \cite{FodaWelsh1999} to prove
Melzer's identities.
In particular, three combinatorial transforms,
$\Tran_1$, $\Tran_2$ and $\Tran_3$, that act on ABF lattice paths,
are introduced. These are combined to define the $\Tran$-transform.
In Section \ref{Sec:ABFfin}, it is shown how to use the
$\Tran$-transform to rederive the four fermionic expressions for finitized
$M(p,p+1)$ characters.
These immediately imply the expressions \eqref{Eq:ABFferms}
for the $M(p,p+1)$ characters themselves.
This novel rederivation of Melzer's identities is an essential
ingredient of this work because its half-lattice generalization
proceeds along similar lines.

Half-lattice paths are defined in Section \ref{Sec:HL}.
In the same section, we also present bosonic and fermionic expressions 
for the (polynomial) generating functions for half-lattice paths
of finite length.
In the limit of infinite length paths, these yield the bosonic character
formula \eqref{Eq:Rocha} for $M(k,2k\pm1)$ characters,
and the fermionic expressions given in Theorem \ref{Thrm:HLferms}.
Section \ref{Sec:Refine} is devoted to proving
the fermionic expressions for the finite length half-lattice paths.
There, the paths are regarded as restricted
rescaled versions of the ABF paths.
Thereupon, a minor variant of the $\Tran$-transform enables
an approach parallel to that of Section \ref{Sec:ABFfin} to be used.
Finally, the bosonic expressions for finite length half-lattice
paths are derived in Section \ref{Sec:BosonicProof},
making use of various identities for $q$-trinomial coefficients
given in the Appendix.

\section{Combinatorial transforms for ABF paths}
\label{Sec:CombTran}

The proof of Melzer's \cite{Melzer1994} four expressions for the
finitized ABF characters that was developed in \cite{FodaWelsh1999},
using ideas of Bressoud \cite{Bressoud1989},
employs combinatorial methods to transform between
generating functions for different sets of paths.
Thereby, the generating function for a trivial case
(essentially the $p=2$ case) is used
to obtain Melzer's expressions in the case of general $p$.
In this section, we rework this proof with a view to refining it
to apply to the case of half-lattice paths.
Although the method presented in this section differs from the
proof given in \cite{FodaWelsh1999}, it is, in essence,
a dual version of that proof, obtained naturally through mapping
$q\to q^{-1}$.

\subsection{ABF paths}
\label{Sec:ABFpaths}

We define an ABF path $\habf$ of length $L$ to be a finite sequence
$\habf=(h_{-1},h_0,h_1,h_2,\ldots,h_L,h_{L+1})$ 
satisfying $h_i\in\ZZ$ and
$|h_{i}-h_{i-1}|=1$ for $0\le i\le L+1$.
An ABF path $\habf$ is said to be $(f,g)$-restricted if
$f\le h_i\le g$ for $0\le i\le L$
(note that this doesn't apply to $h_{-1}$ and $h_{L+1}$).
For $a,b,p,L\in\ZZ$ and $e,f\in\{0,1\}$,
define $\ABFpabLef$ to be the set of all ABF paths
$\habf$ of length $L$ that are $(1,p)$-restricted
with $h_0=a$, $h_L=b$, $h_{-1}=a+1-2e$, $h_{L+1}=b+1-2f$.
These paths are a minor variant on those originally obtained
in \cite{ABF1984}, and are special cases of the RSOS paths
discussed in \cite{FB1985,FLPW2000}.

The \emph{path picture} of an ABF path $\habf\in\ABFpabLef$
is obtained by linking the vertices $(0,\habf_0)$, $(1,\habf_{1})$,
$(2,\habf_2),\ldots,(L,\habf_L)$ on the plane.
The values of $e$ and $f$ serve to
specify a path presegment and postsegment respectively;
the presegment extends between $(-1,h_{-1})$ and $(0,h_0)$,
while the postsegment extends between $(L,h_L)$ and $(L+1,h_{L+1})$.
The presegment is then in the SE direction if $e=0$,
and in the NE direction if $e=1$;
the postsegment is in the NE direction if $f=0$,
and in the SE direction if $f=1$.

A vertex $(i,\hh_i)$ for $0\le i\le L$ is said to be a peak,
a valley, straight-up or straight-down, depending on whether the
pair of edges that neighbour $(i,\hh_i)$ in this path picture are
in the NE-SE, SE-NE, NE-NE, or SE-SE directions respectively.
Defining paths with presegments and postsegments is convenient
because the manipulations of the ABF paths that we describe below
depend on the nature of the vertices at $(0,a)$ and $(L,b)$
that are thus determined.

The weight $\unwt(\habf)$ of a length $L$ ABF path $\habf$
is defined by
\begin{equation}\label{Eq:ABFwtsDef}
\unwt(\habf)=\frac14 \sum_{i=1}^{L}
i\, |\habf_{i+1}-\habf_{i-1}|.
\end{equation}
The weight $\unwt(\habf)$ is thus half the sum of the $i\in\ZZ$
for which $(i,h_i)$ is a straight vertex.

\begin{lemma}\label{Lem:wtSwitch}
Let $1\le a,b\le p$ and $L\ge0$.
Then define the four paths
$h^{(0,0)}$, $h^{(1,0)}$, $h^{(0,1)}$, $h^{(1,1)}$,
such that each $h^{(e,f)}\in\ABFpabLef$ and
$h^{(e,f)}_i=h^{(e',f')}_i$ for $0\le i\le L$.
Then for $e,f\in\{0,1\}$:
\begin{enumerate}
\item $\unwt(\habf^{(0,f)})=\unwt(\habf^{(1,f)})$;
\item If $b=1$ then $\unwt(\habf^{(e,1)})=\unwt(\habf^{(e,0)})+\frac12L$;
\item If $b=p$ then $\unwt(\habf^{(e,0)})=\unwt(\habf^{(e,1)})+\frac12L$.
\end{enumerate}
\end{lemma}
\Proof The first expression is immediate because, by \eqref{Eq:ABFwtsDef},
changing the direction of the presegment does not affect the weight.
The other two expressions are immediate for $L=0$.
In the case where $L>0$ and $b=1$, necessarily $h^{(e,f)}_{L-1}=2$ and so,
in \eqref{Eq:ABFwtsDef}, the $i=L$ term contributes $L/2$ to the
weight $\unwt(\habf^{(e,1)})$, but nothing to the weight
$\unwt(\habf^{(e,0)})$.  The first expression follows.
In the case where $L>0$ and $b=p$, necessarily $h^{(e,f)}_{L-1}=p-1$,
whence a similar argument gives the second expression.
\cqfd

For $a,b,p,L\in\ZZ$ and $e,f\in\{0,1\}$,
we define the path generating functions
\begin{equation}\label{Eq:ABFgfDef}
\ABFGFe{p}{a,b}{e,f}(L)=
\ABFGFe{p}{a,b}{e,f}(L;q)=
\sum_{\habf\in\ABFpabLef} q^{\unwt(\habf)}.
\end{equation}

Note that the sets
$\ABFsetef{p}{a,b}{0,0}(L)$,
$\ABFsetef{p}{a,b}{1,0}(L)$,
$\ABFsetef{p}{a,b}{0,1}(L)$,
and
$\ABFsetef{p}{a,b}{1,1}(L)$
are all trivially in bijection with one another,
with, if $h$ belongs to one of these sets, the corresponding
element $h'$ of another defined by
$h_i=h'_i$ for $0\le i\le L$.
This observation, together with
Lemma \ref{Lem:wtSwitch}(1) and \eqref{Eq:ABFgfDef}, implies that
\begin{equation}\label{Eq:ABFgfsComp}
\ABFGFe{p}{a,b}{0,f}(L)=
\ABFGFe{p}{a,b}{1,f}(L)
\end{equation}
for $f\in\{0,1\}$.

The following bosonic expression for $\ABFGFe{p}{a,b}{e,f}(L)$
was obtained in \cite{ABF1984}:
\begin{equation}\label{Eq:ABFbosonic}
\begin{split}
\ABFGFe{p}{a,b}{e,f}(L)
&
=q^{\frac14(a-b)(a-b-1+2f)}
\sum_{\lambda=-\infty}^\infty
\biggl(
q^{\lambda (p+1)(\lambda p+b-f)-\lambda pa}
\qbinom{L}{\frac{L+a-b}{2}-(p+1)\lambda}
\\
&\hskip60mm
-q^{(\lambda p+b-f)(\lambda p+\lambda +a)}
\qbinom{L}{\frac{L-a-b}{2}-(p+1)\lambda}
\biggr)
\end{split}
\end{equation}
for $e\in\{0,1\}$.
This immediately leads to
\begin{equation}\label{Eq:ABFbosonicLim}
\lim_{L\to\infty}
\ABFGFe{p}{a,b}{e,f}(L)
=
q^{\frac14(a-b)(a-b-1+2f)}\,
\chi^{p,p+1}_{b-f,a}
\end{equation}
for $e\in\{0,1\}$.
The expression on the right may then be viewed as the generating function
for infinite length ABF paths that are $(1,p)$-restricted and
eventually oscillate between heights $b$ and $b+1-2f$.

In what follows, we obtain fermionic expressions
for $\ABFGFe{p}{a,b}{e,f}(L)$.
Using \eqref{Eq:ABFbosonicLim}, these then yield the fermionic
expressions \eqref{Eq:ABFferms} for the characters $\chi^{p,p+1}_{r,s}$.

\subsection{Vertex word}
\label{Sec:VertexWord}

Each ABF path $h\in\ABFpabLef$ is conveniently encoded
by its \emph{vertex word} $\vword(h)$ which is a word
$v_0v_1v_2\cdots v_{L}$ of length $L+1$ in the symbols $N$ and $S$.%
\footnote{This notion of vertex word for RSOS paths was
originally described in \cite{Welsh2006}.}
The symbols in this word describe the sequence of vertices,
non-straight or straight, of $h$ read from left to right,
beginning with the vertex at $(0,a)$.
Thus $\vword(h)$ depends on the direction of the presegment of $h$,
and also its postsegment, and hence on the values of $e$ and $f$.

Given a word $v$ in the symbols $N$ and $S$, and values $a$ and $e$,
the ABF path $h$ such that $\vword(h)=v$ and $h_0=a$ and $h_{-1}=a+1-2e$
is readily determined by working from left to right
(the values of $b$ and $f$ are determined by $v$, $a$ and $e$).
Thus $h$ is uniquely determined by $\vword(h)$.

\begin{lemma}\label{Lem:wtStrike}
For $1\le a,b\le p$ and $e,f\in\{0,1\}$ and $L\ge0$, let
$\habf\in\ABFpabLef$.
Let the corresponding vertex word $\vword(h)$ have symbols $N$ at
positions $j_0,j_1,j_2,\ldots,j_k$.
Then:
\begin{equation}\label{Eq:wtStrike}
\unwt(\habf)
=
\frac14L(L+1)
-\frac12\sum_{i=0}^{k} j_i.
\end{equation}
\end{lemma}
\Proof
This follows because,
by \eqref{Eq:ABFwtsDef}, $\unwt(\habf)$ is half the
sum of the positions of the $S$ symbols in $\vword(h)$.
\cqfd

For $\habf\in\ABFpabLef$, we define
$\mnos(\habf)$ to be the number of symbols $S$ in $\vword(h)$.
We then define the generating functions
\begin{equation}\label{Eq:ABFMgfDef}
\ABFGFe{p}{a,b}{e,f}(L,m)=
\ABFGFe{p}{a,b}{e,f}(L,m;q)=
\sum_{\substack{\habf\in\ABFpabLef\\
                \mnos(\habf)=m}}
q^{\unwt(\habf)}.
\end{equation}

Note that if $\habf\in\ABFpabLef$ and $\habf'\in\ABFsetef{p}{a,b}{e',f'}(L)$
are such that $\habf_i=\habf'_i$ for $0\le i\le L$ then
$\mnos(\habf)$ and $\mnos(\habf')$ are not necessarily equal when
$e\ne e'$ or $f\ne f'$.
In particular, we have the following result:%
\footnote{Throughout this paper, we use the symbol ``$\equiv$''
to denote congruence modulo 2.}
\begin{lemma}\label{Lem:mSwitch}
Let $1\le a,b\le p$ and $L\ge0$.
Then define the four paths
$h^{(0,0)}$, $h^{(1,0)}$, $h^{(0,1)}$, $h^{(1,1)}$,
such that each $h^{(e,f)}\in\ABFpabLef$ and
$h^{(e,f)}_i=h^{(e',f')}_i$ for $0\le i\le L$.
Then for $e,f\in\{0,1\}$:
\begin{enumerate}
\item $\mnos(\habf^{(e,f)})\equiv L+e+f$;
\item If $a=1$ then $\mnos(\habf^{(0,f)})=\mnos(\habf^{(1,f)})-1$;
\item If $a=p$ then $\mnos(\habf^{(1,f)})=\mnos(\habf^{(0,f)})-1$;
\item If $b=1$ then $\mnos(\habf^{(e,0)})=\mnos(\habf^{(e,1)})-1$;
\item If $b=p$ then $\mnos(\habf^{(e,1)})=\mnos(\habf^{(e,0)})-1$.
\end{enumerate}
\end{lemma}
\Proof If $\vword(h^{(e,f)})$ contains $k+1$ symbols $N$ then
$e+f\equiv k$ because the path changes direction at each $N$.
The first case then follows because $L=k+\mnos(h^{(e,f)})$.
The other cases result from changing the direction of
the presegment or postsegment of the appropriate $\habf^{(e,f)}$.
\cqfd

\begin{lemma}\label{Lem:mSwitch2}
Let $1\le a,b\le p$ and $L\ge0$ and $e,f\in\{0,1\}$.
Then:
\begin{enumerate}
\item If $m\not\equiv L+e+f$ then
  $\ABFGFe{p}{a,b}{e,f}(L,m)=0$;
\item
  $\ABFGFe{p}{a,b}{e,f}(L)
   =\sum_{m\ge0} \ABFGFe{p}{a,b}{e,f}(L,m)$;
\item $\ABFGFe{p}{1,b}{1,f}(L,m)=\ABFGFe{p}{1,b}{0,f}(L,m-1)$;
\item $\ABFGFe{p}{p,b}{0,f}(L,m)=\ABFGFe{p}{p,b}{1,f}(L,m-1)$;
\item $\ABFGFe{p}{a,1}{e,1}(L,m)=q^{L/2}\ABFGFe{p}{a,1}{e,0}(L,m-1)$;
\item $\ABFGFe{p}{a,p}{e,0}(L,m)=q^{L/2}\ABFGFe{p}{a,p}{e,1}(L,m-1)$.
\end{enumerate}
\end{lemma}
\Proof
The first case follows from Lemma \ref{Lem:mSwitch}(1)
and the definition \eqref{Eq:ABFMgfDef}.
The second case follows from
the definitions \eqref{Eq:ABFgfDef} and \eqref{Eq:ABFMgfDef}.
The other cases follow from Lemmas \ref{Lem:wtSwitch} and
\ref{Lem:mSwitch} and the definition \eqref{Eq:ABFMgfDef}.
\cqfd

\begin{lemma}\label{Lem:Seed}
Let $1\le a,b\le p$ and $L,m\ge0$ and $e,f\in\{0,1\}$.
Then:
\begin{enumerate}
\item $\ABFGFe{p}{a,b}{e,f}(0,m)=\delta_{a,b}\,\delta_{m,|e-f|}$;
\item $\ABFGFe{1}{1,1}{e,f}(L,m)=\delta_{L,0}\,\delta_{m,|e-f|}$;
\item $\ABFGFe{p}{a,b}{e,f}(L,0)
             =\delta_{a-e,b-f}\,\delta_{(L+e+f)\bmod2,0}$.
\end{enumerate}
\end{lemma}
\Proof 
If $a\ne b$ then $\ABFsetef{p}{a,b}{e,f}(0)=\emptyset$.
On the other hand, if $a=b$ then $\ABFsetef{p}{a,b}{e,f}(0)$
contains a single element $\habf$,
indicated in Figure \ref{Fig:ZeroLenPaths}.
For this, $\vword(\habf)=N$ if $e=f$, and
$\vword(\habf)=S$ if $e\ne f$.
For these two cases, we then have $\mnos(\habf)=0$ and
$\mnos(h)=1$ respectively.
Then, after noting that $\unwt(\habf)=0$ in both cases,
the first result follows from the definition \eqref{Eq:ABFMgfDef}.

The second result follows from the first after it is noted
that $\ABFsetef{1}{1,1}{e,f}(L)=\emptyset$ for $L>0$.

For the third result, first note that if $\mnos(\habf)=0$
then the segments of the path $\habf$, together with its
presegment and postsegment, necessarily alternate in direction.
As indicated in Figure \ref{Fig:ZigZags},
there can be only one such path,
for which $e=f$ if and only if $L$ is even.
Furthermore, if $L$ is even then necessarily $a=b$,
and if $L$ is odd then necessarily $|a-b|=1$,
with $b=a-1$ if $e=1$ and $b=a+1$ if $e=0$.
The third result follows.
\cqfd

\begin{figure}[ht]
\caption{Paths $h$ of zero length}
\label{Fig:ZeroLenPaths}
\begin{center}
\psset{yunit=0.32cm,xunit=0.28cm}
\begin{pspicture}(0,-0.75)(28,4.5)
\rput[bl](0,0){
%dashed grid
\psset{linewidth=0.25pt,linestyle=dashed, dash=2.5pt 1.5pt,linecolor=gray}
\psline{-}(0,1)(4,1) \psline{-}(0,2)(4,2) \psline{-}(0,3)(4,3)
%axes
\psset{linewidth=0.25pt,fillstyle=none,linestyle=solid,linecolor=black}
\psline{-}(2,1)(2,3)
%path
\psset{linewidth=0.7pt,fillstyle=none,linestyle=solid,linecolor=black}
\psline(1.5,2.5)(2,2)(2.5,2.5)
%annotate
\rput(2.0,0.1){\scriptsize $e=f=0$}
\rput(2.0,-0.75){\scriptsize $v(h)=N$}
}
\rput[bl](8,0){
%dashed grid
\psset{linewidth=0.25pt,linestyle=dashed, dash=2.5pt 1.5pt,linecolor=gray}
\psline{-}(0,1)(4,1) \psline{-}(0,2)(4,2) \psline{-}(0,3)(4,3)
%axes
\psset{linewidth=0.25pt,fillstyle=none,linestyle=solid,linecolor=black}
\psline{-}(2,1)(2,3)
%path
\psset{linewidth=0.7pt,fillstyle=none,linestyle=solid,linecolor=black}
\psline(1.5,1.5)(2,2)(2.5,1.5)
%annotate
\rput(2.0,0.1){\scriptsize $e=f=1$}
\rput(2.0,-0.75){\scriptsize $v(h)=N$}
}
\rput[bl](16,0){
%dashed grid
\psset{linewidth=0.25pt,linestyle=dashed, dash=2.5pt 1.5pt,linecolor=gray}
\psline{-}(0,1)(4,1) \psline{-}(0,2)(4,2) \psline{-}(0,3)(4,3)
%axes
\psset{linewidth=0.25pt,fillstyle=none,linestyle=solid,linecolor=black}
\psline{-}(2,1)(2,3)
%path
\psset{linewidth=0.7pt,fillstyle=none,linestyle=solid,linecolor=black}
\psline(1.5,1.5)(2.5,2.5)
%annotate
\rput(2.0,0.1){\scriptsize $e=1$, $f=0$}
\rput(2.0,-0.75){\scriptsize $v(h)=S$}
}
\rput[bl](24,0){
%dashed grid
\psset{linewidth=0.25pt,linestyle=dashed, dash=2.5pt 1.5pt,linecolor=gray}
\psline{-}(0,1)(4,1) \psline{-}(0,2)(4,2) \psline{-}(0,3)(4,3)
%axes
\psset{linewidth=0.25pt,fillstyle=none,linestyle=solid,linecolor=black}
\psline{-}(2,1)(2,3)
%path
\psset{linewidth=0.7pt,fillstyle=none,linestyle=solid,linecolor=black}
\psline(1.5,2.5)(2.5,1.5)
%annotate
\rput(2.0,0.1){\scriptsize $e=0$, $f=1$}
\rput(2.0,-0.75){\scriptsize $v(h)=S$}
}
\end{pspicture}
\end{center}
\end{figure}
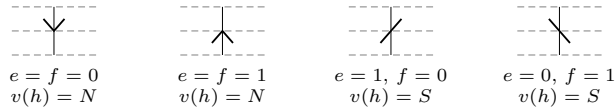
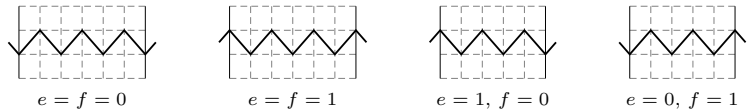
\begin{figure}[ht]
\caption{Paths without straight vertices}
\label{Fig:ZigZags}
\begin{center}
\psset{yunit=0.32cm,xunit=0.28cm}
\begin{pspicture}(-1,-0.5)(38,5.5)
\rput[bl](0,0){
%dashed grid
\psset{linewidth=0.25pt,linestyle=dashed, dash=2.5pt 1.5pt,linecolor=gray}
\psline{-}(0,1)(6,1) \psline{-}(0,2)(6,2)
\psline{-}(0,3)(6,3) \psline{-}(0,4)(6,4)
\psline{-}(1,1)(1,4) \psline{-}(2,1)(2,4) \psline{-}(3,1)(3,4)
\psline{-}(4,1)(4,4) \psline{-}(5,1)(5,4)
%axes
\psset{linewidth=0.25pt,fillstyle=none,linestyle=solid,linecolor=black}
\psline{-}(0,1)(0,4) \psline{-}(6,1)(6,4)
%path
\psset{linewidth=0.7pt,fillstyle=none,linestyle=solid,linecolor=black}
\psline(-0.5,2.5)(0,2)(1,3)(2,2)(3,3)(4,2)(5,3)(6,2)(6.5,2.5)
%annotate
\rput(3.0,0.1){\scriptsize $e=f=0$}
}
\rput[bl](10,0){
%dashed grid
\psset{linewidth=0.25pt,linestyle=dashed, dash=2.5pt 1.5pt,linecolor=gray}
\psline{-}(0,1)(6,1) \psline{-}(0,2)(6,2)
\psline{-}(0,3)(6,3) \psline{-}(0,4)(6,4)
\psline{-}(1,1)(1,4) \psline{-}(2,1)(2,4) \psline{-}(3,1)(3,4)
\psline{-}(4,1)(4,4) \psline{-}(5,1)(5,4)
%axes
\psset{linewidth=0.25pt,fillstyle=none,linestyle=solid,linecolor=black}
\psline{-}(0,1)(0,4) \psline{-}(6,1)(6,4)
%path
\psset{linewidth=0.7pt,fillstyle=none,linestyle=solid,linecolor=black}
\psline(-0.5,2.5)(0,3)(1,2)(2,3)(3,2)(4,3)(5,2)(6,3)(6.5,2.5)
%annotate
\rput(3.0,0.1){\scriptsize $e=f=1$}
}
\rput[bl](20,0){
%dashed grid
\psset{linewidth=0.25pt,linestyle=dashed, dash=2.5pt 1.5pt,linecolor=gray}
\psline{-}(0,1)(5,1) \psline{-}(0,2)(5,2)
\psline{-}(0,3)(5,3) \psline{-}(0,4)(5,4)
\psline{-}(1,1)(1,4) \psline{-}(2,1)(2,4) \psline{-}(3,1)(3,4)
\psline{-}(4,1)(4,4)
%axes
\psset{linewidth=0.25pt,fillstyle=none,linestyle=solid,linecolor=black}
\psline{-}(0,1)(0,4) \psline{-}(5,1)(5,4)
%path
\psset{linewidth=0.7pt,fillstyle=none,linestyle=solid,linecolor=black}
\psline(-0.5,2.5)(0,3)(1,2)(2,3)(3,2)(4,3)(5,2)(5.5,2.5)
%annotate
\rput(2.5,0.1){\scriptsize $e=1$, $f=0$}
}
\rput[bl](29,0){
%dashed grid
\psset{linewidth=0.25pt,linestyle=dashed, dash=2.5pt 1.5pt,linecolor=gray}
\psline{-}(0,1)(5,1) \psline{-}(0,2)(5,2)
\psline{-}(0,3)(5,3) \psline{-}(0,4)(5,4)
\psline{-}(1,1)(1,4) \psline{-}(2,1)(2,4) \psline{-}(3,1)(3,4)
\psline{-}(4,1)(4,4)
%axes
\psset{linewidth=0.25pt,fillstyle=none,linestyle=solid,linecolor=black}
\psline{-}(0,1)(0,4) \psline{-}(5,1)(5,4)
%path
\psset{linewidth=0.7pt,fillstyle=none,linestyle=solid,linecolor=black}
\psline(-0.5,2.5)(0,2)(1,3)(2,2)(3,3)(4,2)(5,3)(5.5,2.5)
%annotate
\rput(2.5,0.1){\scriptsize $e=0$, $f=1$}
}
\end{pspicture}
\end{center}
\end{figure}

\subsection{\texorpdfstring
             {$\Tran_1$-transform}
             {C1-transform}}
\label{Sec:B1trans}

In this section, we specify a method of transforming a path in
$\ABFpabLef$ to one in $\ABFsetef{p+1}{a',b'}{e,f}(L')$ for
certain $a',b',L'$.
This transform is referred to as a $\Tran_1$-transform.
It is closely related to the $\B_1$-transform of \cite{FLPW2000},
which itself was inspired by the ``volcanic uplift'' of \cite{Bressoud1989}.

\begin{figure}[ht]
\caption{Example of $\Tran_1$-transform (here $e=0$ and $f=1$)}
\vskip-0.2cm
\label{Fig:Btrans}
\begin{center}
\psset{yunit=0.32cm,xunit=0.28cm}
\begin{pspicture}(-1,0)(47,10)
\rput[bl](0,0.5){
%dashed grid
\psset{linewidth=0.25pt,linestyle=dashed, dash=2.5pt 1.5pt,linecolor=gray}
\psline{-}(0,2)(17,2) \psline{-}(0,4)(17,4) \psline{-}(0,6)(17,6)
\psline{-}(0,3)(17,3) \psline{-}(0,5)(17,5)
\psline{-}(1,1)(1,7) \psline{-}(2,1)(2,7) \psline{-}(3,1)(3,7)
\psline{-}(4,1)(4,7) \psline{-}(5,1)(5,7) \psline{-}(6,1)(6,7)
\psline{-}(7,1)(7,7) \psline{-}(8,1)(8,7) \psline{-}(9,1)(9,7)
\psline{-}(10,1)(10,7) \psline{-}(11,1)(11,7) \psline{-}(12,1)(12,7)
\psline{-}(13,1)(13,7) \psline{-}(14,1)(14,7) \psline{-}(15,1)(15,7)
\psline{-}(16,1)(16,7)
%axes
\psset{linewidth=0.25pt,fillstyle=none,linestyle=solid,linecolor=black}
\psline{-}(0,1)(17,1)
\psline{-}(0,7)(17,7)
\psline{-}(0,1)(0,7)
\psline{-}(17,1)(17,7)
%numeros
\rput(-1,1){\scriptsize $1$}\rput(-1,2){\scriptsize $2$}
\rput(-1,3){\scriptsize $3$}\rput(-1,4){\scriptsize $4$}
\rput(-1,5){\scriptsize $5$}\rput(-1,6){\scriptsize $6$}
\rput(-1,7){\scriptsize $7$}
\rput(0,0.5){\scriptsize $0$} \rput(2,0.5){\scriptsize $2$}
\rput(4,0.5){\scriptsize $4$} \rput(6,0.5){\scriptsize $6$}
\rput(8,0.5){\scriptsize $8$} \rput(10,0.5){\scriptsize $10$}
\rput(12,0.5){\scriptsize $12$} \rput(14,0.5){\scriptsize $14$}
\rput(16,0.5){\scriptsize $16$}
%path
\psset{linewidth=0.7pt,fillstyle=none,linestyle=solid,linecolor=black}
\psline(-0.5,3.5)(0,3)(3,6)(4,5)(6,7)(12,1)(15,4)(16,3)(17,4)(17.5,3.5)
}
\psline{->}(18.5,4.5)(20.5,4.5)
\rput[bl](23,0){
%dashed grid
\psset{linewidth=0.25pt,linestyle=dashed, dash=2.5pt 1.5pt,linecolor=gray}
\psline{-}(0,2)(24,2) \psline{-}(0,4)(24,4) \psline{-}(0,6)(24,6)
\psline{-}(0,3)(24,3) \psline{-}(0,5)(24,5) \psline{-}(0,7)(24,7)
\psline{-}(1,1)(1,8) \psline{-}(2,1)(2,8) \psline{-}(3,1)(3,8)
\psline{-}(4,1)(4,8) \psline{-}(5,1)(5,8) \psline{-}(6,1)(6,8)
\psline{-}(7,1)(7,8) \psline{-}(8,1)(8,8) \psline{-}(9,1)(9,8)
\psline{-}(10,1)(10,8) \psline{-}(11,1)(11,8) \psline{-}(12,1)(12,8)
\psline{-}(13,1)(13,8) \psline{-}(14,1)(14,8) \psline{-}(15,1)(15,8)
\psline{-}(16,1)(16,8) \psline{-}(17,1)(17,8) \psline{-}(18,1)(18,8)
\psline{-}(19,1)(19,8) \psline{-}(20,1)(20,8) \psline{-}(21,1)(21,8)
\psline{-}(22,1)(22,8) \psline{-}(23,1)(23,8)
%axes
\psset{linewidth=0.25pt,fillstyle=none,linestyle=solid,linecolor=black}
\psline{-}(0,1)(24,1)
\psline{-}(0,8)(24,8)
\psline{-}(0,1)(0,8)
\psline{-}(24,1)(24,8)
%numeros
\rput(-1,1){\scriptsize $1$}\rput(-1,2){\scriptsize $2$}
\rput(-1,3){\scriptsize $3$}\rput(-1,4){\scriptsize $4$}
\rput(-1,5){\scriptsize $5$}\rput(-1,6){\scriptsize $6$}
\rput(-1,7){\scriptsize $7$}\rput(-1,8){\scriptsize $8$}
\rput(0,0.5){\scriptsize $0$} \rput(2,0.5){\scriptsize $2$}
\rput(4,0.5){\scriptsize $4$} \rput(6,0.5){\scriptsize $6$}
\rput(8,0.5){\scriptsize $8$} \rput(10,0.5){\scriptsize $10$}
\rput(12,0.5){\scriptsize $12$} \rput(14,0.5){\scriptsize $14$}
\rput(16,0.5){\scriptsize $16$} \rput(18,0.5){\scriptsize $18$}
\rput(20,0.5){\scriptsize $20$} \rput(22,0.5){\scriptsize $22$}
\rput(24,0.5){\scriptsize $24$}
%path
\psset{linewidth=0.7pt,fillstyle=none,linestyle=solid,linecolor=black}
\psline(-0.5,3.5)(0,3)(4,7)(6,5)(9,8)(16,1)(20,5)(22,3)(24,5)(24.5,4.5)
}
\end{pspicture}
\end{center}
\end{figure}
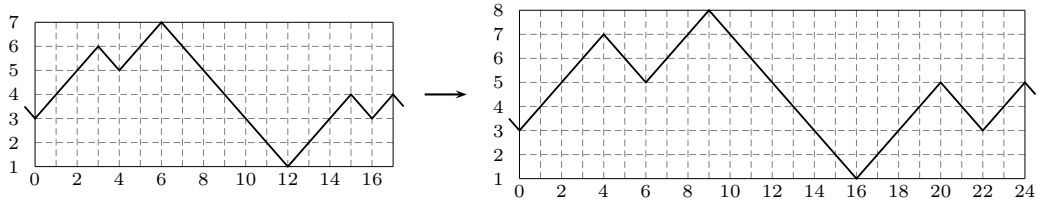

The $\Tran_1$-transform of $\habf\in\ABFpabLef$ depends on $e,f\in\{0,1\}$,
and is readily described in terms of the corresponding vertex word
$\vword(\habf)$.
Let this vertex word have $k+1$ symbols $N$ ($k\ge-1$),
and let their positions be $j_0,j_1,j_2,\ldots,j_k$.
With $a'=a+e$,
the action of the $\Tran_1$-transform on $\habf$ is then defined to result
in the path $\hbone\in\ABFsetef{p+1}{a',b'}{e,f}(L+k)$ whose vertex word
$\vword(\hbone)$ also has $k+1$ symbols $N$,
these being at positions $j_0,j_1+1,j_2+2,\ldots,j_k+k$.
An example of a $\Tran_1$-transform is given in Figure~\ref{Fig:Btrans}.
In the case of a zero length path $\habf$ for which $e\ne f$,
so that $\vword(\habf)$ is $S$, we choose
not to define the action of the $\Tran_1$-transform
(Lemma \ref{Lem:B1trans} below does not then apply to this case).

Note that for the path $\hbone$ that results from the $\Tran_1$-transform,
$\vword(\hbone)$ has no adjacent $NN$ pair of vertices.
This fact will be useful later.

\begin{lemma}\label{Lem:B1trans}
Let $1\le a,b\le p$ and $L\ge0$ and $e,f\in\{0,1\}$,
with $L>0$ if $e\ne f$.
Then let $\hbone\in\ABFsetef{p+1}{a',b'}{e,f}(\Lbone)$ be obtained
from the action of the $\Tran_1$-transform on $\habf\in\ABFpabLef$.
Set $m=\mnos(\habf)$.
Then:
\begin{enumerate}
\item $a'=a+e$ and $b'=b+f$;
\item $\Lbone=2L-m$;
\item $\mnos(\hbone)=L$;
\item $\unwt(\hbone)=\unwt(\habf) + \frac12L(L-m)$.
\end{enumerate}
\end{lemma}
\Proof
That $a'=a+e$ follows from the definition of the $\Tran_1$-transform.
Let $\vword(h)$ contain $k+1$ symbols $N$, and let their
positions be $j_0,j_1,\ldots,j_k$.
Consideration of the distances between the non-straight
vertices in $h$ shows that
\begin{equation}
\begin{split}
b-a&=(1-2e)(-j_0+(j_1-j_0)-(j_2-j_1)+\cdots +(-1)^k(L-j_k))\\
&=(1-2e)(2(-j_0+j_1-j_2+\cdots-(-1)^kj_k)+(-1)^kL).
\end{split}
\end{equation}
Comparing this with the analogous expression for $\hbone$,
and noting that $L'=L+k$, shows that
$(b'-a')-(b-a)=0$ if $k$ is even, and
$(b'-a')-(b-a)=1-2e$ if $k$ is odd.
Using $a'=a+e$ then shows that $b'-b=e$ if $k$ is even,
and $b'-b=1-e$ if $k$ is odd.
Because $e+f\equiv k$, it follows that $b'-b=f$,
as required for the first result.

That $\vword(h)$ contains $k+1$ symbols $N$ implies $m+k=L$.
Then $\Lbone=L+k=L+(L-m)$ gives the second result.
Because $\vword(\hbone)$ also has $k+1$ symbols $N$,
$\mnos(\hbone)=\Lbone-k=L$ gives the third.

Applying \eqref{Eq:wtStrike} to both $\habf$ and $\hbone$,
and noting that $\Lbone-L=L-m$ gives
\begin{equation}
\begin{split}
4\unwt(\hbone)-4\unwt(\habf)
&=\Lbone(\Lbone+1)-L(L+1)-2\sum_{i=0}^{L-m}i\\
&=(\Lbone+L+1)(\Lbone-L)-(L-m+1)(L-m)\\
&=(3L-m+1)(L-m)-(L-m+1)(L-m)\\
&=2L(L-m),
\end{split}
\end{equation}
thereby yielding the final expression of the lemma.
\cqfd

\subsection{Particle insertion}
\label{Sec:Insertion}

In this section, we specify a further method of transforming a path.
This method simply extends the path to the right, by augmenting
it with an alternating sequence of NE and SE segments. 
This insertion process depends on a value $f\in\{0,1\}$ in that the
first of these segments is in the same direction as the
postsegment specified by $f$
(the process is not affected by the direction of the presegment).

Specifically, for $n\ge0$ and a path $\habf^{(0)}\in\ABFpabLef$,
we say that the path $\habf^{(n)}\in\ABFsetef{p}{a,b}{e,f}(L+2n)$
has been obtained by inserting $n$ particles into $\habf^{(0)}$
if the vertex word $\vword(\habf^{(n)})$ is obtained
from $\vword(\habf^{(0)})$ by appending $2n$ symbols $N$.
We refer to this process of inserting $n$ particles
as a $\Tran_2(n)$-transform.

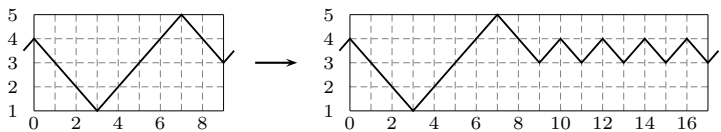
\begin{figure}[ht]
\caption{Example of $\Tran_2$-transform (here $f=0$ and $n=4$)}
\vskip-0.2cm
\label{Fig:B2transF0}
\begin{center}
\psset{yunit=0.32cm,xunit=0.28cm}
\begin{pspicture}(-1,0)(32,7)
\rput[bl](0,0){
%dashed grid
\psset{linewidth=0.25pt,linestyle=dashed, dash=2.5pt 1.5pt,linecolor=gray}
\psline{-}(0,2)(9,2) \psline{-}(0,3)(9,3) \psline{-}(0,4)(9,4)
\psline{-}(1,1)(1,5) \psline{-}(2,1)(2,5) \psline{-}(3,1)(3,5)
\psline{-}(4,1)(4,5) \psline{-}(5,1)(5,5) \psline{-}(6,1)(6,5)
\psline{-}(7,1)(7,5) \psline{-}(8,1)(8,5)
%axes
\psset{linewidth=0.25pt,fillstyle=none,linestyle=solid,linecolor=black}
\psline{-}(0,1)(9,1)
\psline{-}(0,5)(9,5)
\psline{-}(0,1)(0,5)
\psline{-}(9,1)(9,5)
%numeros
\rput(-1,1){\scriptsize $1$}\rput(-1,2){\scriptsize $2$}
\rput(-1,3){\scriptsize $3$}\rput(-1,4){\scriptsize $4$}
\rput(-1,5){\scriptsize $5$}
\rput(0,0.5){\scriptsize $0$} \rput(2,0.5){\scriptsize $2$}
\rput(4,0.5){\scriptsize $4$} \rput(6,0.5){\scriptsize $6$}
\rput(8,0.5){\scriptsize $8$}
%path
\psset{linewidth=0.7pt,fillstyle=none,linestyle=solid,linecolor=black}
\psline(-0.5,3.5)(0,4)(3,1)(7,5)(9,3)(9.5,3.5)
}
\psline{->}(10.5,3)(12.5,3)
\rput[bl](15,0){
%dashed grid
\psset{linewidth=0.25pt,linestyle=dashed, dash=2.5pt 1.5pt,linecolor=gray}
\psline{-}(0,2)(17,2) \psline{-}(0,3)(17,3) \psline{-}(0,4)(17,4)
\psline{-}(1,1)(1,5) \psline{-}(2,1)(2,5) \psline{-}(3,1)(3,5)
\psline{-}(4,1)(4,5) \psline{-}(5,1)(5,5) \psline{-}(6,1)(6,5)
\psline{-}(7,1)(7,5) \psline{-}(8,1)(8,5) \psline{-}(9,1)(9,5)
\psline{-}(10,1)(10,5) \psline{-}(11,1)(11,5) \psline{-}(12,1)(12,5)
\psline{-}(13,1)(13,5) \psline{-}(14,1)(14,5) \psline{-}(15,1)(15,5)
\psline{-}(16,1)(16,5)
%axes
\psset{linewidth=0.25pt,fillstyle=none,linestyle=solid,linecolor=black}
\psline{-}(0,1)(17,1)
\psline{-}(0,5)(17,5)
\psline{-}(0,1)(0,5)
\psline{-}(17,1)(17,5)
%numeros
\rput(-1,1){\scriptsize $1$}\rput(-1,2){\scriptsize $2$}
\rput(-1,3){\scriptsize $3$}\rput(-1,4){\scriptsize $4$}
\rput(-1,5){\scriptsize $5$}
\rput(0,0.5){\scriptsize $0$} \rput(2,0.5){\scriptsize $2$}
\rput(4,0.5){\scriptsize $4$} \rput(6,0.5){\scriptsize $6$}
\rput(8,0.5){\scriptsize $8$} \rput(10,0.5){\scriptsize $10$}
\rput(12,0.5){\scriptsize $12$} \rput(14,0.5){\scriptsize $14$}
\rput(16,0.5){\scriptsize $16$}
%path
\psset{linewidth=0.7pt,fillstyle=none,linestyle=solid,linecolor=black}
\psline(-0.5,3.5)(0,4)(3,1)(7,5)(9,3)(10,4)(11,3)(12,4)(13,3)(14,4)(15,3)
(16,4)(17,3)(17.5,3.5)
}
\end{pspicture}
\end{center}
\end{figure}
\begin{figure}[ht]
\caption{Example of $\Tran_2$-transform (here $f=1$ and $n=4$)}
\vskip-0.2cm
\label{Fig:B2transF1}
\begin{center}
\psset{yunit=0.32cm,xunit=0.28cm}
\begin{pspicture}(-1,0)(32,7)
\rput[bl](0,0){
%dashed grid
\psset{linewidth=0.25pt,linestyle=dashed, dash=2.5pt 1.5pt,linecolor=gray}
\psline{-}(0,2)(9,2) \psline{-}(0,3)(9,3) \psline{-}(0,4)(9,4)
\psline{-}(1,1)(1,5) \psline{-}(2,1)(2,5) \psline{-}(3,1)(3,5)
\psline{-}(4,1)(4,5) \psline{-}(5,1)(5,5) \psline{-}(6,1)(6,5)
\psline{-}(7,1)(7,5) \psline{-}(8,1)(8,5)
%axes
\psset{linewidth=0.25pt,fillstyle=none,linestyle=solid,linecolor=black}
\psline{-}(0,1)(9,1)
\psline{-}(0,5)(9,5)
\psline{-}(0,1)(0,5)
\psline{-}(9,1)(9,5)
%numeros
\rput(-1,1){\scriptsize $1$}\rput(-1,2){\scriptsize $2$}
\rput(-1,3){\scriptsize $3$}\rput(-1,4){\scriptsize $4$}
\rput(-1,5){\scriptsize $5$}
\rput(0,0.5){\scriptsize $0$} \rput(2,0.5){\scriptsize $2$}
\rput(4,0.5){\scriptsize $4$} \rput(6,0.5){\scriptsize $6$}
\rput(8,0.5){\scriptsize $8$}
%path
\psset{linewidth=0.7pt,fillstyle=none,linestyle=solid,linecolor=black}
\psline(-0.5,3.5)(0,4)(3,1)(7,5)(9.5,2.5)
}
\psline{->}(10.5,3)(12.5,3)
\rput[bl](15,0){
%dashed grid
\psset{linewidth=0.25pt,linestyle=dashed, dash=2.5pt 1.5pt,linecolor=gray}
\psline{-}(0,2)(17,2) \psline{-}(0,3)(17,3) \psline{-}(0,4)(17,4)
\psline{-}(1,1)(1,5) \psline{-}(2,1)(2,5) \psline{-}(3,1)(3,5)
\psline{-}(4,1)(4,5) \psline{-}(5,1)(5,5) \psline{-}(6,1)(6,5)
\psline{-}(7,1)(7,5) \psline{-}(8,1)(8,5) \psline{-}(9,1)(9,5)
\psline{-}(10,1)(10,5) \psline{-}(11,1)(11,5) \psline{-}(12,1)(12,5)
\psline{-}(13,1)(13,5) \psline{-}(14,1)(14,5) \psline{-}(15,1)(15,5)
\psline{-}(16,1)(16,5)
%axes
\psset{linewidth=0.25pt,fillstyle=none,linestyle=solid,linecolor=black}
\psline{-}(0,1)(17,1)
\psline{-}(0,5)(17,5)
\psline{-}(0,1)(0,5)
\psline{-}(17,1)(17,5)
%numeros
\rput(-1,1){\scriptsize $1$}\rput(-1,2){\scriptsize $2$}
\rput(-1,3){\scriptsize $3$}\rput(-1,4){\scriptsize $4$}
\rput(-1,5){\scriptsize $5$}
\rput(0,0.5){\scriptsize $0$} \rput(2,0.5){\scriptsize $2$}
\rput(4,0.5){\scriptsize $4$} \rput(6,0.5){\scriptsize $6$}
\rput(8,0.5){\scriptsize $8$} \rput(10,0.5){\scriptsize $10$}
\rput(12,0.5){\scriptsize $12$} \rput(14,0.5){\scriptsize $14$}
\rput(16,0.5){\scriptsize $16$}
%path
\psset{linewidth=0.7pt,fillstyle=none,linestyle=solid,linecolor=black}
\psline(-0.5,3.5)(0,4)(3,1)(7,5)(10,2)(11,3)(12,2)(13,3)(14,2)(15,3)
(16,2)(17,3)(17.5,2.5)
}
\end{pspicture}
\end{center}
\end{figure}
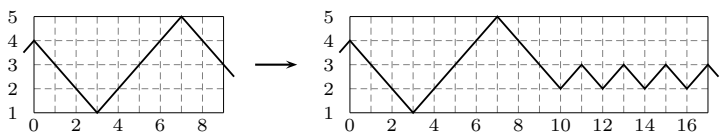

In comparing this process with the analogous process of
\cite{FodaWelsh1999}, we see that the insertion takes place
at the right end of the path rather than at the left.
This is in accord with the earlier claim that the process
described here is the dual of that of \cite{FodaWelsh1999}.

\begin{lemma}\label{Lem:Insertion}
With $1\le a,b\le p$ and $L^{(0)}\ge0$ and $e,f\in\{0,1\}$,
let $\habf^{(n)}\in\ABFsetef{p}{a,b}{e,f}(L')$ be obtained from the
$\Tran_2(n)$-transform of $\habf^{(0)}\in\ABFsetef{p}{a,b}{e,f}(L^{(0)})$.
Then:
\begin{enumerate}
\item $L'=L^{(0)}+2n$;
\item $\mnos(\habf^{(n)})=\mnos(\habf^{(0)})$;
\item $\unwt(\habf^{(n)})=\unwt(\habf^{(0)})$.
\end{enumerate}
\end{lemma}
\Proof
The first statement follows from the definition.
The second follows because the number of $S$ vertices
in $\vword(\habf^{(0)})$ and $\vword(\habf^{(n)})$ are equal.
The third statement follows from the definition \eqref{Eq:ABFwtsDef},
after noting that the insertion adds no straight vertices.
\cqfd

\subsection{Particle moves}
\label{Sec:ParticleMoves}

\begin{figure}[ht]
\caption{Particle moves}
\label{Fig:BasicMoves}
\begin{center}
\psset{yunit=0.32cm,xunit=0.28cm}
\begin{pspicture}(0,-0.5)(40,4.5)
\rput[bl](0,0){
%dashed grid
\psset{linewidth=0.25pt,linestyle=dashed, dash=2.5pt 1.5pt,linecolor=gray}
\psline{-}(0,1)(6,1) \psline{-}(0,2)(6,2) \psline{-}(0,3)(6,3)
\psline{-}(1,1)(1,3) \psline{-}(2,1)(2,3) \psline{-}(3,1)(3,3)
\psline{-}(4,1)(4,3) \psline{-}(5,1)(5,3)
%path
\psset{linewidth=0.7pt,fillstyle=none,linestyle=solid,linecolor=black}
\psline(1,3)(3,1)(4,2)(5,1)
%annotate
\rput(2.0,0.3){\scriptsize $x$}
}
\psline{->}(7,2)(9,2)
\rput[bl](10,0){
%dashed grid
\psset{linewidth=0.25pt,linestyle=dashed, dash=2.5pt 1.5pt,linecolor=gray}
\psline{-}(0,1)(6,1) \psline{-}(0,2)(6,2) \psline{-}(0,3)(6,3)
\psline{-}(1,1)(1,3) \psline{-}(2,1)(2,3) \psline{-}(3,1)(3,3)
\psline{-}(4,1)(4,3) \psline{-}(5,1)(5,3)
%path
\psset{linewidth=0.7pt,fillstyle=none,linestyle=solid,linecolor=black}
\psline(1,3)(2,2)(3,3)(5,1)
%annotate
\rput(2.0,0.3){\scriptsize $x$}
}
\rput[bl](24,0){
%dashed grid
\psset{linewidth=0.25pt,linestyle=dashed, dash=2.5pt 1.5pt,linecolor=gray}
\psline{-}(0,1)(6,1) \psline{-}(0,2)(6,2) \psline{-}(0,3)(6,3)
\psline{-}(1,1)(1,3) \psline{-}(2,1)(2,3) \psline{-}(3,1)(3,3)
\psline{-}(4,1)(4,3) \psline{-}(5,1)(5,3)
%path
\psset{linewidth=0.7pt,fillstyle=none,linestyle=solid,linecolor=black}
\psline(1,1)(3,3)(4,2)(5,3)
%annotate
\rput(2.0,0.3){\scriptsize $x$}
}
\psline{->}(31,2)(33,2)
\rput[bl](34,0){
%dashed grid
\psset{linewidth=0.25pt,linestyle=dashed, dash=2.5pt 1.5pt,linecolor=gray}
\psline{-}(0,1)(6,1) \psline{-}(0,2)(6,2) \psline{-}(0,3)(6,3)
\psline{-}(1,1)(1,3) \psline{-}(2,1)(2,3) \psline{-}(3,1)(3,3)
\psline{-}(4,1)(4,3) \psline{-}(5,1)(5,3)
%path
\psset{linewidth=0.7pt,fillstyle=none,linestyle=solid,linecolor=black}
\psline(1,1)(2,2)(3,1)(5,3)
%annotate
\rput(2.0,0.3){\scriptsize $x$}
}
\end{pspicture}
\end{center}
\end{figure}
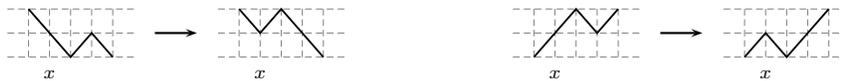

In this section, we specify two types of local deformation of a path.
These deformations are known as \emph{moves}.
In each case, a particular sequence of four segments of
a path is changed to a different sequence, the remainder of
the path being unchanged.
These two moves are depicted in Figure \ref{Fig:BasicMoves}:
the portion of the path to the left of the arrow is
changed to the portion on the right.
These moves apply whenever $x>0$ and $x+2<L$.
They also apply in the cases $x=0$ and $x+2=L$ if
the path's presegment and postsegment respectively accord with
the corresponding segments in Figure \ref{Fig:BasicMoves}.
Specifically, for $x=0$, the first move applies if $e=0$,
and the second applies if $e=1$.
Similarly, for $x+2=L$, the first move applies if $f=1$,
and the second applies if $f=0$.
(If both $x=0$ and $L=2$ then the first move applies if both
$e=0$ and $f=1$, and the second applies if both $e=1$ and $f=0$.)

The particle moves described in Figure \ref{Fig:BasicMoves} above are
easily described in terms of the vertex words: each merely exchanges
a sequence of three vertices $SNN$ for the sequence $NNS$.
Thus a particle is identified with a pair $NN$ of adjacent vertices,
and a move is viewed as such a pair passing beyond an $S$ vertex.%
\footnote{This interpretation is the dual of that given in \cite{Welsh2006}.}
This view of a particle accords with the process of particle
insertion described in Section \ref{Sec:Insertion}.
There, the insertion of each particle corresponds simply to
appending a pair $NN$ to the vertex word.

\begin{lemma}\label{Lem:BasicMoves}
For $1\le a,b\le p$ and $L\ge0$ and $e,f\in\{0,1\}$,
let $\habf'\in\ABFpabLef$ be obtained from $\habf\in\ABFpabLef$
through one of the above moves.
Then:
\begin{enumerate} % want roman!
\item $\unwt(\habf')=\unwt(\habf)+1$;
\item $\mnos(\habf')=\mnos(\habf)$.
\end{enumerate}
\end{lemma}
\Proof
If $\vword(\habf)$ has $N$ symbols at positions
$j_0,j_1,\ldots,j_k$, the effect of a move is to decrease
the values of two of these by one each.
The first result then follows from \eqref{Eq:wtStrike}.
Because $\vword(\habf)$ and $\vword(\habf')$
have the same number of symbols $S$, the second result follows.
\cqfd

If a particle, in the guise of an $NN$ pair, having moved to the left
beyond a single straight vertex $S$, finds another vertex $S$ to its
left, it may move again beyond this $S$.
Having done so, it may do the same through subsequent vertices $S$,
until it encounters another $N$.
If this $N$ has another $N$ to its left, then the original particle
is said to be blocked, and cannot move further.
However, if instead it has an $S$ to its left, the sequence
$SNNN$ of vertices appears.
We then re-identify the particle with the leftmost pair $NN$ therein,
whereupon it can continue to move to the left.
For example, the following sequence of moves is attributed to
the movement of a single particle:
\begin{equation}
\begin{split}
&
SSNSSNS\boldN\boldN
\longrightarrow
SSNSS\boldN\boldN NS
\longrightarrow
SSNS\boldN\boldN SNS\\
&\qquad\qquad
\longrightarrow
SS\boldN\boldN NSSNS
\longrightarrow
S\boldN\boldN SNSSNS
\longrightarrow
\boldN\boldN SSNSSNS.
\end{split}
\end{equation}

For each particle, the number of vertices $S$ to its right
in $\vword(h)$ is said to be the particle's \emph{excitation}.
Each move a particle makes thus increases its excitation by 1.
Within a sequence of exactly $2k$ adjacent symbols $N$, there
are precisely $k$ particles, each having the same excitation.
Only the particle on the left is not blocked by one of the others.
However, that on the right can perform a backward move,
decreasing its excitation and that of the path.
There are also precisely $k$ particles within a sequence of
exactly $2k+1$ adjacent symbols $N$,
each having the same excitation.
Here, however, when performing backward moves,
different $NN$ pairs are viewed as particles.
Thus, for odd-length sequences of $N$ vertices, which actual
pairs correspond to the particles is ambiguous.
Nonetheless, each move is well-defined,
and is attributed to a specific particle whose excitation changes.

If $h$ contains $n$ particles, let the excitation of the
$i$th, counting from the left, be $\lambda_i$.
Then $\lambda_1\ge\lambda_2\ge\cdots\ge\lambda_n\ge0$,
and therefore
$\lambda=(\lambda_1,\lambda_2,\ldots,\lambda_n)$
is a partition having at most $n$ parts.
We refer to it as the \emph{excitation partition} of $h$.
As usual, the weight $|\lambda|$ of a partition
$\lambda=(\lambda_1,\lambda_2,\ldots,\lambda_n)$
is defined by $|\lambda|=\lambda_1+\lambda_2+\cdots+\lambda_n$.

\subsection{Particle waves}
\label{Sec:ParticleWaves}

Consider a path $\hbone$ whose vertex word $\vword(\hbone)$
has no adjacent pair $NN$, and thus contains no particle.
For $n>0$, let $\hbnnn$ be obtained from $\hbone$ by acting
on it with the $\Tran_2(n)$-transform.
Because there are no particles to its left,
the leftmost particle in $\hbnnn$ is able
to make as many moves as there are symbols $S$ in $\vword(\hbnnn)$.
This maximal number is $m'=\mnos(\hbnnn)$.
Similarly, the second leftmost particle in $\hbnnn$, if there is one,
can move beyond the same vertices $S$ as the first,
until if it makes the same number of moves as the first,
it is blocked and cannot move further.
Continuing in this way, we see that if the $i$th leftmost
particle in $\hbnnn$ makes $\lambda_i$ moves then
$m'\ge\lambda_{1}\ge\lambda_{2}\ge\cdots\ge\lambda_{n}\ge0$.
Thus, by proceeding in this way, we see that every
excitation partition $\lambda=(\lambda_1,\ldots,\lambda_n)$
for which $\lambda_1\le m'$
can be obtained by moving the particles in $\hbnnn$.
If $h'$ results from moving these particles according to the
parts of a partition $\lambda$, we say that it results from
the action of the $\Tran_3(\lambda)$-transform on $\hbnnn$.

\begin{lemma}\label{Lem:BasicWaves}
Let $1\le a,b\le p$ and $L^{(0)}\ge0$ and $e,f\in\{0,1\}$
and $\hbone\in\ABFsetef{p}{a,b}{e,f}(L^{(0)})$.
Then, for $n\ge0$ and $\lambda$ a partition having at most $n$ parts
with $\lambda_1\le\mnos(\hbone)$,
let $\habf'\in\ABFsetef{p}{a,b}{e,f}(L')$ be obtained from
$\hbone$ through the action of the $\Tran_2(n)$-transform
followed by the $\Tran_3(\lambda)$-transform.
Then:
\begin{enumerate}
\item $L'=L^{(0)}+2n$;
\item $\mnos(\habf')=\mnos(\hbone)$;
\item $\unwt(\habf')=\unwt(\hbone)+|\lambda|$.
\end{enumerate}
\end{lemma}
\Proof
This follows immediately from Lemmas \ref{Lem:Insertion}
and \ref{Lem:BasicMoves}.
\cqfd

\begin{note}\label{Note:BasicWaves}
The path $\habf'$ obtained from $\hbone$ as in the
above lemma is easily obtained using the vertex words.
Namely, $\vword(\habf')$ is obtained from $\vword(\hbone)$ by,
for each $i\le n$, inserting an $NN$ pair so that
it has exactly $\lambda_i$ symbols $S$ to its right.
\end{note}

The three aspects of path manipulation that are described above,
are combined to define the $\Tran$-transform.
For $\habf\in\ABFpabLef$, the result of acting on $\habf$
with the combined action of the $\Tran_1$-transform followed
by the $\Tran_2(n)$-transform followed by the
$\Tran_3(\lambda)$-transform for $\lambda$ a partition with at most
$n$ parts,
is termed the $\Tran(n,\lambda)$-transform.

In what follows, we use a succession of $\Tran$-transforms to
obtain the generating function for certain sets of paths.
The particles manipulated during different $\Tran$-transforms
are said to belong to different species.
Only those particles of the same species are subject to the
fermionic exclusion property, which results from the
``blocking'' described above.
A path obtained from a succession of $\Tran$-transforms is thus seen
to contain a specific number of particles of various species.
In our proofs below, these particle counts will be given by
$n_1$, $n_2$, $n_3,\cdots$, with $n_1$ the number of particles
inserted during the final $\Tran$-transform.
This approach is valid for each of the four cases,
these leading to the four cases of \eqref{Eq:ABFferms}
and the four cases of Theorem \ref{Thrm:HLferms}.
In \cite{Bressoud1989,Warnaar1996b}, an alternative
non-recursive (somewhat) approach to identifying particles
in a path is given.
Therein, particles are viewed as triangular deformations
in the path, with the differently sized triangles
corresponding to the different species.
However, this approach can only be applied directly to two of
the four cases (one instance was considered in \cite{BfMW2012}), 
a deficiency from which the current approach doesn't suffer.

The $\Tran$-transform is used
in both Section \ref{Sec:ABFfin} below,
to obtain Melzer's expressions for the generating functions of the
ABF paths, and in Section \ref{Sec:Refine}, to obtain fermionic
expressions for the generating functions of the half-lattice paths.

%%%%%%%%%%%%%%%%%%%%%%%%%%%%%%%%%%%%%%%%%%%%%%%%%%%%%%%%%%%%%%%%%%%%%%
%%%%%%%%%%%%%%%%%%%%%%%%%%%%%%%%%%%%%%%%%%%%%%%%%%%%%%%%%%%%%%%%%%%%%%

\section{Revisiting Melzer's identities:
         Finitized fermionic expressions for ABF paths}
\label{Sec:ABFfin}

\subsection{Transforming the generating function}

In this subsection, we use the $\Tran$-transform to describe
a bijection involving different sets of ABF paths.
To specify this, define
$\mathcal{P}_{n,m}$ to be the set of all partitions
$\lambda=(\lambda_1,\lambda_2,\ldots,\lambda_n)$
with $m\ge\lambda_1\ge\lambda_2\ge\cdots\ge\lambda_n\ge0$.
Below, we will use the fact that \cite{Andrews1976}
\begin{equation}\label{Eq:PartitionGF}
\sum_{\lambda\in\mathcal{P}_{n,m}} q^{|\lambda|}=
\qbinom{n+m}{n}.
\end{equation}

\begin{lemma}\label{Lem:Btransform}
Let $1\le a,b\le p$ and $L,L'\ge0$ and $e,f\in\{0,1\}$,
with $L>0$ if $e\ne f$.
Then there is a bijection between the sets
\begin{equation}\label{Eq:Btransform}
\ABFsetef{p+1}{a+e,b+f}{e,f}(L',L)
\longleftrightarrow
\bigcup_{n\ge0}
\ABFsetef{p}{a,b}{e,f}(L,2L-L'+2n)
\times
\mathcal{P}_{n,L},
\end{equation}
under which, if $h'\in\ABFsetef{p+1}{a+e,b+f}{e,f}(L',L)$
maps to the triple $(h,n,\lambda)$,
where $h\in\ABFsetef{p}{a,b}{e,f}(L,m)$ with $m=2L-L'+2n$,
and $\lambda\in\mathcal{P}_{n,L}$,
then $h'$ is the result of the $\Tran(n,\lambda)$-transform
acting on $h$, and
\begin{equation}\label{Eq:BtransformWt}
\unwt(h')=
\unwt(h)+\frac12 L(L-m)+|\lambda|.
\end{equation}
\end{lemma}

\Proof
Let $h\in\ABFsetef{p}{a,b}{e,f}(L,m)$, $n\ge0$ and
$\lambda\in\mathcal{P}_{n,L}$,
and let $h'$ be obtained from $h$ by the action of the
$\Tran(n,\lambda)$-transform.
Lemma \ref{Lem:B1trans} and Lemma \ref{Lem:BasicWaves} show
that $h'\in\ABFsetef{p+1}{a+e,b+f}{e,f}(L',L)$ where
$L'=2L-m+2n$.

We now claim that for each
$h'\in\ABFsetef{p+1}{a+e,b+f}{e,f}(L',L)$
there is a unique triple $(h,n,\lambda)$ with
$h\in\ABFsetef{p}{a,b}{e,f}(L,2L-L'+2n)$,
such that $h'$ arises from the action of the
$\Tran(n,\lambda)$-transform on $h$.
To see this,
first note that the path $\hbone$ which would arise from the
action of the $\Tran_1$-transform on $h$ has a vertex word with no
adjacent pair $NN$.
Thus, in view of Note \ref{Note:BasicWaves},
$\hbone$ is determined uniquely by $h'$, with its vertex word
obtained from that of $h'$ by repeatedly removing pairs $NN$
from the latter until no such pairs remain.
The value of $n$ is the number of such pairs,
while their excitation partition gives $\lambda$.

Because $\hbone$ would be obtained from $h$ through the action of
the $\Tran_1$-transform, if $\vword(\hbone)$ has symbols $N$ at positions
$j_0,j_1,j_2,\ldots j_k$ ($k=L'-m\ge-1$) then if $k\ge0$,
the vertex word of $\vword(h)$ has symbols at positions
$j_0,j_1-1,j_2-2,\ldots,j_k-k$.
If $k=-1$ then both $\vword(\hbone)$ and $\vword(h)$
comprise only symbols $S$.
In particular, in all cases, $h$ is determined uniquely.
The Lemma is then proved, with \eqref{Eq:BtransformWt} following
from Corollary \ref{Lem:B1trans}(4) and Lemma \ref{Lem:BasicWaves}(3).
\cqfd

\begin{corollary}\label{Cor:Btransform}
Let $1\le a,b\le p$ and $L,L'\ge0$ and $e,f\in\{0,1\}$.
Then:
\begin{equation}
\ABFGFe{p+1}{a+e,b+f}{e,f}(L',L)
=
\sum_{n\ge0}
q^{\frac12 L(L-m)}
\qbinom{n+L}{n}
\ABFGFe{p}{a,b}{e,f}(L,m),
\end{equation}
where $m$ is obtained from $n$ via $m=2L-L'+2n$.
\end{corollary}

\Proof In the cases other than where $L=0$ and $e\ne f$,
this follows from equating the generating functions of the
two sides of \eqref{Eq:Btransform}, taking \eqref{Eq:BtransformWt}
into account, and using \eqref{Eq:PartitionGF}.

Now consider the case where $L=0$ and $e\ne f$.
For the LHS, Lemma \ref{Lem:Seed}(3) shows that
$\ABFGFe{p+1}{a+e,b+f}{e,f}(L',0)=1$
if $L'$ is odd and $a=b$,
and is zero otherwise.
For the RHS, Lemma \ref{Lem:Seed}(1) shows that
$\ABFGFe{p}{a,b}{e,f}(0,m)=1$
if $m=1$ and $a=b$,
and is zero otherwise.
Because $L'=2L-m+2n$, and thus $L'=2n-1$ here,
the RHS is also $1$ if $L'$ is odd and $a=b$,
and zero otherwise.
\cqfd

\subsection{Melzer's expressions}
\label{Sec:Melzer}

We now show how Melzer's fermionic expressions for the generating
functions $\ABFGFe{p}{a,b}{e,f}(L)$ of the finite length ABF paths
may be obtained by recursive use of Corollary \ref{Cor:Btransform}.

\begin{theorem}\label{Thrm:MelzerFin}
\cite{Melzer1994}
Let $1\le a,b\le p$ with $p\ge3$,
and set $\boldC=\boldC^{(p-2)}$.
Then, for $e,f\in\{0,1\}$, we have the following expression for
$\ABFGFe{p}{a,b}{e,f}(L)$
in which the values of $\ell$ and $\Delta_i$ are as given
in Table~\ref{Tab:ABFL} below:
\begin{equation}\label{Eq:ABFL}
\begin{split}
\ABFGFe{p}{a,b}{e,f}(L)
=\sum_{\sboldn\in\ZZ^{p-1}_{\ge0}|m_0=L}
\hskip-3mm
&q^{\frac14\sboldm\sboldC{\sboldm}^T-\frac12 m_{\ell}}
\prod_{i=1}^{p-2}
\qbinom{n_i+m_i}{n_i}\,,\\
\text{where}\quad
m_i=2\sum_{i<k<p}
(k&-i)n_k-\Delta_i
\quad\text{for}\quad
0\le i<p.
\end{split}
\end{equation}
Here, the sum is over all non-negative integer vectors
$\boldn=(n_1,n_2,\ldots,n_{p-1})$, with
$\boldmp=(m_0,m_1,\ldots,m_{p-2})$ obtained from $\boldn$
as indicated, and $\boldm=(m_1,m_2,\ldots,m_{p-2})$.
Note that by virtue of the restriction $m_0=L$,
the sum is, in effect, finite.
Also note that $m_{p-1}=0$ in each case.

\begin{table}[ht]
\caption{Parameters for the four cases of the expressions \eqref{Eq:ABFL}.} 
\vskip.2cm
\label{Tab:ABFL}
\begin{center}
\begin{tabular}{|l|c||c|c|c|c| }\hline
\mystrut& \textup(e,f\textup)& $a$& $b$&$\ell$ & $ \Delta_i$\\
\hline\hline
\mystrut{\rm (a)}&\textup(1,1\textup)& $\ne 1$ & $\ne1$& $a-1$
  & $\pabs{a-1-i}+\pabs{b-1-i}$ \\
\hline
\mystrut{\rm (b)}&\textup(0,1\textup)& $\ne p$ &$\ne1$& $ p-a$
  & $ \pabs{p-a-i}+\pabs{b-1-i} +p-1-i$ \\
\hline
\mystrut{\rm (c)}&\textup(0,0\textup)& $\ne p$ & $\ne p$ &$ p-a$
  & $ \pabs{p-a-i}+\pabs{p-b-i}$ \\
\hline
\mystrut{\rm (d)}&\textup(1,0\textup)& $\ne 1$ &$\ne p$ & $ a-1$
  & $\pabs{a-1-i}+\pabs{p-b-i} +p-1-i$ \\
\hline
\end{tabular}
\end{center}
\end{table}
\end{theorem}

The four expressions \eqref{Eq:ABFferms} with Table \ref{Tab:ABF}
are an immediate corollary
of these, following by taking the limit $L\to\infty$
through taking $n_1\to\infty$,
and using \eqref{Eq:ABFbosonicLim}.

In what follows, each of the four expressions of Theorem \ref{Thrm:MelzerFin}
is proved by recursively applying Corollary \ref{Cor:Btransform}
to express the generating function
$\ABFGFe{p}{a,b}{e,f}(L,m)$,
in terms of the trivial $\ABFGFe{1}{1,1}{1-e,1-f}(m_{p-1},m_p)$,
which is given by Lemma \ref{Lem:Seed}.
The four cases arise from the choices $e,f\in\{0,1\}$.
By repeated application of Corollary \ref{Cor:Btransform},
a succession of generating functions
$\ABFGFe{p+1-i}{a_i,b_i}{e_i,f_i}(m_{i-1},m_i)$ is obtained
for $i=1,2,\ldots,p$.
This sequence is determined by the two sequences
$e_1,e_2,\ldots,e_{p-1}$ and $f_1,f_2,\ldots,f_{p-1}$,
each element of which is $0$ or $1$.
However, within each of these sequences, the value may
change only once (why this is so will become apparent below),
and $a_p=1$ and $b_p=1$ will be achieved only if
the first sequence contains exactly $(a-1)$ 1s,
and the second contains exactly $(b-1)$ 1s.
Thus in each case, only two sequences are possible.
They are given explicitly in the second column of the
following list
(although, it isn't necessary to specify $e_p$ and $f_p$,
it is convenient to do so in order to treat uniformly
the applications of Corollary \ref{Cor:Btransform} below):
\begin{subequations}\label{Eq:LR12}
\begin{align}
\label{Eq:L1}
\text{Case L1:}\quad
&
e_i=
\begin{cases}
1&\text{for $1\le i<a$,}\\
0&\text{for $a\le i\le p$;}
\end{cases}
&&
a_i=
\begin{cases}
a-i+1&\text{for $1\le i\le a$,}\\
1&\text{for $a\le i\le p$;}
\end{cases}
\\
\label{Eq:L2}
\text{Case L0:}\quad
&
e_i=
\begin{cases}
0&\text{for $1\le i\le p-a$,}\\
1&\text{for $p-a<i\le p$;}
\end{cases}
&&
a_i=
\begin{cases}
a&\text{for $1\le i\le p-a+1$,}\\
p-i+1&\text{for $p-a+1\le i\le p$;}
\end{cases}
\\
\label{Eq:R1}
\text{Case R1:}\quad
&
f_i=
\begin{cases}
1&\text{for $1\le i<b$,}\\
0&\text{for $b\le i\le p$;}
\end{cases}
&&
b_i=
\begin{cases}
b-i+1&\text{for $1\le i\le b$,}\\
1&\text{for $b\le i\le p$;}
\end{cases}
\\
\label{Eq:R2}
\text{Case R0:}\quad
&
f_i=
\begin{cases}
0&\text{for $1\le i\le p-b$,}\\
1&\text{for $p-b<i\le p$;}
\end{cases}
&&
b_i=
\begin{cases}
b&\text{for $1\le i\le p-b+1$,}\\
p-i+1&\text{for $p-b+1\le i\le p$.}
\end{cases}
\end{align}
\end{subequations}
The values $a_i$ and $b_i$ stated in this list give
the startpoints and endpoints of
the sequences of paths enumerated.
As Corollary \ref{Cor:Btransform} indicates,
these values are determined by the $e_i$ and $f_i$ using
\begin{equation}
a_i=a-\sum_{j=1}^{i-1} e_j,
\qquad
b_i=b-\sum_{j=1}^{i-1} f_j
\end{equation}
for $1\le i\le p$.
Note then that $a_i=a_{i+1}+e_i$ and $b_i=b_{i+1}+f_i$
for $1\le i<p$.
Also note that $a_1=a$, $b_1=b$ and $a_p=b_p=1$.

The four cases of Theorem \ref{Thrm:MelzerFin} are now
obtained by combining each of the Cases L1 and L0 with
each of the Cases R1 and R0.
Here we explicitly give proofs of all four cases
(albeit briefly for all but the first).
This contrasts with the presentation given in \cite{FodaWelsh1999},
where, because of the up-down symmetry,
two of the cases could be obtained simply from the other two.
Although this symmetry could be exploited likewise to
prove Theorem \ref{Thrm:MelzerFin} here,
it no longer exists once we refine our approach to the cases
of the half-lattice paths, and thus we require the details of
all four cases.

\subsubsection{System A}\label{Sec:System1}

For $1<a,b\le p$, consider the sequence
of $\Tran$-transforms governed by Cases L1 and R1.

Corollary \ref{Cor:Btransform} implies that for each
$i=1,2,\ldots,p-1$,
\begin{equation}\label{Eq:Brecurse}
\ABFGFe{p+1-i}{a_i,b_i}{e_i,f_i}(m_{i-1},m_i)
=\sum_{n_i\ge0}
q^{\frac12m_i(m_i-m)}
\qbinom{n_i+m_i}{n_i}
\ABFGFe{p-i}{a_{i+1},b_{i+1}}{e_i,f_i}(m_{i},m),
\end{equation}
where $m=2m_i+2n_i-m_{i-1}$.
In the $i$th case, we replace the variable $m$ in \eqref{Eq:Brecurse}
with
\begin{equation}\label{Eq:Msub1}
m=m_{i+1}+\delta_{i,a-1}+\delta_{i,b-1}.
\end{equation}
We now express
$\ABFGFe{p-i}{a_{i+1},b_{i+1}}{e_i,f_i}(m_{i},m)$
in terms of
$\ABFGFe{p-i}{a_{i+1},b_{i+1}}{e_{i+1},f_{i+1}}(m_{i},m_{i+1})$.
Firstly, we obtain
\begin{equation}\label{Eq:BrecurseX1}
\ABFGFe{p-i}{a_{i+1},b_{i+1}}{e_{i},f_{i}}
                     (m_{i},m_{i+1}+\delta_{i,a-1}+\delta_{i,b-1})
=\ABFGFe{p-i}{a_{i+1},b_{i+1}}{e_{i+1},f_{i}}
                     (m_{i},m_{i+1}+\delta_{i,b-1}),
\end{equation}
which follows from Lemma \ref{Lem:mSwitch2}(3) in the $i=a-1$ case
because then $a_{i+1}=1$, $e_i=1$ and $e_{i+1}=0$, and follows
trivially in the $i\ne a-1$ case because then $e_{i+1}=e_i$.
We then obtain
\begin{equation}\label{Eq:BrecurseX2}
\ABFGFe{p-i}{a_{i+1},b_{i+1}}{e_{i+1},f_{i}}
                    (m_{i},m_{i+1}+\delta_{i,b-1})
=q^{\frac12\delta_{i,b-1}m_{b-1}}
\ABFGFe{p-i}{a_{i+1},b_{i+1}}{e_{i+1},f_{i+1}}(m_{i},m_{i+1}),
\end{equation}
which follows from Lemma \ref{Lem:mSwitch2}(5) in the $i=b-1$ case
because then $b_{i+1}=1$, $f_i=1$ and $f_{i+1}=0$, and follows
trivially in the $i\ne b-1$ case because then $f_{i+1}=f_i$.

Combining the $i=1,2,\ldots,p-1$ cases of
\eqref{Eq:Brecurse}, \eqref{Eq:BrecurseX1} and \eqref{Eq:BrecurseX2}
results in
\begin{equation}\label{Eq:Ball1}
\ABFGFe{p}{a,b}{1,1}(m_0,m_1)
=
\!\!\!
\sum_{\sboldn\in\ZZ^{p-1}_{\ge0}}
\!\!
q^{-\frac12m_{a-1}
   +\frac12\sum_{i=1}^{p-1}m_i(m_i-m_{i+1})}
\ABFGFe{1}{1,1}{0,0}(m_{p-1},m_p)
\prod_{i=1}^{p-1}
\qbinom{n_i+m_i}{n_i}\!,
\end{equation}
where the values of $m_i$ are obtained recursively
from $\boldn=(n_1,n_2,\ldots,n_{p-1})$ using
\begin{equation}\label{Eq:MNsys1}
m_{i+1}=2m_{i}+2n_{i}-m_{i-1}-\delta_{i,a-1}-\delta_{i,b-1}
\qquad(1\le i<p).
\end{equation}
Because $\ABFGFe{1}{1,1}{0,0}(m_{p-1},m_p)=\delta_{m_{p-1},0}\delta_{m_p,0}$
by Lemma \ref{Lem:Seed}, and $\qbinom{n_{p-1}}{n_{p-1}}=1$, this results in
\begin{equation}\label{Eq:Ball1eval}
\ABFGFe{p}{a,b}{1,1}(m_0,m_1)
=\sum_{\sboldn\in\ZZ^{p-1}_{\ge0}}
q^{-\frac12m_{a-1}
%   +\frac14\sboldm\sboldC^{(p-2)}{\sboldm}^T}
   +\frac12\sum_{i=1}^{p-1}m_i(m_i-m_{i+1})}
\prod_{i=1}^{p-2}
\qbinom{n_i+m_i}{n_i},
\end{equation}
with $\boldn$ constrained such that \eqref{Eq:MNsys1}
yields $m_{p-1}=m_p=0$.
The constraints \eqref{Eq:MNsys1} are readily solved to yield
\begin{equation}\label{Eq:MNsol1}
m_i=2\sum_{i<k<p}(k-i)n_k-\pabs{a-1-i}-\pabs{b-1-i}
\qquad(0\le i<p).
\end{equation}
Then, using Lemma \ref{Lem:mSwitch2}(2) in the form
\begin{equation}
\ABFGFe{p}{a,b}{1,1}(m_0)=\sum_{m_1\ge0}
\ABFGFe{p}{a,b}{1,1}(m_0,m_1),
\end{equation}
proves (\ref{Eq:ABFL}a).

\subsubsection{System B}\label{Sec:System2}

For $1\le a<p$ and $1<b\le p$, consider the sequence
of $\Tran$-transforms governed by Cases L0 and R1.

Proceeding as for system A, but using
\begin{equation}\label{Eq:Msub2}
m=m_{i+1}+\delta_{i,p-a}+\delta_{i,b-1}
\end{equation}
instead of \eqref{Eq:Msub1},
we obtain the following analogue of \eqref{Eq:Ball1}:
\begin{equation}\label{Eq:Ball2}
\ABFGFe{p}{a,b}{0,1}(m_0,m_1)
=
\!\!\!
\sum_{\sboldn\in\ZZ^{p-1}_{\ge0}}
\!\!
q^{-\frac12m_{p-a}
   +\frac12\sum_{i=1}^{p-1}m_i(m_i-m_{i+1})}
\ABFGFe{1}{1,1}{1,0}(m_{p-1},m_p)
\prod_{i=1}^{p-1}
\qbinom{n_i+m_i}{n_i}\!,
\end{equation}
where the values of $m_i$ are obtained recursively
from $\boldn=(n_1,n_2,\ldots,n_{p-1})$ using
\begin{equation}\label{Eq:MNsys2}
m_{i+1}=2m_{i}+2n_{i}-m_{i-1}-\delta_{i,p-a}-\delta_{i,b-1}
\qquad(1\le i<p).
\end{equation}
Because $\ABFGFe{1}{1,1}{1,0}(m_{p-1},m_p)=\delta_{m_{p-1},0}\delta_{m_p,1}$
by Lemma \ref{Lem:Seed}, and $\qbinom{n_{p-1}}{n_{p-1}}=1$, this results in
\begin{equation}\label{Eq:Ball2eval}
\ABFGFe{p}{a,b}{0,1}(m_0,m_1)
=\sum_{\sboldn\in\ZZ^{p-1}_{\ge0}}
q^{-\frac12m_{p-a}
%   +\frac14\sboldm\sboldC^{(p-2)}{\sboldm}^T}
   +\frac12\sum_{i=1}^{p-1}m_i(m_i-m_{i+1})}
\prod_{i=1}^{p-2}
\qbinom{n_i+m_i}{n_i},
\end{equation}
with $\boldn$ constrained such that \eqref{Eq:MNsys2}
yields $m_{p-1}=0$ and $m_p=1$.
The constraints \eqref{Eq:MNsys2} are readily solved to yield
\begin{equation}\label{Eq:MNsol2}
m_i=2\sum_{i<k<p}(k-i)n_k -\pabs{p-a-i} -\pabs{b-1-i} -p+1+i
\qquad(0\le i<p).
\end{equation}
Then, summing \eqref{Eq:Ball2eval} over $m_1$,
using Lemma \ref{Lem:mSwitch2}(2),
proves (\ref{Eq:ABFL}b).

\subsubsection{System C}\label{Sec:System3}

For $1\le a,b<p$, consider the sequence
of $\Tran$-transforms governed by Cases L0 and R0.

Again proceeding as for system A, but using
\begin{equation}\label{Eq:Msub3}
m=m_{i+1}+\delta_{i,p-a}+\delta_{i,p-b}
\end{equation}
here instead of \eqref{Eq:Msub1},
we obtain the following analogue of \eqref{Eq:Ball1}:
\begin{equation}\label{Eq:Ball3}
\ABFGFe{p}{a,b}{0,0}(m_0,m_1)
=
\!\!\!
\sum_{\sboldn\in\ZZ^{p-1}_{\ge0}}
\!\!
q^{-\frac12m_{p-a}
   +\frac12\sum_{i=1}^{p-1}m_i(m_i-m_{i+1})}
\ABFGFe{1}{1,1}{1,1}(m_{p-1},m_p)
\prod_{i=1}^{p-1}
\qbinom{n_i+m_i}{n_i}\!,
\end{equation}
where the values of $m_i$ are obtained recursively
from $\boldn=(n_1,n_2,\ldots,n_{p-1})$ using
\begin{equation}\label{Eq:MNsys3}
m_{i+1}=2m_{i}+2n_{i}-m_{i-1}-\delta_{i,p-a}-\delta_{i,p-b}
\qquad(1\le i<p).
\end{equation}
Because $\ABFGFe{1}{1,1}{1,1}(m_{p-1},m_p)=\delta_{m_{p-1},0}\delta_{m_p,0}$
by Lemma \ref{Lem:Seed}, and $\qbinom{n_{p-1}}{n_{p-1}}=1$, this results in
\begin{equation}\label{Eq:Ball3eval}
\ABFGFe{p}{a,b}{0,0}(m_0,m_1)
=\sum_{\sboldn\in\ZZ^{p-1}_{\ge0}}
q^{-\frac12m_{p-a}
%   +\frac14\sboldm\sboldC^{(p-2)}{\sboldm}^T}
   +\frac12\sum_{i=1}^{p-1}m_i(m_i-m_{i+1})}
\prod_{i=1}^{p-2}
\qbinom{n_i+m_i}{n_i},
\end{equation}
with $\boldn$ constrained such that \eqref{Eq:MNsys3}
yields $m_{p-1}=m_p=0$.
The constraints \eqref{Eq:MNsys3} are readily solved to yield
\begin{equation}\label{Eq:MNsol3}
m_i=2\sum_{i<k<p}(k-i)n_k -\pabs{p-a-i} -\pabs{p-b-i}
\qquad(0\le i<p).
\end{equation}
Then, summing \eqref{Eq:Ball3eval} over $m_1$,
using Lemma \ref{Lem:mSwitch2}(2),
proves (\ref{Eq:ABFL}c).

\subsubsection{System D}\label{Sec:System4}

For $1<a\le p$ and $1\le b<p$, consider the sequence
of $\Tran$-transforms governed by Cases L1 and R0.

Again proceeding as for system A, but using
\begin{equation}\label{Eq:Msub4}
m=m_{i+1}+\delta_{i,a-1}+\delta_{i,p-b}
\end{equation}
here instead of \eqref{Eq:Msub1},
we obtain the following analogue of \eqref{Eq:Ball1}:
\begin{equation}\label{Eq:Ball4}
\ABFGFe{p}{a,b}{1,0}(m_0,m_1)
=
\!\!\!
\sum_{\sboldn\in\ZZ^{p-1}_{\ge0}}
\!\!
q^{-\frac12m_{a-1}
   +\frac12\sum_{i=1}^{p-1}m_i(m_i-m_{i+1})}
\ABFGFe{1}{1,1}{0,1}(m_{p-1},m_p)
\prod_{i=1}^{p-1}
\qbinom{n_i+m_i}{n_i}\!,
\end{equation}
where the values of $m_i$ are obtained recursively
from $\boldn=(n_1,n_2,\ldots,n_{p-1})$ using
\begin{equation}\label{Eq:MNsys4}
m_{i+1}=2m_{i}+2n_{i}-m_{i-1}-\delta_{i,a-1}-\delta_{i,p-b}
\qquad(1\le i<p).
\end{equation}
Because $\ABFGFe{1}{1,1}{0,1}(m_{p-1},m_p)=\delta_{m_{p-1},0}\delta_{m_p,1}$
by Lemma \ref{Lem:Seed}, and $\qbinom{n_{p-1}}{n_{p-1}}=1$, this results in
\begin{equation}\label{Eq:Ball4eval}
\ABFGFe{p}{a,b}{1,0}(m_0,m_1)
=\sum_{\sboldn\in\ZZ^{p-1}_{\ge0}}
q^{-\frac12m_{a-1}
%   +\frac14\sboldm\sboldC^{(p-2)}{\sboldm}^T}
   +\frac12\sum_{i=1}^{p-1}m_i(m_i-m_{i+1})}
\prod_{i=1}^{p-2}
\qbinom{n_i+m_i}{n_i},
\end{equation}
with $\boldn$ constrained such that \eqref{Eq:MNsys4}
yields $m_{p-1}=0$ and $m_p=1$.
The constraints \eqref{Eq:MNsys4} are readily solved to yield
\begin{equation}\label{Eq:MNsol4}
m_i=2\sum_{i<k<p}(k-i)n_k -\pabs{a-1-i} -\pabs{p-b-i} -p+1+i
\qquad(0\le i<p).
\end{equation}
Then, summing \eqref{Eq:Ball4eval} over $m_1$,
using Lemma \ref{Lem:mSwitch2}(2),
proves (\ref{Eq:ABFL}d).

%%%%%%%%%%%%%%%%%%%%%%%%%%%%%%%%%%%%%%%%%%%%%%%%%%%%%%%%%%%%%%%%%%%%%%
%%%%%%%%%%%%%%%%%%%%%%%%%%%%%%%%%%%%%%%%%%%%%%%%%%%%%%%%%%%%%%%%%%%%%%

\section{Half-lattice paths}\label{Sec:HL}

\subsection{Half-lattice path definition}
\label{Sec:HLpaths}

A half-lattice path $\hh$ of length $L\in\HZZ_{\ge0}$
is a finite sequence
$\hh=(\hh_{-1/2},\hh_0,\hh_{1/2},\hh_1,\linebreak[1]\hh_{3/2},\ldots,
\hh_L,\hh_{L+1/2})$
satisfying $\hh_x\in\HZZ$ and
$|\hh_x-\hh_{x-1/2}|=\tfrac12$ for each $x\in\HZZp$,
with the \emph{additional} restriction that if
$\hh_x=\hh_{x+1}\in\ZZ$, then $\hh_{x+1/2}=\hh_x+1/2$.
A half-lattice path $\hh$ is said to be $(f,g)$-restricted if
$f\le \hh_x\le g$ for $0\le x\le L$.
For $t,\ha,\hb,L\in\HZZ$, define $\HLtabLef$ to be the set of all
length $L$ half-lattice paths $\hh$ that are $(1,t)$-restricted
with $\hh_0=\ha$, $\hh_L=\hb$, $\hh_{-1/2}=\ha+1/2-e$,
$\hh_{L+1/2}=\hb+1/2-f$.

\begin{figure}[ht]
\caption{Half-lattice path $\hh\in\mathcal{H}_{4,3}^{e,f;4}(25)$}
\vskip-0.2cm
\label{TypicalHalfHat}
\begin{center}
\psset{yunit=0.32cm,xunit=0.28cm}
\begin{pspicture}(0,-0.5)(50,8.5)
%dotted grid
\psset{linewidth=0.25pt,linestyle=dotted, dotsep=1.0pt,linecolor=gray}
\psline{-}(0,2)(50,2) \psline{-}(0,4)(50,4) \psline{-}(0,6)(50,6)
%dashed grid
\psset{linewidth=0.25pt,linestyle=dashed, dash=2.5pt 1.5pt,linecolor=gray}
\psline{-}(0,3)(50,3) \psline{-}(0,5)(50,5)
\psline{-}(1,1)(1,7) \psline{-}(2,1)(2,7) \psline{-}(3,1)(3,7)
\psline{-}(4,1)(4,7) \psline{-}(5,1)(5,7) \psline{-}(6,1)(6,7)
\psline{-}(7,1)(7,7) \psline{-}(8,1)(8,7) \psline{-}(9,1)(9,7)
\psline{-}(10,1)(10,7) \psline{-}(11,1)(11,7) \psline{-}(12,1)(12,7)
\psline{-}(13,1)(13,7) \psline{-}(14,1)(14,7) \psline{-}(15,1)(15,7)
\psline{-}(16,1)(16,7) \psline{-}(17,1)(17,7) \psline{-}(18,1)(18,7)
\psline{-}(19,1)(19,7) \psline{-}(20,1)(20,7) \psline{-}(21,1)(21,7)
\psline{-}(22,1)(22,7) \psline{-}(23,1)(23,7) \psline{-}(24,1)(24,7)
\psline{-}(25,1)(25,7) \psline{-}(26,1)(26,7) \psline{-}(27,1)(27,7)
\psline{-}(28,1)(28,7) \psline{-}(29,1)(29,7) \psline{-}(30,1)(30,7)
\psline{-}(31,1)(31,7) \psline{-}(32,1)(32,7) \psline{-}(33,1)(33,7)
\psline{-}(34,1)(34,7) \psline{-}(35,1)(35,7) \psline{-}(36,1)(36,7)
\psline{-}(37,1)(37,7) \psline{-}(38,1)(38,7) \psline{-}(39,1)(39,7)
\psline{-}(40,1)(40,7) \psline{-}(41,1)(41,7) \psline{-}(42,1)(42,7)
\psline{-}(43,1)(43,7) \psline{-}(44,1)(44,7) \psline{-}(45,1)(45,7)
\psline{-}(46,1)(46,7) \psline{-}(47,1)(47,7) \psline{-}(48,1)(48,7)
\psline{-}(49,1)(49,7)
%axes
\psset{linewidth=0.25pt,fillstyle=none,linestyle=solid,linecolor=black}
%\psline{->}(0,1)(46.5,1)
\psline{-}(0,1)(50,1)
\psline{-}(0,7)(50,7)
\psline{-}(0,1)(0,7)
\psline{-}(50,1)(50,7)
%numeros
\rput(-0.5,1){\scriptsize $1$}\rput(-0.5,3){\scriptsize $2$}
\rput(-0.5,5){\scriptsize $3$}\rput(-0.5,7){\scriptsize $4$}
\rput(0,0.5){\scriptsize $0$} \rput(2,0.5){\scriptsize $1$}
\rput(4,0.5){\scriptsize $2$} \rput(6,0.5){\scriptsize $3$}
\rput(8,0.5){\scriptsize $4$} \rput(10,0.5){\scriptsize $5$}
\rput(12,0.5){\scriptsize $6$} \rput(14,0.5){\scriptsize $7$}
\rput(16,0.5){\scriptsize $8$} \rput(18,0.5){\scriptsize $9$}
\rput(20,0.5){\scriptsize $10$} \rput(22,0.5){\scriptsize $11$}
\rput(24,0.5){\scriptsize $12$} \rput(26,0.5){\scriptsize $13$}
\rput(28,0.5){\scriptsize $14$} \rput(30,0.5){\scriptsize $15$}
\rput(32,0.5){\scriptsize $16$} \rput(34,0.5){\scriptsize $17$}
\rput(36,0.5){\scriptsize $18$} \rput(38,0.5){\scriptsize $19$}
\rput(40,0.5){\scriptsize $20$} \rput(42,0.5){\scriptsize $21$}
\rput(44,0.5){\scriptsize $22$} \rput(46,0.5){\scriptsize $23$}
\rput(48,0.5){\scriptsize $24$} \rput(50,0.5){\scriptsize $25$}
%path
\psset{linewidth=0.7pt,fillstyle=none,linestyle=solid,linecolor=black}
\psline(0,7)(1,6)(2,5)(3,4)(4,3)(5,2)(6,1)(7,2)(8,1)(9,2)(10,3)(11,4)
 (12,5)(13,4)(14,3)(15,4)(16,5)(17,6)(18,5)(19,6)(20,5)(21,6)(22,5)
 (23,6)(24,5)(25,6)(26,7)(27,6)(28,5)(29,6)(30,5)(31,6)(32,5)(33,6)
 (34,5)(35,6)(36,7)(37,6)(38,5)(39,6)(40,5)(41,6)(42,7)(43,6)(44,5)
 (45,6)(46,5)(47,6)(48,5)(49,6)(50,5)
\end{pspicture}
\end{center}
\end{figure}
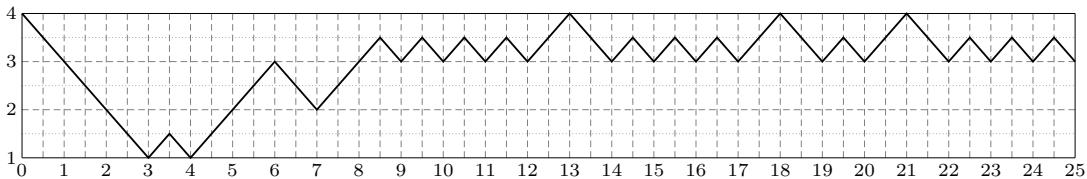

The \emph{path picture} of a half-lattice path $\hh\in\HLtabLef$
is obtained by linking the vertices
$(0,\hh_0), (1/2,\hh_{1/2}), (1,\hh_1),$ $\ldots,(L,\hh_L)$ on the plane.
The examples in Figsure \ref{TypicalDualBij} and \ref{TypicalHalfHat}
pertain to the $t=7/2$ and $t=4$ cases respectively.
When convenient, we also specify a path \emph{presegment} linking
$(-1/2,\hh_{-1/2})$ and $(0,\hh_0)$, and a
path \emph{postsegment} linking
$(L,\hh_{L})$ and $(L+1/2,\hh_{L+1/2})$.
The presegment is then in the SE direction if $e=0$,
and in the NE direction if $e=1$;
the postsegment is in the NE direction if $f=0$,
and in the SE direction if $f=1$.

A vertex $(x,\hh_x)$, for $x\in\HZZ$ with $0\le x\le L$, is said to
be a peak, a valley, straight-up or straight-down, depending on whether
the pair of edges that neighbour $(x,\hh_x)$ in this path picture
are in the NE-SE, SE-NE, NE-NE, or SE-SE directions respectively.
The specification of the path's presegment and postsegment
then determines the nature of the vertices at the path's
startpoint $(0,\ha)$ and endpoint $(L,\hb)$, respectively.
Note that the additional restriction above implies that
valleys can occur only at integer heights.

The weight $\hhunwt(\hh)$ of a half-lattice path $\hh\in\HLtabLef$
is defined by
\begin{equation}
\label{Eq:WtsDef}
\hhunwt(\hh)=\frac12 \sum_{i=1}^{2L}
i\, |\hh_{(i+1)/2}-\hh_{(i-1)/2}|,
\end{equation}
Thus $\hhunwt(\hh)$ is half the sum of the
$x\in\tfrac12{\ZZ_{\ge0}}$ for which $(x,\hh_x)$ is a straight-vertex.

We then define the generating function
\begin{equation}\label{Eq:HLGFfindef}
\HLGFe{t}{a,b}{e,f}(L)=
\HLGFe{t}{a,b}{e,f}(L;q)=
\sum_{\hh\in\HLtabLef} q^{\hhunwt(\hh)}.
\end{equation}

It is useful to note that for $a\in\ZZ$,
\begin{equation}\label{Eq:HLGFeq}
\HLGFe{t}{a,b}{0,f}(L)
=\HLGFe{t}{a,b}{1,f}(L).
\end{equation}
This follows because for $a\in\ZZ$, the two sets
$\HLsetef{t}{a,b}{0,f}$ and $\HLsetef{t}{a,b}{1,f}$ are
in bijection with $\hh\in\HLsetef{t}{a,b}{0,f}$ mapping
to $\hh'\in\HLsetef{t}{a,b}{1,f}$ with
$\hh'_x=\hh_x$ for $0\le x\le L$;
and then $\hhunwt(\hh)=\hhunwt(\hh')$.
For $a\in\ZZph$, \eqref{Eq:HLGFeq} doesn't hold, in general.

In what follows, we obtain bosonic expressions and fermionic
expressions for $\HLGFe{t}{a,b}{e,f}(L;q)$,
the former only for $a\in\ZZ$,
this being sufficient for our purposes.

\subsection{\texorpdfstring
             {Bosonic expressions and $q$-trinomials}
             {Bosonic expressions and q-trinomials}}
\label{Sec:BosonicGF}

The polynomials $Y^{n;t}_{a,b}(L)= Y^{n;t}_{a,b}(L;q)$
are defined for $t\in\HZZ$ and $L,a,b,n\in\ZZ$ by
\begin{equation}\label{Eq:Ydef}
Y^{n;t}_{a,b}(L;q)
=
\!
\sum_{\lambda=-\infty}^\infty
\!
\bigl(
q^{\lambda^2tt'+\lambda(t'b-2ta)}\mytri{n}{a-b-t'\lambda}{L}
-
q^{(\lambda t+b)(\lambda t'+2a)}\mytri{n}{-a-b-t'\lambda}{L}
\bigr),
\end{equation}
where $t'=2t+1$, and $\mytri{n}{d}{L}$ are the $q$-analogues
of trinomial coefficients defined by
\begin{equation}
\label{Eq:TriDefX}
\mytri{n}{d}{L}
=
\mytriq{n}{d}{L}{q}=
\sum_{k=0}^{\lfloor(L-d)/2\rfloor}
q^{k(k+d-n)}
\frac{(q)_L}{(q)_k(q)_{k+d}(q)_{L-2k-d}}.
\end{equation}
These $q$-trinomial coefficients $\mytri{n}{d}{L}$ were first
introduced by Andrews and Baxter \cite{AndrewsBaxter1987}.
Some of their properties are given in \elrm{Appendix }\ref{App:Trinoms}.

The next theorem, which is proved in Section \ref{Sec:BosonicProof}
below, shows that for $a\in\ZZ$, the generating functions
$\HLGFe{t}{a,b}{e,f}(L)$ are, up to normalisation,
given by $Y^{f;t}_{a,b}(L)$ when $b\in\ZZ$ or $f=0$,
and by a sum of two such terms when $b\in\ZZph$ and $f=1$.

\begin{theorem}\label{Thrm:FinGen}
Let $t\in\HZZ$ and $a,b,L\in\ZZ$ with $1\le a,b\le t$.
If $e,f\in\{0,1\}$, then
\begin{subequations}
\begin{equation}
\label{Eq:FinGen}
\HLGFe{t}{a,b}{e,f}(L)
=q^{\frac12(a-b)(a-b-\frac12)+\frac12fL}\,Y^{f;t}_{a,b}(L).
\end{equation}
If $b<t$ and $e\in\{0,1\}$, then
\begin{align}
\label{Eq:FinGenHb1}
\HLGFe{t}{a,b+1/2}{e,1}(L+1/2)
&=q^{\frac12(a-b)(a-b-\frac12)}\,Y^{0;t}_{a,b}(L)
+q^{L+1+\frac12(a-b)(a-b-\frac52)}\,Y^{1;t}_{a,b+1}(L),
\\
\label{Eq:FinGenHb2}
\HLGFe{t}{a,b+1/2}{e,0}(L+1/2)
&=q^{\frac12L+\frac14+\frac12(a-b)(a-b-\frac12)}\,Y^{0;t}_{a,b}(L).
\end{align}
\end{subequations}
\end{theorem}

\subsection{Infinite length half-lattice paths and Virasoro characters}

Applying the $L\to\infty$ properties \eqref{Eq:U0limit} and
\eqref{Eq:U1limit} of the $q$-trinomials to \eqref{Eq:Ydef}
and comparing the result with \eqref{Eq:Rocha} leads to:
\begin{align}
\label{Eq:Y0lim}
\lim_{L\to\infty} Y^{0;t}_{a,b}(L)
&=
\chi^{t,2t+1}_{b,2a},\\
\label{Eq:Y1lim}
\lim_{L\to\infty} Y^{1;t}_{a,b}(L)
&=
\chi^{t,2t+1}_{b,2a} + q^{a-b} \chi^{t,2t+1}_{b-1,2a}.
\end{align}
Theorem \ref{Thrm:FinGen} then immediately implies:
\begin{corollary}\label{Cor:HLGFvir}
If $t\in\HZZ$, $e\in\{0,1\}$ and $a,b\in\ZZ$ with
$1\le a\le t$ and $1\le b<t$
then
\begin{subequations}
\begin{align}
\label{Eq:HLGFvir1}
\lim_{L\to\infty}\HLGFe{t}{a,b}{e,0}(L)
&=q^{\frac12(a-b)(a-b-\frac12)}
\,\chi^{t,2t+1}_{b,2a}
,\\
\label{Eq:HLGFvir2}
\lim_{L\to\infty}\HLGFe{t}{a,b+1/2}{e,1}(L)
&=q^{\frac12(a-b)(a-b-\frac12)}
\,\chi^{t,2t+1}_{b,2a}
,\\
\label{Eq:HLGFvir3}
\lim_{L\to\infty}q^{-\frac12L}\HLGFe{t}{a,b+1/2}{e,0}(L)
&=q^{\frac12(a-b)(a-b-\frac12)+\frac14}
\,\chi^{t,2t+1}_{b,2a}
,\\
\label{Eq:HLGFvir4}
\lim_{L\to\infty}q^{-\frac12L}\HLGFe{t}{a,b}{e,1}(L)
&=q^{\frac12(a-b)(a-b-\frac12)}
\Bigl(
\chi^{t,2t+1}_{b,2a}
+q^{a-b}\chi^{t,2t+1}_{b-1,2a}
\Bigr)
\qquad(b>1).
\end{align}
\end{subequations}
\end{corollary}

The functions on the left sides of \eqref{Eq:HLGFvir1}
and \eqref{Eq:HLGFvir2} are readily interpreted as the
generating functions of
infinite-length half-lattice paths as follows.
For $t\in\HZZ$ and $a,b\in\ZZ$ with $1\le a,b\le t$,
define $\HLtabinf$ to be the set of half-lattice paths
$\hh=(\hh_0,\hh_{1/2},\hh_1,\hh_{3/2},\ldots)$
that are $(1,t)$-restricted
with $\hh_0=\ha$, and are $\hb$-tailed in that for each $\hh$,
there exists $L$
for which $\hh_i\in\{\hb,\hb+\frac12\}$ for all $i\in\HZZ$
with $i\ge L$.
Let $L(\hh)$ be the smallest such $L$, and define
\begin{equation}
\label{Eq:WtsInfDef}
\hhunwt(\hh)=\frac12 \sum_{i=1}^{2L(\hh)}
i\, |\hh_{(i+1)/2}-\hh_{(i-1)/2}|.
\end{equation}
Then define the generating function
\begin{equation}\label{Eq:HLGFinfdef}
\HLGFinf{t}{a,b}(q)
=\sum_{\hh\in\HLtabinf} q^{\hhunwt(\hh)}.
\end{equation}

\begin{theorem}\label{Thrm:HLvir}
Let $t\in\HZZ$ and $a,b\in\ZZ$ with $1\le a\le t$ and $1\le b<t$.
Then
\begin{equation}\label{Eq:HLvir}
q^{-\frac12(a-b)(a-b-\frac12)}
\HLGFinf{t}{a,b}(q)=\chi^{t,2t+1}_{b,2a}\,.
\end{equation}
\end{theorem}

\Proof
From \eqref{Eq:HLGFfindef} and \eqref{Eq:HLGFinfdef},
we see that
\begin{equation}\label{Eq:HLGFinf}
\HLGFinf{t}{a,b}(q)
=\lim_{L\to\infty} \HLGFe{t}{a,b}{e,0}(L;q)
=\lim_{L\to\infty} \HLGFe{t}{a,b+1/2}{e,1}(L;q).
\end{equation}
for $e\in\{0,1\}$.
Corollary \ref{Cor:HLGFvir} thus yields
\eqref{Eq:HLvir}
\cqfd

This result, which shows that,
up to normalisation, $\chi^{t,2t+1}_{b,2a}$ is
the generating function for infinite length half-lattice paths,
was established in \cite{BfMW2012} by the alternative
method of constructing
a weight-preserving bijection between those
paths and the RSOS paths \cite{FB1985,FLPW2000} of $M(p,2p\pm1)$.

By virtue of \eqref{Eq:FinGen}, \eqref{Eq:HLGFvir1}
and \eqref{Eq:RochaSwitch},
the $t\in\ZZ$ and $t\in\ZZph$ cases of $Y^{0;t}_{a,b}(L;q)$
are finitizations of characters $\chi^{p,p'}_{r,s}$
for which $p'=2t+1$ and $p'=2t$ respectively.
This, in particular, accounts for the appearance in
\cite[eqn.~(9.4)]{BMcP1998}
of the two different finitizations of $\chi^{p,p'}_{r_0,s_0}$ defined in
\cite[eqns.~(3.10) and (5.7)]{BMcP1998}.

It is intriguing to note the related fact that,
for the $\chi^{p,2p-1}_{r,s}$ characters,
the startpoint and tail configuration of the half-lattice
paths being enumerated are determined by $r$ and $s$ respectively.
This is in contrast to the $\chi^{p,2p+1}_{r,s}$ case,
where they are determined by $s$ and $r$ respectively.
The latter statement also applies to the RSOS paths
\cite{FB1985,FLPW2000}
for all $\chi^{p,p'}_{r,s}$.

\subsection{Fermionic expressions for half-lattice paths}

In this section, we state fermionic expressions for the
half-lattice path generating functions $\HLGFe{t}{a,b}{e,f}(L;q)$.
These expressions will be proved in Section \ref{Sec:Refine} below,
by extending the combinatorial techniques of
Sections \ref{Sec:CombTran} and \ref{Sec:ABFfin}.

\begin{theorem}\label{Thrm:Yfinferms}
Let $a,b,t,L\in\HZZ$ with $1\le a,b\le t$ and $L+a+b\in\ZZ$, 
and set $\boldC=\boldC^{(2t-3)}$.
Then, for $e,f\in\{0,1\}$, we have the following expression for
$\HLGFe{t}{a,b}{e,f}(L)$ in which the values of
$\ell$, $\Delta_i$,
$\boldQL=(\QL_1,\ldots,\QL_{2t-2})$ and $\boldQR=(\QR_1,\ldots,\QR_{2t-2})$,
are as given in Table~\ref{tabHLL} below:
\begin{equation}\label{Eq:Y1234}
\begin{split}
\HLGFe{t}{a,b}{e,f}(L)
=
&\sum_{\sboldn\in\ZZ^{2t-2}_{\ge0}|m_0=2L}
\hskip-5mm
q^{\frac18\sboldm\sboldC\sboldm^T
   -\frac14 m_{\ell}
   +\frac12\sboldn\cdot\sboldQR}
\prod_{i=1}^{2t-3}
\qbinom{n_i+\hat m_i}{n_i}^\prime\,,\\
\text{where}\quad
m_i=2\!\!\sum_{i<k<2t-1}
(k-i)n_k -\Delta_i
\quad&\text{for}\quad
0\le i\le2t-2,
\quad\text{and}\quad 
\hat m_i=\frac12(m_i-\QL_i-\QR_i)
\quad\text{for}\quad
1\le i\le2t-2.
\end{split}
\end{equation}
Here, the sum is over all non-negative integer vectors
$\boldn=(n_1,n_2,\ldots,n_{2t-2})$, with
$\boldmp=(m_0,m_1,\ldots,m_{2t-3})$ obtained from $\boldn$
as indicated, and $\boldm=(m_1,m_2,\ldots,m_{2t-3})$.
Note that by virtue of the restriction $m_0=2L$,
the sum is, in effect, finite.
Also note that $m_{2t-2}=0$ in each case, and for each $i<2t-2$,
$\hat m_i$ is an integer equal to either
$\frac12m_i$, $\frac12(m_i-1)$ or $\frac12m_i-1$.

\begin{table}[ht]
\caption{Parameters for the four cases of the expressions \eqref{Eq:Y1234}.} 
\vskip.2cm
\label{tabHLL}
\begin{center}
\begin{tabular}{|l|c||c|c|c|c|c|c| }\hline
\mystrut&\textup(e,f\textup)& $a$& $b$&$\ell$
  & $\Delta_i$ & $\boldQL$ & $\boldQR$ \\
\hline\hline
\mystrut{\rm (a)}&\textup(1,1\textup)
  & $\ne 1$ & $\ne1$& $2a-2$
  & $\pabs{2a-2-i}+\pabs{2b-2-i}$
  & $\boldQ^{(2a-1,2t-2)}$
  & $\boldQ^{(2b-1,2t-2)}$
\\
\hline
\mystrut{\rm (b)}&\textup(0,1\textup)
  & $\ne t$ &$\ne1$& $2t-2a$
  & $\pabs{2t-2a-i} +\pabs{2b-2-i} +2t-2-i$
  & $\boldR^{(2a-1,2t-2)}$
  & $\boldQ^{(2b-1,2t-2)}$
\\
\hline
\mystrut{\rm (c)}&\textup(0,0\textup)
  & $\ne t$ & $\ne t$ &$2t-2a$
  & $\pabs{2t-2a-i}+\pabs{2t-2b-i}$
  & $\boldR^{(2a-1,2t-2)}$
  & $\boldR^{(2b-1,2t-2)}$
\\
\hline
\mystrut{\rm (d)}&\textup(1,0\textup)
  & $\ne 1$ &$\ne t$ & $ 2a-2$
  & $\pabs{2a-2-i} +\pabs{2t-2b-i} +2t-2-i$
  & $\boldQ^{(2a-1,2t-2)}$
  & $\boldR^{(2b-1,2t-2)}$
\\
\hline
\end{tabular}
\end{center}
\end{table}
\end{theorem}

Theorem \ref{Thrm:HLferms} follows from Theorem \ref{Thrm:Yfinferms}
by taking the $L\to\infty$ limit of \eqref{Eq:Y1234} through
taking $n_1\to\infty$, and using
\eqref{Eq:HLGFvir1} and \eqref{Eq:HLGFvir2}.
Using \eqref{Eq:HLGFvir3} in the same way leads to expressions
equivalent to those already obtained.
On the other hand,
using \eqref{Eq:HLGFvir4} in the same way leads to fermionic
expressions for the combination
$\chi^{t,2t+1}_{b,2a}+q^{a-b}\chi^{t,2t+1}_{b-1,2a}$
of two characters.
We don't give these expressions explicitly, but they are easily obtained.
Note that for the cases \eqref{Eq:HLGFvir3} and \eqref{Eq:HLGFvir4},
in taking the $L\to\infty$ limit of \eqref{Eq:Y1234},
the $q^{-L/2}$ factor in the former is compensated for by
a $q^{L/2}$ factor arising from the 1st component of
$\frac12\boldn\cdot\boldQR$ in the exponent in \eqref{Eq:Y1234}.

%%%%%%%%%%%%%%%%%%%%%%%%%%%%%%%%%%%%%%%%%%%%%%%%%%%%%%%%%%%%%%%%%%%%%%
%%%%%%%%%%%%%%%%%%%%%%%%%%%%%%%%%%%%%%%%%%%%%%%%%%%%%%%%%%%%%%%%%%%%%%

\section{Non-unitary Melzer-type identities:
         Refining the combinatorial transforms}
\label{Sec:Refine}

A refinement of the strategy used above is now used to obtain
generating functions for the half-lattice paths.
First note that because the half-lattice paths are defined on
an half-integer lattice, they may be obtained by shrinking the
ABF paths by a factor of 2, and then excluding all paths that
have valleys at half-integer heights.
We will use the equivalent approach of first excluding ABF paths
that have valleys at even heights before shrinking them.
We say that an ABF path is \emph{valley-restricted} if
it has no valley at an even height.

The key to obtaining the fermionic generating functions for
the valley-restricted ABF paths is to note that for both of the
particle moves depicted in Figure \ref{Fig:BasicMoves},
a valley at an even or odd height is exchanged for one at
a height of the opposite parity (the same is also true of a peak).
Thus, to retain a valley-restricted path,
it is necessary for each particle to make moves in steps of two.
Therefore in the valley-restricted case, a particle can make, roughly,
half as many moves as in the ABF case.
A minor complication arises because the $\Tran_2(n)$-transform
sometimes inserts particles with their valleys lying at an
even height, as in the example of Figure \ref{Fig:B2transF1}.
This requires their initial position to be excluded.
Care must then be taken, however, when $n=0$,
for naively excluding the initial position in this case,
would exclude a valid path.
As will be seen, obviating this necessitates use of the modified
$q$-binomial defined by \eqref{Eq:qModBinomialDef}.

\subsection{Valley-restricted ABF paths}

Let $h\in\ABFpabLMef$, and let $\vword(\habf)=v_0v_1\cdots v_L$
contain $k+1$ symbols $N$ (here $k\ge -1$).
Because $\vword(\habf)$ contains $m$ symbols $S$ then $m+k=L$.
Let the indices of the symbols $N$ in
$\vword(\habf)$ be $j_0,j_1,j_2,\ldots,j_k$.
For $e=0$, the valleys are then at positions $j_0,j_2,j_4,\ldots,$
while for $e=1$, the valleys are at positions $j_1,j_3,j_5,\ldots.$
Then the number of valleys in $\habf$ that are at even height
is given by $\evenval(\habf)$, where we define
\begin{equation}\label{Eq:evenvalDef}
\evenval(\habf)=
\#\{i\,|\,0\le i\le k, i\equiv e, j_i\not\equiv a\}.
\end{equation}
We are, of course, particularly interested in the paths $h$
for which $\evenval(\habf)=0$,
these being the valley-restricted paths.

Thereupon, for $e,f\in\{0,1\}$, we define the generating functions
\begin{equation}\label{Eq:RABFgfDef}
\RABFGFe{p}{a,b}{e,f}(L)=
\RABFGFe{p}{a,b}{e,f}(L;q)=
\sum_{\substack{\habf\in\ABFpabLMef\\
                \evenval(\habf)=0}}
q^{\unwt(\habf)}
\end{equation}
and
\begin{equation}\label{Eq:RABFMgfDef}
\RABFGFe{p}{a,b}{e,f}(L,m)=
\RABFGFe{p}{a,b}{e,f}(L,m;q)=
\sum_{\substack{\habf\in\ABFpabLMef\\
                \evenval(\habf)=0}}
q^{\unwt(\habf)}.
\end{equation}
In what follows, we require analogues of Lemmas \ref{Lem:mSwitch2}
and \ref{Lem:Seed} for these generating functions.
\begin{lemma}\label{Lem:RmSwitch2}
Let $1\le a,b\le p$ and $L\ge0$ and $e,f\in\{0,1\}$.
Then:
\begin{enumerate}
\item If $m\not\equiv L+e+f$ then
  $\RABFGFe{p}{a,b}{e,f}(L,m)=0$;
\item
  $\RABFGFe{p}{a,b}{e,f}(L)
   =\sum_{m\ge0} \RABFGFe{p}{a,b}{e,f}(L,m)$;
\item $\RABFGFe{p}{1,b}{1,f}(L,m)=\RABFGFe{p}{1,b}{0,f}(L,m-1)$;
\item $\RABFGFe{p}{p,b}{0,f}(L,m)=\RABFGFe{p}{p,b}{1,f}(L,m-1)$;
\item $\RABFGFe{p}{a,1}{e,1}(L,m)=q^{L/2}\RABFGFe{p}{a,1}{e,0}(L,m-1)$;
\item $\RABFGFe{p}{a,p}{e,0}(L,m)=q^{L/2}\RABFGFe{p}{a,p}{e,1}(L,m-1)$.
\end{enumerate}
\end{lemma}
\Proof
The first two cases follow from Lemma \ref{Lem:mSwitch}(1)
and the definitions \eqref{Eq:ABFgfDef} and \eqref{Eq:ABFMgfDef}.
The other cases follow from Lemmas \ref{Lem:wtSwitch} and
\ref{Lem:mSwitch} and the definition \eqref{Eq:ABFMgfDef}.
\cqfd

\begin{lemma}\label{Lem:RSeed}
Let $1\le a,b\le p$ and $L,m\ge0$ and $e,f\in\{0,1\}$.
Then:
\begin{enumerate}
\item $\RABFGFe{p}{a,b}{e,f}(0,m)=
  \begin{cases}
     0 & \text{if $e=f=0$ and $a$ is even,} \\
     \delta_{a,b}\,\delta_{m,|e-f|} & \text{otherwise};
  \end{cases}
$
\item $\RABFGFe{1}{1,1}{e,f}(L,m)=\delta_{L,0}\,\delta_{m,|e-f|}$;
\item $\RABFGFe{p}{a,b}{e,f}(L,0)=
  \begin{cases}
     0 & \text{if $a$ is even and $e=0$,}\\
     0 & \text{if $a$ is odd and $e=1$ and $L>0$,} \\
     \delta_{a-e,b-f}\,\delta_{(L+e+f)\bmod2,0} & \text{otherwise}.
  \end{cases}
$
\end{enumerate}
\end{lemma}
\Proof 
If $a\ne b$ then $\ABFsetef{p}{a,b}{e,f}(0)=\emptyset$.
On the other hand, if $a=b$ then $\ABFsetef{p}{a,b}{e,f}(0)$
contains a single element $\habf$
for which $\vword(\habf)=N$ if $e=f$, and
$\vword(\habf)=S$ if $e\ne f$.
For these two cases, we then have $\mnos(\habf)=0$ and
$\mnos(h)=1$ respectively,
with $\evenval(\habf)\ne0$ only if $e=f=0$ and $a$ even.
Then, after noting that $\unwt(\habf)=0$ in both cases,
the first result follows from the definition \eqref{Eq:RABFMgfDef}.

The second result follows from the first after it is noted
that $\ABFsetef{1}{1,1}{e,f}(L)=\emptyset$ for $L>0$.

For the third result, first note that if $\mnos(\habf)=0$
then the segments of the path $\habf$, together with its
presegment and postsegment, necessarily alternate in direction.
As indicated in Figure \ref{Fig:ZigZags},
there can only be one such path,
for which $e=f$ if and only if $L$ is even.
Furthermore, if $L$ is even then necessarily $a=b$,
and if $L$ is odd then necessarily $|a-b|=1$,
with $b=a-1$ if $e=1$ and $b=a+1$ if $e=0$.
As seen in Figure \ref{Fig:ZigZags},
the valleys of such a path would all occur at height $a-e$,
yet if (and only if) $L=0$, $a$ is odd and $e=1$,
there are none at even height.
The third result follows.

\subsection{\texorpdfstring
             {Refining the $\Tran$-transform}
             {Refining the C-transform}}

Let the path $\hbone\in\ABFsetef{p+1}{a+e,b+f}{e,f}(L')$ result from the
action of the $\Tran_1$-transform on $\habf\in\ABFpabLef$.
If the indices of the symbols $N$ in
$\vword(\habf)$ are $j_0,j_1,j_2,\ldots,j_k$
then, by the definition in Section \ref{Sec:B1trans}, those of
$\vword(\hbone)$ are $j_0,j_1+1,j_2+2,\ldots,j_k+k$
(again we avoid considering the case where $L=0$ and $e\ne f$).
We see from \eqref{Eq:evenvalDef} that
$\evenval(\habf')=\evenval(\habf)$.
In particular, if either $\habf$ or $\hbone$ has no valley at an
even height, then the same is true of the other.

So now assume that $\hbone$ is valley-restricted.
The $\Tran_2(n)$-transform maps $\hbone$ to a path $\hbnnn$ by
appending to the former an alternating sequence of peaks and valleys,
$n$ of each.
We see that the newly created valleys are each at height $b-f$
(with the newly created peaks each at height $b-f+1$).
Thus, if $b-f$ is even, the path $\hbnnn$ itself should not
be included when enumerating the valley-restricted paths.
However, paths obtained from $\hbnnn$ by moving particles
should be included.

As indicated above, each particle move,
although maintaining the number of valleys,
changes the parity of the height of one valley.
Therefore, if an ABF path is valley-restricted,
then moving the particles in steps of two maintains the restriction.
Thus for $\hbnnn$ obtained above, in the case where $b-f$ is odd,
the action of $\Tran_3(\lambda)$-transform on $\hbnnn$ results in
a valley-restricted path $\habf'$ if each part $\lambda_i$ of $\lambda$
is even.
On the other hand, if $b-f$ is even, then valley-restricted
paths are obtained by moving each particle in $\hbnnn$
by a single initial step, and thereafter in steps of two.
Thus, in this case, the $\Tran_3(\lambda)$-transform acting
on $\hbnnn$ results in a valley-restricted path $\habf'$ if each
$\lambda_i$ is odd.
Figures \ref{Fig:B2transF0} and \ref{Fig:B2transF1} provide
examples for which $b-f$ is odd and even respectively.

\subsection{Transforming the valley-restricted generating function}

We now use this refined $\Tran$-transform to describe a bijection
involving sets of restricted paths.

\begin{lemma}\label{Lem:RBtransform}
Let $1\le a,b\le p$ and $L>0$ and $L'\ge0$ and $e,f\in\{0,1\}$,
and set $\QL=(a+1)\bmod2$ and $\QR=(b+1)\bmod2$.
Then there is a bijection between the sets
\begin{equation}\label{Eq:RBtransform}
\RABFsetef{p+1}{a+e,b+f}{e,f}(L',L)
\longleftrightarrow
\bigcup_{n\ge0}
\RABFsetef{p}{a,b}{e,f}(L,2L-L'+2n)
\times
\mathcal{P}_{n,(L-\QL-\QR)/2},
\end{equation}
under which,
if $h'\in\RABFsetef{p+1}{a+e,b+f}{e,f}(L',L)$ maps to $(h,n,\mu)$,
where $h\in\RABFsetef{p}{a,b}{e,f}(L,m)$ with $m=2L-L'+2n$,
and $\mu\in\mathcal{P}_{n,(L-\QL-\QR)/2}$, then
\begin{equation}\label{Eq:RBtransformWt}
\unwt_{f}(h')=
\unwt_{f}(h)+\frac12 L(L-m)+2|\mu|+n\QR.
\end{equation}
\end{lemma}

\Proof
The sets
$\RABFsetef{p+1}{a+e,b+f}{e,f}(L',L)$
and
$\RABFsetef{p}{a,b}{e,f}(L,m)$ of paths
are subsets of
$\ABFsetef{p+1}{a+e,b+f}{e,f}(L',L)$
and
$\ABFsetef{p}{a,b}{e,f}(L,m)$
respectively.
We claim that the bijective map of Lemma \ref{Lem:Btransform}
remains a bijection under restriction to these subsets
with $\lambda\in\mathcal{P}_{n,L}$
there obtained from $\mu\in\mathcal{P}_{n,(L-\QL-\QR)/2}$
by setting $\lambda_i=2\mu_i+\QR$ for $1\le i\le n$.
Note then that $\lambda_i\equiv\QR$ for $1\le i\le n$.

Throughout this proof, let $\habf\in\ABFsetef{p}{a,b}{e,f}(L,2L-L'+2n)$
and let $\habf'\in\ABFsetef{p+1}{a+e,b+f}{e,f}(L',L)$ result from
the action of the $\Tran(n,\lambda)$-transform on $\habf$.
Also, let the path $h^{(0)}\in\ABFsetef{p+1}{a+e,b+f}{e,f}(L'-2n)$
be obtained by the action of the $\Tran_1$-transform on $\habf$,
and $h^{(n)}\in\ABFsetef{p+1}{a+e,b+f}{e,f}(L')$ be obtained
from the action of the $\Tran_2(n)$-transform on $h^{(0)}$.
The path $\habf'$ is then obtained by the
$\Tran_3(\lambda)$-transform acting on $h^{(n)}$.
We claim that
$\habf'\in\RABFsetef{p+1}{a+e,b+f}{e,f}(L',L)$
if and only if both
$\habf\in\RABFsetef{p}{a,b}{e,f}(L,2L-L'+2n)$
and $\lambda_i\equiv\QR$ for $1\le i\le n$.

Assuming the latter, the above discussion shows that the
path $h^{(0)}$ is valley-restricted.
It also shows that each of the $n$ particles in $h^{(n)}$
has a valley at height $b$.
Because each move switches the parity of a valley,
if $\lambda_i\equiv\QR$ for $1\le i\le n$,
all the valleys in $\habf'$ are at odd height,
thus establishing the `if' part of the above claim.

For the `only if' part,
first consider $\lambda_i\not\equiv\QR$ for some $i$.
The above discussion immediately shows that
the $i$th particle in $h'$ has a valley at even height.
Secondly, if $h\not\in\RABFsetef{p}{a,b}{e,f}(L,2L-L'+2n)$
then $h$ has a valley at even height, and thus so does $h^{(0)}$.
In $h^{(0)}$, this valley has no adjacent non-straight vertex,
and is thus not part of a particle.
This feature remains in $h^{(n)}$.
It also remains however the particles in $h^{(n)}$ are moved.
Thus $h'\not\in\RABFsetef{p+1}{a+e,b+f}{e,f}(L',L)$ and the `only if'
part of the above claim is established.

The Lemma is then proved, with \eqref{Eq:RBtransformWt} following
from \eqref{Eq:BtransformWt}
after noting that $|\lambda|=2|\mu|+n\QR$.
\cqfd

\begin{corollary}\label{Cor:RBtransform}
Let $1\le a,b\le p$ and $L,L'\ge0$ and $e,f\in\{0,1\}$,
and set $\QL=(a+1)\bmod2$ and $\QR=(b+1)\bmod2$.
Then:
\begin{equation}\label{Eq:RBtransformGF}
\RABFGFe{p+1}{a+e,b+f}{e,f}(L',L)
=
\sum_{n\ge0}
q^{\frac12 L(L-m)+n\QR}
\qbinomq{n+\frac12(L-\QL-\QR)}{n}{q^2}'
\RABFGFe{p}{a,b}{e,f}(L,m),
\end{equation}
where $m$ is obtained from $n$ via $m=2L-L'+2n$.
\end{corollary}

\Proof In the $L>0$ cases,
this follows from equating the generating functions of the
two sides of \eqref{Eq:RBtransform}, using \eqref{Eq:RBtransformWt},
and \eqref{Eq:PartitionGF} in the form
\begin{equation}\label{Eq:RPartitionGF}
\sum_{\mu\in\mathcal{P}_{n,(L-\QL-\QR)/2}} q^{2|\mu|}=
\qbinomq{n+\frac12(L-\QL-\QR)}{n}{q^2}.
\end{equation}

The $L=0$ case is somewhat tricky and requires the consideration
of various subcases.
In each of these subcases, we use
Lemma \ref{Lem:RSeed}(3,1) to evaluate
$\RABFGFe{p+1}{a+e,b+f}{e,f}(L',0)$
and
$\RABFGFe{p}{a,b}{e,f}(0,m)$
on the two sides of \eqref{Eq:RBtransformGF}.
For $a\ne b$, we immediately see that both sides of
\eqref{Eq:RBtransformGF} are zero.
Thus we only consider the $a=b$ cases hereafter.

In the $e\ne f$ cases,
$\RABFGFe{p+1}{a+e,b+f}{e,f}(L',0)=1$
if both $L'$ and $a$ are odd, and is zero otherwise.
In the $e\ne f$ cases,
$\RABFGFe{p}{a,b}{e,f}(0,m)=1$
if $m=1$, and is zero otherwise.
Because $m=2n-L'$ here,
the RHS is non-zero only if $L'$ is odd.
Then, for $a$ odd,
so that $\QL=\QR=0$,
we see that the RHS is 1,
whereas for $a$ even,
so that $\QL=\QR=1$,
we see that the RHS is $q^n\qbinomq{n-1}{n}{q^2}'$
for $n=(L'+1)/2>0$,
and is thus zero in this case.

For the $e=f=0$ cases,
$\RABFGFe{p+1}{a+e,b+f}{e,f}(L',0)=1$
if $L'$ is even and $a$ is odd, and is zero otherwise.
For the $e=f=0$ cases,
$\RABFGFe{p}{a,b}{e,f}(0,m)=1$
if $m=0$ and $a$ is odd, and is zero otherwise.
Because $m=2n-L'$ here,
if $L'$ is even and $a$ is odd,
we see that the RHS of \eqref{Eq:RBtransformGF} is also $1$
after noting that $\QL=\QR=0$,
and is zero otherwise.

For the cases $e=f=1$, we consider separately the odd and even
cases of $a$.
For $a$ odd,
$\RABFGFe{p+1}{a+e,b+f}{e,f}(L',0)=1$
if $L'$ is even, and zero if $L'$ is odd.
For the cases $e=f=1$ with $a$ odd,
$\RABFGFe{p}{a,b}{e,f}(0,m)=1$
if $m=0$, and zero otherwise.
Because $m=2n-L'$ here,
if $L'$ is even,
we see that the RHS of \eqref{Eq:RBtransformGF} is also $1$
after noting that $\QL=\QR=0$,
and is zero otherwise.

Finally, for $e=f=1$ and $a$ even,
$\RABFGFe{p+1}{a+e,b+f}{e,f}(L',0)=1$
if $L'=0$, and is zero otherwise.
For $e=f=1$ and $a$ even,
$\RABFGFe{p}{a,b}{e,f}(0,m)=1$
if $m=0$, and is zero otherwise.
Because $m=2n-L'$ here,
we see that for the RHS of \eqref{Eq:RBtransformGF} to be non-zero
requires $L'$ to be even.
However, $a$ and $b$ being even yields $\QL=\QR=1$,
and therefore the $q$-binomial on the RHS of
\eqref{Eq:RBtransformGF} takes the form $\qbinomq{n-1}{n}{q^2}'$.
In this case, the definition of the modified $q$-binomial
ensures that the $n=0$ term contributes $1$ if and only if $L'=0$.
The proof of Corollary \ref{Cor:RBtransform} is then complete.
\cqfd

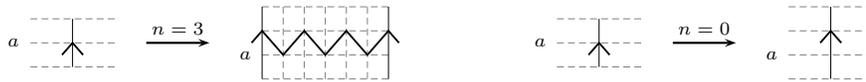
\begin{figure}[ht]
\caption{Explaining the necessity of the modified $q$-binomial}
\label{Fig:Tricksy}
\begin{center}
\psset{yunit=0.32cm,xunit=0.28cm}
\begin{pspicture}(0,0)(40,5.5)
\rput[bl](0,0.5){
%dashed grid
\psset{linewidth=0.25pt,linestyle=dashed, dash=2.5pt 1.5pt,linecolor=gray}
\psline{-}(0,1)(4,1) \psline{-}(0,2)(4,2) \psline{-}(0,3)(4,3)
%axes
\psset{linewidth=0.25pt,fillstyle=none,linestyle=solid,linecolor=black}
\psline{-}(2,1)(2,3)
%path
\psset{linewidth=0.7pt,fillstyle=none,linestyle=solid,linecolor=black}
\psline(1.5,1.5)(2,2)(2.5,1.5)
%annotate
\rput(-0.8,2){\scriptsize $a$}
}
\psline{->}(5.5,2.5)(8.5,2.5)
\rput(7.0,3.1){\scriptsize $n=3$}
\rput[bl](11,0){
%dashed grid
\psset{linewidth=0.25pt,linestyle=dashed, dash=2.5pt 1.5pt,linecolor=gray}
\psline{-}(0,1)(6,1) \psline{-}(0,2)(6,2)
\psline{-}(0,3)(6,3) \psline{-}(0,4)(6,4)
\psline{-}(1,1)(1,4) \psline{-}(2,1)(2,4) \psline{-}(3,1)(3,4)
\psline{-}(4,1)(4,4) \psline{-}(5,1)(5,4)
%axes
\psset{linewidth=0.25pt,fillstyle=none,linestyle=solid,linecolor=black}
\psline{-}(0,1)(0,4) \psline{-}(6,1)(6,4)
%path
\psset{linewidth=0.7pt,fillstyle=none,linestyle=solid,linecolor=black}
\psline(-0.5,2.5)(0,3)(1,2)(2,3)(3,2)(4,3)(5,2)(6,3)(6.5,2.5)
%annotate
\rput(-0.8,2){\scriptsize $a$}
}
\rput[bl](25,0.5){
%dashed grid
\psset{linewidth=0.25pt,linestyle=dashed, dash=2.5pt 1.5pt,linecolor=gray}
\psline{-}(0,1)(4,1) \psline{-}(0,2)(4,2) \psline{-}(0,3)(4,3)
%axes
\psset{linewidth=0.25pt,fillstyle=none,linestyle=solid,linecolor=black}
\psline{-}(2,1)(2,3)
%path
\psset{linewidth=0.7pt,fillstyle=none,linestyle=solid,linecolor=black}
\psline(1.5,1.5)(2,2)(2.5,1.5)
%annotate
\rput(-0.8,2){\scriptsize $a$}
}
\psline{->}(30.5,2.5)(33.5,2.5)
\rput(32.0,3.1){\scriptsize $n=0$}
\rput[bl](36,0){
%dashed grid
\psset{linewidth=0.25pt,linestyle=dashed, dash=2.5pt 1.5pt,linecolor=gray}
\psline{-}(0,1)(4,1) \psline{-}(0,2)(4,2)
\psline{-}(0,3)(4,3) \psline{-}(0,4)(4,4)
%
%axes
\psset{linewidth=0.25pt,fillstyle=none,linestyle=solid,linecolor=black}
\psline{-}(2,1)(2,4)
%path
\psset{linewidth=0.7pt,fillstyle=none,linestyle=solid,linecolor=black}
\psline(1.5,2.5)(2,3)(2.5,2.5)
%annotate
\rput(-0.8,2){\scriptsize $a$}
}
\end{pspicture}
\end{center}
\end{figure}

\begin{note}\label{Note:RBtransform}
Examination of the proof of the above Corollary shows that
the use of the modified $q$-binomial in \eqref{Eq:RBtransformGF},
is necessary when $L=0$, $e=f=1$ and $a=b$ are even
\upshape(and thus $\QL=\QR=1$\upshape),
but the standard $q$-binomial suffices in all other cases.
In terms of the action of the $\Tran$-transform, the reason for this is
exemplified in Figure \ref{Fig:Tricksy}: for $n>0$, the resulting path
is not valley-restricted
and thus should be excluded,
whereas the opposite is the case for $n=0$.
\end{note}

\subsection{Refining Melzer's expressions}
\label{Sec:RMelzer}

We now use Corollary \ref{Cor:RBtransform},
together with Lemmas \ref{Lem:RmSwitch2} and \ref{Lem:RSeed},
to produce fermionic expressions for $\RABFGFe{p}{a,b}{e,f}(L)$,
for each pair $e,f\in\{0,1\}$.
These four expressions are stated in the following Theorem.

\begin{theorem}\label{Thrm:HLfinfermsX}
Let $p,a,b,L\in\ZZ$ with $1\le a,b\le p$ and $L+a+b\in2\ZZ$, and set
$\boldC=\boldC^{(p-2)}$.
Then, for $e,f\in\{0,1\}$, we have the following expression for
$\RABFGFe{p}{a,b}{e,f}(L)$
in which the values of
$\ell$, $\Delta_i$,
$\boldQL=(\QL_1,\ldots,\QL_{p-1})$ and $\boldQR=(\QR_1,\ldots,\QR_{p-1})$,
are as given in Table~\ref{tab:HLLL} below:
\begin{equation}\label{Eq:RABFGF1234}
\begin{split}
\RABFGFe{p}{a,b}{e,f}(L)
=
\hskip-3mm
\sum_{\sboldn\in\ZZ^{p-1}_{\ge0}|m_0=L}
\hskip-3mm
&q^{\frac14\sboldm\sboldC\sboldm^T
   -\frac12 m_{\ell}
   +\sboldn\cdot\sboldQR}
\prod_{i=1}^{p-2}
\qbinomq{n_i+\hat m_i}{n_i}{q^2}^\prime\,,
\\
\text{where}\quad
m_i=2\!\!\sum_{i<k<2t-1}
(k-i)n_k -\Delta_i
\quad&\text{for}\quad
0\le i<p,
\quad\text{and}\quad 
\hat m_i=\frac12(m_i-\QL_i-\QR_i)
\quad\text{for}\quad
1\le i<p.
\end{split}
\end{equation}
Here, the sum is over all non-negative integer vectors
$\boldn=(n_1,n_2,\ldots,n_{p-1})$, with
$\boldmp=(m_0,m_1,\ldots,m_{p-2})$ obtained from $\boldn$
as indicated, and $\boldm=(m_1,m_2,\ldots,m_{p-2})$.
Note that by virtue of the restriction $m_0=L$,
the sum is, in effect, finite.
Also note that $m_{p-1}=0$ in each case, and for each $i<p-1$,
$\hat m_i$ is an integer equal to either
$\frac12m_i$, $\frac12(m_i-1)$ or $\frac12m_i-1$.

\begin{table}[ht]
\caption{Parameters for the four cases
                of the expressions \eqref{Eq:RABFGF1234}.} 
\vskip.2cm
\label{tab:HLLL}
\begin{center}
\begin{tabular}{|l|c||c|c|c|c|c|c| }\hline
\mystrut&\textup(e,f\textup)& $a$ & $b$ & $\ell$
  & $\Delta_i$ & $\boldQL$ & $\boldQR$ \\
\hline\hline
\mystrut{\rm (a)}&\textup(1,1\textup)
  & $\ne 1$ & $\ne1$& $a-1$
  & $\pabs{a-1-i}+\pabs{b-1-i} $  & $\boldQ^{(a,p-1)}$
  & $\boldQ^{(b,p-1)}$
\\
\hline
\mystrut{\rm (b)}&\textup(0,1\textup)
 & $\ne p$ &$\ne1$& $p-a$
 & $\pabs{p-a-i} +\pabs{b-1-i} +p-1-i$ & $\boldR^{(a,p-1)}$
 & $\boldQ^{(b,p-1)}$
\\
\hline
\mystrut{\rm (c)}& \textup(0,0\textup)
 & $\ne p$ & $\ne p$ &$ p-a$
 & $ \pabs{p-a-i}+\pabs{p-b-i}$ & $\boldR^{(a,p-1)}$
 & $\boldR^{(b,p-1)}$
\\
\hline
\mystrut{\rm (d)}&\textup(1,0\textup)
 & $\ne 1$ &$\ne p$ & $ a-1$
 & $\pabs{a-1-i} +\pabs{p-b-i} +p-1-i$ & $\boldQ^{(a,p-1)}$
 & $\boldR^{(b,p-1)}$
\\
\hline
\end{tabular}
\end{center}
\end{table}
\end{theorem}

Theorem \ref{Thrm:Yfinferms} immediately follows from
Theorem \ref{Thrm:HLfinfermsX} because, by virtue of the definitions
\eqref{Eq:HLGFfindef} and \eqref{Eq:RABFgfDef},
for $t,\hat a,\hat b,L\in\HZZ$ with $1\le \hat a,\hat b\le t$,
\begin{equation}\label{Eq:Restricted2Half}
\HLGFe{t}{\hat a,\hat b}{e,f}(L,q)
=\RABFGFe{2t-1}{2\hat a-1,2\hat b-1}{e,f}(2L,q^{1/2}).
\end{equation}

The expressions in Theorem \ref{Thrm:HLfinfermsX} are proved in a similar
way to those of Theorem \ref{Thrm:MelzerFin} in Section \ref{Sec:Melzer}.
In particular, expressions for
$\RABFGFe{p}{a,b}{e,f}(L,m)$ are first obtained for
the four cases of $e,f\in\{0,1\}$ using
the Cases specified by \eqref{Eq:LR12},
with Case L1 (resp.~L0) used for the $e=1$ (resp.~$e=0$) cases,
and Case R1 (resp.~R0) used for the $f=1$ (resp.~$f=0$) cases.
Thus, in accordance with Table \ref{tab:HLLL},
we use $\boldQL=\boldQ^{(a,p-1)}$ for $e=1$,
$\boldQL=\boldR^{(a,p-1)}$ for $e=0$,
$\boldQR=\boldQ^{(b,p-1)}$ for $f=1$,
and
$\boldQR=\boldR^{(b,p-1)}$ for $f=0$.
Comparison with \eqref{Eq:LR12} then shows that, in each case,
\begin{equation}\label{Eq:QLQR1}
\QL_i=(a_{i+1}+1)\bmod 2,
\qquad
\QR_i=(b_{i+1}+1)\bmod 2,
\end{equation}
for $1\le i<p$.
In particular, note that $\QL_{p-1}=\QR_{p-1}=0$ in each case.

\subsubsection{System A}\label{Sec:RSystem1}

For $1<a,b\le p$, consider the sequence
of $\Tran$-transforms governed by Cases L1 and R1.

Proceeding as in Section \ref{Sec:System1},
Corollary \ref{Cor:RBtransform} implies that for each $i=1,2,\ldots,p-1$,
\begin{equation}\label{Eq:RBrecurse}
\RABFGFe{p+1-i}{a_i,b_i}{e_i,f_i}(m_{i-1},m_i)
=\sum_{n_i\ge0}
q^{\frac12m_i(m_i-m)+n_i\QR_i}
\qbinomq{n_i+\frac12(m_i-\QL_i-\QR_i)}{n_i}{q^2}'
\RABFGFe{p-i}{a_{i+1},b_{i+1}}{e_i,f_i}(m_{i},m),
\end{equation}
where $m=2m_i+2n_i-m_{i-1}$, after noting \eqref{Eq:QLQR1}.
In the $i$th case, we replace the variable $m$ in \eqref{Eq:RBrecurse}
with
\begin{equation}\label{Eq:RMsub1}
m=m_{i+1}+\delta_{i,a-1}+\delta_{i,b-1}.
\end{equation}
We now express
$\RABFGFe{p-i}{a_{i+1},b_{i+1}}{e_i,f_i}(m_{i},m)$
in terms of
$\RABFGFe{p-i}{a_{i+1},b_{i+1}}{e_{i+1},f_{i+1}}(m_{i},m_{i+1})$.
Firstly, we obtain
\begin{equation}\label{Eq:RBrecurseX1}
\RABFGFe{p-i}{a_{i+1},b_{i+1}}{e_{i},f_{i}}
                     (m_{i},m_{i+1}+\delta_{i,a-1}+\delta_{i,b-1})
=\RABFGFe{p-i}{a_{i+1},b_{i+1}}{e_{i+1},f_{i}}
                     (m_{i},m_{i+1}+\delta_{i,b-1}),
\end{equation}
which follows from Lemma \ref{Lem:RmSwitch2}(3) in the $i=a-1$ case
because then $a_{i+1}=1$, $e_i=1$ and $e_{i+1}=0$, and follows
trivially in the $i\ne a-1$ case because then $e_{i+1}=e_i$.
We then obtain
\begin{equation}\label{Eq:RBrecurseX2}
\RABFGFe{p-i}{a_{i+1},b_{i+1}}{e_{i+1},f_{i}}
                    (m_{i},m_{i+1}+\delta_{i,b-1})
=q^{\frac12\delta_{i,b-1}m_{b-1}}
\RABFGFe{p-i}{a_{i+1},b_{i+1}}{e_{i+1},f_{i+1}}(m_{i},m_{i+1}),
\end{equation}
which follows from Lemma \ref{Lem:RmSwitch2}(5) in the $i=b-1$ case
because then $b_{i+1}=1$, $f_i=1$ and $f_{i+1}=0$, and follows
trivially in the $i\ne b-1$ case because then $f_{i+1}=f_i$.

Combining the $i=1,2,\ldots,p-1$ cases of
\eqref{Eq:RBrecurse}, \eqref{Eq:RBrecurseX1} and \eqref{Eq:RBrecurseX2}
results in
\begin{equation}\label{Eq:RBall1}
\RABFGFe{p}{a,b}{1,1}(m_0,m_1)
=\!\sum_{\sboldn\in\ZZ^{p-1}_{\ge0}}
\!
q^{-\frac12m_{a-1}+\sum_{i=1}^{p-1}\frac12m_i(m_i-m_{i+1})+n_i\QR_i}
\,
\RABFGFe{1}{1,1}{0,0}(m_{p-1},m_p)
\prod_{i=1}^{p-1}
\qbinomq{n_i+\frac12(m_i-\QL_i-\QR_i)}{n_i}{q^2}',
\end{equation}
where the values of $m_i$ are obtained recursively
from $\boldn=(n_1,n_2,\ldots,n_{p-1})$ using \eqref{Eq:MNsys1}.
Because $\RABFGFe{1}{1,1}{0,0}(m_{p-1},m_p)
                    =\delta_{m_{p-1},0}\delta_{m_p,0}$,
by Lemma \ref{Lem:RSeed},
and $\qbinomq{n_{p-1}-\frac12(\QL_{p-1}+\QR_{p-1})}{n_{p-1}}{q^2}'=1$
(because $\QL_{p-1}=\QR_{p-1}=0$),
this results in
\begin{equation}\label{Eq:RBall1eval}
\RABFGFe{p}{a,b}{1,1}(m_0,m_1)
=\!\sum_{\sboldn\in\ZZ^{p-1}_{\ge0}}
\!
q^{-\frac12m_{a-1}+\sum_{i=1}^{p-1}\frac12m_i(m_i-m_{i+1})+n_i\QR_i}
\,
\prod_{i=1}^{p-2}
\qbinomq{n_i+\frac12(m_i-\QL_i-\QR_i)}{n_i}{q^2}',
\end{equation}
with $\boldn$ constrained such that \eqref{Eq:MNsys1}
yields $m_{p-1}=m_p=0$.
These constraints again yield \eqref{Eq:MNsol1}.

Then, using Lemma \ref{Lem:RmSwitch2}(2) in the form
\begin{equation}
\RABFGFe{p}{a,b}{1,1}(m_0)=\sum_{m_1\ge0}
\RABFGFe{p}{a,b}{1,1}(m_0,m_1),
\end{equation}
proves (\ref{Eq:RABFGF1234}a).

\subsubsection{System B}\label{Sec:RSystem2}

For $1\le a<p$ and $1<b\le p$, consider the sequence
of $\Tran$-transforms governed by Cases L0 and R1.

Proceeding as for system A, but using
\begin{equation}\label{Eq:RMsub2}
m=m_{i+1}+\delta_{i,p-a}+\delta_{i,b-1}
\end{equation}
instead of \eqref{Eq:RMsub1},
we obtain the following analogue of \eqref{Eq:RBall1}:
\begin{equation}\label{Eq:RBall2}
\RABFGFe{p}{a,b}{0,1}(m_0,m_1)
=\!\sum_{\sboldn\in\ZZ^{p-1}_{\ge0}}
\!
q^{-\frac12m_{p-a}+\sum_{i=1}^{p-1}\frac12m_i(m_i-m_{i+1})+n_i\QR_i}
\,
\RABFGFe{1}{1,1}{1,0}(m_{p-1},m_p)
\prod_{i=1}^{p-1}
\qbinomq{n_i+\frac12(m_i-\QL_i-\QR_i)}{n_i}{q^2}',
\end{equation}
where the values of $m_i$ are obtained recursively
from $\boldn=(n_1,n_2,\ldots,n_{p-1})$ using \eqref{Eq:MNsys2}.
Because $\RABFGFe{1}{1,1}{1,0}(m_{p-1},m_p)
                    =\delta_{m_{p-1},0}\delta_{m_p,1}$,
by Lemma \ref{Lem:RSeed},
and $\qbinomq{n_{p-1}-\frac12(\QL_{p-1}+\QR_{p-1})}{n_{p-1}}{q^2}'=1$,
this results in
\begin{equation}\label{Eq:RBall2eval}
\RABFGFe{p}{a,b}{0,1}(m_0,m_1)
=\!\sum_{\sboldn\in\ZZ^{p-1}_{\ge0}}
\!
q^{-\frac12m_{p-a}+\sum_{i=1}^{p-1}\frac12m_i(m_i-m_{i+1})+n_i\QR_i}
\,
\prod_{i=1}^{p-2}
\qbinomq{n_i+\frac12(m_i-\QL_i-\QR_i)}{n_i}{q^2}',
\end{equation}
with $\boldn$ constrained such that \eqref{Eq:MNsys2}
yields $m_{p-1}=0$ and $m_p=1$.
These constraints again yield \eqref{Eq:MNsol2}.
Then, summing \eqref{Eq:RBall2eval} over $m_1$,
using Lemma \ref{Lem:RmSwitch2}(2),
proves (\ref{Eq:RABFGF1234}b)

\subsubsection{System C}\label{Sec:RSystem3}

For $1\le a,b<p$, consider the sequence
of $\Tran$-transforms governed by Cases L0 and R0.

Proceeding as for system A, but using
\begin{equation}\label{Eq:RMsub3}
m=m_{i+1}+\delta_{i,p-a}+\delta_{i,p-b}
\end{equation}
instead of \eqref{Eq:RMsub1},
we obtain the following analogue of \eqref{Eq:RBall1}:
\begin{equation}\label{Eq:RBall3}
\RABFGFe{p}{a,b}{0,0}(m_0,m_1)
=\!\sum_{\sboldn\in\ZZ^{p-1}_{\ge0}}
\!
q^{-\frac12m_{p-a}+\sum_{i=1}^{p-1}\frac12m_i(m_i-m_{i+1})+n_i\QR_i}
\,
\RABFGFe{1}{1,1}{1,1}(m_{p-1},m_p)
\prod_{i=1}^{p-1}
\qbinomq{n_i+\frac12(m_i-\QL_i-\QR_i)}{n_i}{q^2}',
\end{equation}
where the values of $m_i$ are obtained recursively
from $\boldn=(n_1,n_2,\ldots,n_{p-1})$ using \eqref{Eq:MNsys3}.
Because $\RABFGFe{1}{1,1}{1,1}(m_{p-1},m_p)
                    =\delta_{m_{p-1},0}\delta_{m_p,0}$,
by Lemma \ref{Lem:RSeed},
and $\qbinomq{n_{p-1}-\frac12(\QL_{p-1}+\QR_{p-1})}{n_{p-1}}{q^2}'=1$,
this results in
\begin{equation}\label{Eq:RBall3eval}
\RABFGFe{p}{a,b}{0,0}(m_0,m_1)
=\!\sum_{\sboldn\in\ZZ^{p-1}_{\ge0}}
\!
q^{-\frac12m_{p-a}+\sum_{i=1}^{p-1}\frac12m_i(m_i-m_{i+1})+n_i\QR_i}
\,
\prod_{i=1}^{p-2}
\qbinomq{n_i+\frac12(m_i-\QL_i-\QR_i)}{n_i}{q^2}',
\end{equation}
with $\boldn$ constrained such that \eqref{Eq:MNsys3}
yields $m_{p-1}=m_p=0$.
These constraints again yield \eqref{Eq:MNsol3}.
Then, summing \eqref{Eq:RBall3eval} over $m_1$,
using Lemma \ref{Lem:RmSwitch2}(2),
proves (\ref{Eq:RABFGF1234}c).

\subsubsection{System D}\label{Sec:RSystem4}

For $1<a\le p$ and $1\le b<p$, consider the sequence
of $\Tran$-transforms governed by Cases L1 and R0.

Proceeding as for system A, but using
\begin{equation}\label{Eq:RMsub4}
m=m_{i+1}+\delta_{i,a-1}+\delta_{i,p-b}
\end{equation}
instead of \eqref{Eq:RMsub1},
we obtain the following analogue of \eqref{Eq:RBall1}:
\begin{equation}\label{Eq:RBall4}
\RABFGFe{p}{a,b}{1,0}(m_0,m_1)
=\!\sum_{\sboldn\in\ZZ^{p-1}_{\ge0}}
\!
q^{-\frac12m_{a-1}+\sum_{i=1}^{p-1}\frac12m_i(m_i-m_{i+1})+n_i\QR_i}
\,
\RABFGFe{1}{1,1}{0,1}(m_{p-1},m_p)
\prod_{i=1}^{p-1}
\qbinomq{n_i+\frac12(m_i-\QL_i-\QR_i)}{n_i}{q^2}',
\end{equation}
where the values of $m_i$ are obtained recursively
from $\boldn=(n_1,n_2,\ldots,n_{p-1})$ using \eqref{Eq:MNsys4}.
Because $\RABFGFe{1}{1,1}{0,1}(m_{p-1},m_p)
                    =\delta_{m_{p-1},0}\delta_{m_p,1}$,
by Lemma \ref{Lem:RSeed},
and $\qbinomq{n_{p-1}-\frac12(\QL_{p-1}+\QR_{p-1})}{n_{p-1}}{q^2}'=1$,
this results in
\begin{equation}\label{Eq:RBall4eval}
\RABFGFe{p}{a,b}{1,0}(m_0,m_1)
=\!\sum_{\sboldn\in\ZZ^{p-1}_{\ge0}}
\!
q^{-\frac12m_{a-1}+\sum_{i=1}^{p-1}\frac12m_i(m_i-m_{i+1})+n_i\QR_i}
\,
\prod_{i=1}^{p-2}
\qbinomq{n_i+\frac12(m_i-\QL_i-\QR_i)}{n_i}{q^2}',
\end{equation}
with $\boldn$ constrained such that \eqref{Eq:MNsys4}
yields $m_{p-1}=0$ and $m_p=1$.
These constraints again yield \eqref{Eq:MNsol4}.
Then, summing \eqref{Eq:RBall4eval} over $m_1$,
using Lemma \ref{Lem:RmSwitch2}(2),
proves (\ref{Eq:RABFGF1234}d).

\subsection{\texorpdfstring
             {The modified $q$-binomials are not always necessary}
             {The modified q-binomials are not always necessary}}

Note \ref{Note:RBtransform} indicates that in the iterations
\eqref{Eq:RBrecurse},
the modified form of the $q$-binomial is necessary only if
$e_i=f_i=1$ and $a_{i+1}=b_{i+1}$ with these even.
Thus, \eqref{Eq:RABFGF1234}
requires the modified $q$-binomial only for some values of
$a$ and $b$, and then only for certain $i$.
On inspecting \eqref{Eq:L1} and \eqref{Eq:R1},
we find that for $(e,f)=(1,1)$,
they are only required if $a=b>2$, and then only for those $i$
for which $i<a$ and $i\equiv a$.
Inspection of \eqref{Eq:L2} and \eqref{Eq:R1},
shows that  for $(e,f)=(0,1)$,
they are only required if $a>1$ and $b=p$, and then only for those $i$
for which $i>p-a$ and $i\equiv p$.
Inspection of \eqref{Eq:L2} and \eqref{Eq:R2},
shows that for $(e,f)=(0,0)$,
they are only required if $a>1$ and $b>1$, and then only for those $i$
for which $i>\max\{p-a,p-b\}$ and $i\equiv p$.
Inspection of \eqref{Eq:L1} and \eqref{Eq:R2},
shows that  for $(e,f)=(1,0)$,
they are only required if $a=p$ and $b>1$, and then only for those $i$
for which $i>p-b$ and $i\equiv p$.

Using \eqref{Eq:Restricted2Half}, we can transfer this observation
to Theorems \ref{Thrm:HLferms} and \ref{Thrm:Yfinferms}.

\begin{note}\label{Note:Mod}
In Theorems \ref{Thrm:HLferms} and \ref{Thrm:Yfinferms},
the modified $q$-binomial can be replaced by the standard binomial
in many cases. Here, we list those cases where the modified $q$-binomial
is required.
For (\ref{Eq:Y1234}a), it is required only if $a=b\ge2$,
and then only for those $i<2a-1$ for which $i\not\equiv2a$.
Consequently, for (\ref{Eq:HLferms}a), it is never required.
For (\ref{Eq:Y1234}b), it is required only if $a>1$ and $b=t$,
and then only for those $i>2(t-a)$ for which $i\not\equiv2t$.
Consequently, for (\ref{Eq:HLferms}b), it is required
only if $a>1$ and $r=t-\frac12$,
and then only for those $i>2(t-a)$ for which $i\not\equiv2t$.
For (\ref{Eq:Y1234}c), it is required only if $a>1$ and $b>1$,
and then only for those $i>\max\{2(t-a),2(t-b)\}$ for which $i\not\equiv2t$.
Consequently, for (\ref{Eq:HLferms}c), it is required
only if $a>1$ and $r>1$,
and then only for those $i>\max\{2(t-a),2(t-r)\}$ for which $i\not\equiv2t$.
For (\ref{Eq:Y1234}d), it is required only if $a=t$ and $b>1$,
and then only for those $i>2(t-b)$ for which $i\not\equiv2t$.
Consequently, for (\ref{Eq:HLferms}d), it is required
only if $a=t$ and $r>1$,
and then only for those $i>2(t-r)$ for which $i\not\equiv2t$.
\end{note}

\section{Proving the bosonic generating function for half-lattice paths}
\label{Sec:BosonicProof}

In this section, we prove the bosonic expressions
for the generating functions $\HLGFe{t}{a,b}{e,f}(L)$
for finite length half-lattice paths,
that were stated in Theorem \ref{Thrm:FinGen}.

Let $t\in\HZZ$ and $a,b,L\in\ZZ_{>0}$ with $1\le a\le\lfloor t\rfloor$.
In view of \eqref{Eq:HLGFeq},
the following results are independent of $e\in\{0,1\}$.
For $1<b<\lfloor t\rfloor$,
after noting that for each path $\hh\in\HLtabLef$,
the pair $(h_{L-1},h_{L-1/2})$ must be one of
$(b-1,b-1/2)$,
$(b,b+1/2)$ or
$(b+1,b+1/2)$,
we obtain the recurrence relations
\begin{align}
\label{Eq:Xrecurse1}
\HLGFe{t}{a,b}{e,0}(L)
&=q^{L-\frac14}
\HLGFe{t}{a,b-1}{e,0}(L-1)
+\HLGFe{t}{a,b}{e,0}(L-1)
+q^{\frac12 L-\frac14}\HLGFe{t}{a,b+1}{e,1}(L-1),\\
\label{Eq:Xrecurse2}
\HLGFe{t}{a,b}{e,1}(L)
&=q^{\frac12L-\frac14}
\HLGFe{t}{a,b-1}{e,0}(L-1)
+q^{\frac12L}\HLGFe{t}{a,b}{e,0}(L-1)
+q^{L-\frac14}\HLGFe{t}{a,b+1}{e,1}(L-1).
\end{align}
These expressions also apply in the cases $b=1$ and $b=\lfloor t\rfloor$
on imposing the boundary conditions
\begin{align}
\label{Eq:Xboundary0}
\HLGFe{t}{a,0}{e,0}(L)
&= 0,\\
\label{Eq:Xboundary1}
\HLGFe{t}{a,t+1/2}{e,1}(L)
&= 0
\qquad\text{for } t\in\ZZph,\\
\label{Eq:Xboundary2}
\HLGFe{t}{a,t}{e,0}(L)
+q^{\frac12 L+\frac14}\HLGFe{t}{a,t+1}{e,1}(L)
&= 0
\qquad\text{for } t\in\ZZ.
\end{align}
In addition, for $L=0$, we have the initial condition
\begin{equation}
\label{Eq:Xinitial}
\HLGFe{t}{a,b}{e,0}(0)
=\HLGFe{t}{a,b}{e,1}(0)
=\delta_{a,b}.
\end{equation}

Our strategy for proving \eqref{Eq:FinGen} is to show that the
polynomials on its right side satisfy the relations
\eqref{Eq:Xrecurse1}--\eqref{Eq:Xinitial}.
So, for $e,f\in\{0,1\}$ and \emph{all} $a,b,L\in\ZZ$, define
$\oHLGFe{t}{a,b}{e,f}(L)$ by
\begin{equation}
\label{Eq:oFinGen}
\oHLGFe{t}{a,b}{e,f}(L)
=q^{\frac12(a-b)(a-b-\frac12)+\frac12fL}\,Y^{f;t}_{a,b}(L).
\end{equation}

The recurrence relations \eqref{Eq:Urec1} and \eqref{Eq:Urec2}
for the $q$-trinomial coefficients $\mytri{n}{d}{L}$
imply that for $L>0$,
\begin{align}
\label{Eq:Yrec1}
Y^{n;t}_{a,b}(L)
&=
q^{L+a-b-n}Y^{n;t}_{a,b-1}(L-1)
+Y^{n;t}_{a,b}(L-1)
+q^{L-a+b}Y^{n+1;t}_{a,b+1}(L-1),\\
\label{Eq:Yrec2}
Y^{n;t}_{a,b}(L)
&=
q^{a-b-n+1}Y^{n-1;t}_{a,b-1}(L-1)
+Y^{n-1;t}_{a,b}(L-1)
+q^{L-a+b}Y^{n;t}_{a,b+1,n}(L-1).
\end{align}
These imply that for all $a,b,L\in\ZZ$ with $L>0$,
eqns.~\eqref{Eq:Xrecurse1} and \eqref{Eq:Xrecurse2}
hold with $\oHLGFe{t}{a,b}{e,f}(L)$ in place of $\HLGFe{t}{a,b}{e,f}(L)$.

From the definition \eqref{Eq:Ydef},
\begin{equation}\label{Eq:Yrec1pf}
\begin{split}
Y^{0;t}_{a,0}(L)
&= \sum_{\lambda=-\infty}^\infty
q^{\lambda t(\lambda t'-2a)}\mytri{0}{a-t'\lambda}{L}
- q^{\lambda t(\lambda t'+2a)}\mytri{0}{-a-t'\lambda}{L}\\
&= \sum_{\lambda=-\infty}^\infty
q^{\lambda t(\lambda t'-2a)}\mytri{0}{a-t'\lambda}{L}
- q^{\lambda t(\lambda t'-2a)}\mytri{0}{a-t'\lambda}{L}\\
&=0,
\end{split}
\end{equation}
after, in the second term of the first line,
changing $\lambda\to-\lambda$ and then using \eqref{Eq:TriMinus}.
Thus $\oHLGFe{t}{a,0}{e,0}(L)=0$.

Changing $\lambda\to-(\lambda+1)$ in the second term
of \eqref{Eq:Ydef} and using \eqref{Eq:TriMinus} yields
\begin{equation}\label{Eq:Yswitch}
\begin{split}
Y^{n;t}_{a,b}(L)
&= \sum_{\lambda=-\infty}^\infty
q^{\lambda^2tt'+\lambda t'b-2\lambda ta}\mytri{n}{a-b-t'\lambda}{L}\\
&\qquad-
q^{\lambda^2tt'+\lambda t'(2t-b+n)-2\lambda ta+(t-b)(t'-2a)+n(t'-a-b)}
        \mytri{n}{a+b-t'-t'\lambda}{L}.
\end{split}
\end{equation}
In the case $t\in\ZZph$, after recalling that $t'=2t+1$,
we immediately obtain
$\oHLGFe{t}{a,t+1/2}{e,1}(L)=0$.

Making use of \eqref{Eq:Yswitch} for $b=t+n$ in both cases $n=0,1$, we
obtain
\begin{equation}\label{Eq:Yswitch2}
\begin{split}
&Y^{0;t}_{a,t}(L)+q^{L+1-a+t}Y^{1;t}_{a,t+1}(L)\\
&= \sum_{\lambda=-\infty}^\infty
q^{\lambda^2tt'+\lambda tt'-2\lambda ta}\bigl(
\mytri{0}{a-t-t'\lambda}{L}
-\mytri{0}{a-t-1-t'\lambda}{L}\\
&\hskip30mm
+q^{L+1-a+t+\lambda t'}\mytri{1}{a-t-1-t'\lambda}{L}
-q^{L}\mytri{1}{a-t-t'\lambda}{L}\bigr)\\
&=0,
\end{split}
\end{equation}
having, for the final line, made use of the $n=0$ case of \eqref{Eq:Urec1234}.
It follows that
$\oHLGFe{t}{a,t}{e,0}(L)+q^{\frac12L+\frac14}\oHLGFe{t}{a,t+1}{e,1}(L)=0$.

For $1\le a,b\le\lfloor t\rfloor$,
necessarily $|a\pm b|<t'$.
The definition \eqref{Eq:Ydef} then implies that
$Y^{n;t}_{a,b}(0)=\delta_{a,b}$ and therefore
$\oHLGFe{t}{a,b}{e,0}(0)=\oHLGFe{t}{a,b}{e,1}(0)=\delta_{a,b}$.

Expression \eqref{Eq:FinGen} is now proved because
$\HLGFe{t}{a,b}{e,f}(L)$ is determined uniquely by
\eqref{Eq:Xrecurse1}--\eqref{Eq:Xinitial} for $f\in\{0,1\}$,
and $\oHLGFe{t}{a,b}{e,f}(L)$ satisfies the same relations.

To prove \eqref{Eq:FinGenHb1} and \eqref{Eq:FinGenHb2},
let $t\in\HZZ$ and $a,b,L\in\ZZ_{\ge0}$ with $1\le a\le\lfloor t\rfloor$.
For $1\le b<\lfloor t\rfloor$,
after noting that for each path
$\hh\in\HLsetef{t}{a,b+1/2}{e,1}(L+1/2)$,
either $\hh_{L}=b$ or $\hh_{L}=b+1$,
we obtain
\begin{equation}
\label{Eq:Xrecurse3}
\HLGFe{t}{a,b+1/2}{e,1}(L+1/2)
=
\HLGFe{t}{a,b}{e,0}(L)
+q^{\frac12 L+\frac14}\HLGFe{t}{a,b+1}{e,1}(L).
\end{equation}
This relation also applies in the case $b=t-1/2$ when $t\in\ZZph$,
by virtue of \eqref{Eq:Xboundary1}.
Expression~\eqref{Eq:FinGenHb1} then results from applying
\eqref{Eq:FinGen} to each of the terms on the right side.

For $1\le b\le\lfloor t\rfloor$, each path
$\hh\in\HLsetef{t}{a,b+1/2}{e,0}(L+1/2)$
necessarily has $\hh_{L}=b$ because of the restriction on
valley positions.
Therefore
\begin{equation}
\label{Eq:Xrecurse4}
\HLGFe{t}{a,b+1/2}{e,0}(L+1/2)
=q^{\frac12 L+\frac14}\HLGFe{t}{a,b}{e,0}(L).
\end{equation}
Expression~\eqref{Eq:FinGenHb2} then results from \eqref{Eq:FinGen}.

%%%%%%%%%%%%%%%%%%%%%%%%%%%%%%%%%%%%%%%%%%%%%%%%%%%%%%%%%%%%%%%%%%%%%%
%%%%%%%%%%%%%%%%%%%%%%%%%%%%%%%%%%%%%%%%%%%%%%%%%%%%%%%%%%%%%%%%%%%%%%

\section{Discussion}
\label{Sec:Diss}

The half-lattice paths analysed in this paper
provide a combinatorial model for the characters
of the conformal minimal models $M(k,2k\pm1)$ that is compatible
with the $\phi_{2,1}$ and $\phi_{1,5}$ perturbations of these theories.
The similarity of these paths to those of the ABF models
enable techniques used in those cases to be applied here
to yield novel fermionic expressions for the generating functions
of both finite and infinite length paths, the latter
generating functions being the characters themselves.
In particular, the four known fermionic expressions for the characters
of $M(p,p+1)$ that are compatible with the $\phi_{1,3}$ perturbation
have analogues for these $M(k,2k\pm1)$ characters.
It is especially interesting to note that, as revealed by
the form of these fermionic expressions,
the characters of the latter are obtained from those of
the former on restricting the excitations of the
quasi-particles to alternate states.

In \cite{JacobMathieu2007}, it was shown that half-lattice paths,
with a weighting function dual to that used here,
may be used to provide a combinatorial model for the
graded parafermion models $\mathcal Z_k$ \cite{CRS1998}.
After mapping $q\to q^{-1}$, the $t\in\ZZ$ cases of
Theorem \ref{Thrm:Yfinferms} then apply to these models,
yielding fermionic expressions that extend those given in
\cite{JacobMathieu2007}.

In future work, we will also explore extending the
half-lattice path combinatorial models
for the $\phi_{2,1}$ and $\phi_{1,5}$ perturbations to other $M(p,p')$.
Such an extension may be expected to admit
a generalisation of the combinatorial $\Tran$-transform
which is analogous to that for the $\phi_{1,3}$ perturbation
described by \eqref{Eq:Cflow},
and whose action corresponds to the renormalisation group
flows described in \cite[eqns.~(3.6)--(3.8)]{DDT2000}.
We anticipate that this extension will involve a banding
structure for the half-lattice paths akin to that introduced in
\cite{FLPW2000} for the RSOS paths of \cite{FB1985}.
However, our initial investigations seem to indicate, curiously,
that such an extension is not possible for all pairs of $p$ and $p'$.

Finally, it is fair to stress that the description of the
$M(k,2k\pm 1)$ states by half-lattice paths is somewhat ad hoc.
This link should eventually be framed in a broader context by
exhibiting a direct relationship between the
$M(k,2k\pm 1)$ minimal models and an integrable lattice model
with an underlying $A^{(2)}_2$ structure.
With this in mind, the non-compact nature of the continuum limit
of regime III of the Izergin-Korepin model \cite{IzerginKorepin1981}
unravelled in \cite{VJS2014} is intriguing.

%%%%%%%%%%%%%%%%%%%%%%%%%%%%%%%%%%%%%%%%%%%%%%%%%%%%%%%%%%%%%%%%%%%%%%
%%%%%%%%%%%%%%%%%%%%%%%%%%%%%%%%%%%%%%%%%%%%%%%%%%%%%%%%%%%%%%%%%%%%%%

\section*{Acknowledgements}

This work was supported by the Natural Sciences and
Engineering Research Council of Canada (NSERC).

%%%%%%%%%%%%%%%%%%%%%%%%%%%%%%%%%%%%%%%%%%%%%%%%%%%%%%%%%%%%%%%%%%%%%%
%%%%%%%%%%%%%%%%%%%%%%%%%%%%%%%%%%%%%%%%%%%%%%%%%%%%%%%%%%%%%%%%%%%%%%

\begin{appendix}

\section{\texorpdfstring
             {Identities involving $q$-trinomials}
             {Identities involving q-trinomials}}
\label{App:Trinoms}

\subsection{Definition}
The $q$-trinomials $\mytri{n}{d}{L}$ may be defined by:
\begin{subequations}\label{Eq:TriDef}
\begin{align}
\label{Eq:TriDef1}
\mytri{n}{d}{L}
=\mytriq{n}{d}{L}{q}
&=\sum_{k=0}^{\lfloor(L-d)/2\rfloor}
q^{k(k+d-n)}
\frac{(q)_L}{(q)_k(q)_{k+d}(q)_{L-2k-d}}\\
\label{Eq:TriDef2}
&=q^{-\frac14(d-n)^2}
\sum_{r=0}^{L-|d|}
q^{\frac14(L-n-r)^2}
\frac{(q)_L}{(q)_{(L-d-r)/2}(q)_{(L+d-r)/2}(q)_r},
\end{align}
\end{subequations}
where the final expression follows from the previous by
setting $k=(L-d-r)/2$ so that
$4k(k+d-n)=(L-n-r)^2-(d-n)^2$.
It follows from \eqref{Eq:TriDef2} that
\begin{equation}
\label{Eq:TriMinus}
\mytri{n}{-d}{L}=q^{-nd}\mytri{n}{d}{L}.
\end{equation}

\subsection{Recurrence relations}
Of the many recurrence relations enjoyed by the $q$-trinomial
coefficients $\mytri{n}{d}{L}$, the following four prove
useful in the current work: for $L>0$,
\begin{subequations}\label{Eq:Urec}
\begin{align}
\mytri{n}{d}{L}
&=
q^{L-d}\mytri{n+1}{d-1}{L-1}+\mytri{n}{d}{L-1}+q^{L+d-n}\mytri{n}{d+1}{L-1}
\label{Eq:Urec1}\\
&=
q^{L-d}\mytri{n}{d-1}{L-1}+\mytri{n-1}{d}{L-1}+q^{d-n+1}\mytri{n-1}{d+1}{L-1}
\label{Eq:Urec2}\\
&=
q^{L-d}\mytri{n}{d-1}{L-1}+\mytri{n}{d}{L-1}+q^{L-n-1}\mytri{n+1}{d+1}{L-1}
\label{Eq:Urec3}\\
&=
\mytri{n-1}{d-1}{L-1}+q^d\mytri{n-1}{d}{L-1}+q^{L+d-n}\mytri{n}{d+1}{L-1}
\label{Eq:Urec4}.%\\
\end{align}
\end{subequations}
The first two of these are obtained after substituting
$(q)_L$ in \eqref{Eq:TriDef1} for, respectively,
\begin{subequations}
\begin{align}
(q)_L&=(q)_{L-1}((1-q^{L-2k-d})+q^{L-2k-d}(1-q^{k+d})+q^{L-k}(1-q^k))\\
\text{and}\quad
(q)_L&=(q)_{L-1}((1-q^{k})+q^k(1-q^{L-2k-d})+q^{L-k-d}(1-q^{k+d})).
\end{align}
\end{subequations}
Identities \eqref{Eq:Urec3} and \eqref{Eq:Urec4} are obtained
from \eqref{Eq:Urec1} and \eqref{Eq:Urec2} respectively,
on exchanging $d\to -d$, and using \eqref{Eq:TriMinus}.

The first of the next pair of identities results from combining
\eqref{Eq:Urec1} and \eqref{Eq:Urec4}; the second results similarly
from combining \eqref{Eq:Urec2} and \eqref{Eq:Urec3}
(or from exchanging $d\to-d$ in the first):
\begin{subequations}\label{Eq:UrecPair}
\begin{align}
\label{Eq:Urec14}
q^{L+1-d}\mytri{n+1}{d-1}{L}
+\mytri{n}{d}{L}
&=
\mytri{n-1}{d-1}{L}
+q^{d}\mytri{n-1}{d}{L},\\
\label{Eq:Urec23}
q^{L-n}\mytri{n+1}{d+1}{L}
+\mytri{n}{d}{L}
&=
q^{d-n+1}\mytri{n-1}{d+1}{L}
+\mytri{n-1}{d}{L}.
\end{align}
\end{subequations}
Exchanging $d\to d+1$ in \eqref{Eq:Urec14} and subtracting
\eqref{Eq:Urec23} multiplied by $q^n$ yields:
\begin{equation}
\label{Eq:Urec1234}
\mytri{n}{d+1}{L}
+q^{L-d}\mytri{n+1}{d}{L}
=
q^L\mytri{n+1}{d+1}{L}
+q^{n}\mytri{n}{d}{L}
+(1-q^n)
\mytri{n-1}{d}{L}.
\end{equation}

\subsection{\texorpdfstring
             {$q$-trinomial limits}
             {q-trinomial limits}}

In the $n=0$ case of \eqref{Eq:TriDef1}, we have the important
$L\to\infty$ limit
\begin{equation}
\label{Eq:U0limit}
\lim_{L\to\infty}
\mytri{0}{d}{L}=
\sum_{k=0}^{\infty}
q^{k(k+d)}
\frac{1}{(q)_k(q)_{k+d}}\\
=\frac1{(q)_\infty},
\end{equation}
having used the Durfee rectangle identity
\cite[eqn.~following (1.6.4)]{GasperRahman1990}.

To obtain $\lim_{L\to\infty}\mytri{n}{d}{L}$ for $n>0$,
first take $L\to\infty$ in \eqref{Eq:Urec4}. This gives:
\begin{equation}
\label{Eq:UlimitRec}
\lim_{L\to\infty}
\mytri{n+1}{d}{L}
=
\lim_{L\to\infty}
\mytri{n}{d-1}{L}
+
q^{d}\lim_{L\to\infty}
\mytri{n}{d}{L}.
\end{equation}
In particular, the $n=0$ case yields:
\begin{equation}
\label{Eq:U1limit}
\lim_{L\to\infty}
\mytri{1}{d}{L}=\frac{1+q^d}{(q)_\infty}.
\end{equation}

\subsection{\texorpdfstring
             {$q$-trinomial references}
             {q-trinomial references}}

The $q$-trinomials $U_n(L,d)$ were first defined in
\cite{AndrewsBaxter1987}.
Identities \eqref{Eq:TriMinus}, \eqref{Eq:Urec1}, \eqref{Eq:Urec3},
\eqref{Eq:U0limit} and \eqref{Eq:U1limit} also appear
in \cite{AndrewsBaxter1987}
(as eqns.~(2.15), (2.29), (2.28), (2.48), and (2.49) resp.).
The $n=0$ case of \eqref{Eq:Urec1234}
(the only case of which we make use)
appears in \cite{SeatonScott1997}.

\end{appendix}

%%%%%%%%%%%%%%%%%%%%%%%%%%%%%%%%%%%%%%%%%%%%%%%%%%%%%%%%%%%%%%%%%%%%%%%%%%%
%%%%%%%%%%%%%%%%%%%%%%%%%%%%%%%%%%%%%%%%%%%%%%%%%%%%%%%%%%%%%%%%%%%%%%%%%%%

% The following bibliography has been generated using

%\bibliographystyle{elsarticle-num}
%\bibliography{/home/trevor/tex/taw}

\begin{thebibliography}{10}
\expandafter\ifx\csname url\endcsname\relax
  \def\url#1{\texttt{#1}}\fi
\expandafter\ifx\csname urlprefix\endcsname\relax\def\urlprefix{URL }\fi
\expandafter\ifx\csname href\endcsname\relax
  \def\href#1#2{#2} \def\path#1{#1}\fi

\bibitem{Melzer1994}
E.~Melzer, Fermionic character sums and the corner transfer matrix, Int. J.
  Mod. Phys. {\rm A} 9 (1994) 1115--1136.

\bibitem{Warnaar1996a}
S.~O. Warnaar, Fermionic solution of the {Andrews}-{Baxter}-{Forrester} model.
  {I}. {U}nification of {CTM} and {TBA} methods, J. Stat. Phys. 82 (1996)
  657--685.

\bibitem{Warnaar1996b}
S.~O. Warnaar, Fermionic solution of the {A}ndrews-{B}axter-{F}orrester model.
  {II}. {P}roof of {M}elzer's polynomial identities, J. Stat. Phys. 84 (1996)
  49--83.

\bibitem{FLPW2000}
O.~Foda, K.~S.~M. Lee, Y.~Pugai, T.~A. Welsh, Path generating transforms,
  Contemp. Math. 254 (2000) 157--186.

\bibitem{FodaWelsh1999}
O.~Foda, T.~A. Welsh, Melzer's identities revisited, Contemp. Math. 248 (1999)
  207--234.

\bibitem{FodaWelsh2000}
O.~Foda, T.~A. Welsh, On the combinatorics of {F}orrester-{B}axter models, in:
  M.~Kashiwara, T.~Miwa (Eds.), Physical Combinatorics, Kyoto 1999, Vol. 191 of
  Progress in Mathematics, Birkhauser, Boston, 2000, pp. 49--103.

\bibitem{Welsh2005}
T.~A. Welsh, Fermionic expressions for minimal model {V}irasoro characters,
  Mem. Amer. Math. Soc. 175~(827) (2005).

\bibitem{JacobMathieu2009}
P.~Jacob, P.~Mathieu, Particles in {RSOS} paths, J. Phys. {\rm A} 42 (2009)
  122001--16.

\bibitem{FB1985}
P.~J. Forrester, R.~J. Baxter, Further exact solutions of the eight-vertex
  {SOS} model and generalizations of the {R}ogers-{R}amanujan identities, J.
  Stat. Phys. 38 (1985) 435--472.

\bibitem{ABF1984}
G.~E. Andrews, R.~J. Baxter, P.~J. Forrester, Eight-vertex {SOS} model and
  generalized {R}ogers-{R}amanujan-type identities, J. Stat. Phys. 35 (1984)
  193--266.

\bibitem{Rocha-Caridi1985}
A.~Rocha-Caridi, Vacuum vector representations of the {V}irasoro algebra, in:
  J.~Lepowsky, S.~Mandelstam, I.~M. Singer (Eds.), Vertex Operators in
  Mathematics and Physics, Springer, New York, 1985, pp. 451--473.

\bibitem{KKMM1993b}
R.~Kedem, T.~Klassen, B.~McCoy, E.~Melzer, Fermionic sum representations for
  conformal field theory characters, Phys. Lett. {\rm B} 307 (1993) 68--76.

\bibitem{BMc1998}
A.~Berkovich, B.~M. McCoy, {R}ogers-{R}amanujan identities: A century of
  progress from mathematics to physics, Doc. Math. J. DMV, Extra Volume ICM III
  (1998) 163--172.

\bibitem{Huse1984}
D.~A. Huse, Exact exponents for infinitely many new multicritical points, Phys.
  Rev. {\rm B} 30 (1984) 3908--3915.

\bibitem{BMcP1998}
A.~Berkovich, B.~M. McCoy, P.~A. Pearce, The perturbations $\phi_{2,1}$ and
  $\phi_{1,5}$ of the minimal models ${M}(p,p')$ and the trinomial analogue of
  {B}ailey's lemma, Nucl. Phys. {\rm B} 519 (1998) 597--625.

\bibitem{BMc2000}
A.~Berkovich, B.~M. McCoy, The perturbation $\phi_{2,1}$ of the ${M}(p,p+1)$
  models of conformal field theory and related polynomial-character identities,
  Ramanujan J. 4 (2000) 353--383.

\bibitem{Berkovich1994}
A.~Berkovich, Fermionic counting of {RSOS}-states and {V}irasoro character
  formulas for the unitary minimal series ${M}(\nu,\nu+1)$. {E}xact results,
  Nucl. Phys. {\rm B} 431 (1994) 315--348.

\bibitem{Schilling1996a}
A.~Schilling, Polynomial fermionic forms for the branching functions of the
  rational coset conformal field theories
  $\widehat{su}(2)_m\times\widehat{su}(2)_n/\widehat{su}(2)_{M+N}$, Nucl. Phys.
  {\rm B} 459 (1996) 393--436.

\bibitem{Warnaar1999}
S.~O. Warnaar, $q$-trinomial identities, J. Math. Phys. 40 (1999) 2514--2530.

\bibitem{JacobMathieu2007}
P.~Jacob, P.~Mathieu, A new path description for the $\mathcal{M}(k+1,2k+3)$
  models and the dual $\mathcal{Z}_k$ graded parafermions,
  J. Stat. Mech. (2007) P11005, (43 pages).

\bibitem{Blondeau-Fournier2011}
O.~Blondeau-Fournier, Approche combinatoire des mod\`eles minimaux en th\'eorie
  des champs conformes: connexion avec les chemins sur r\'eseau demi-entier,
  Master's thesis, Universit\'e Laval, Qu\'ebec, Canada (2011).

\bibitem{BMc1996lmp}
A.~Berkovich, B.~M. McCoy, Continued fractions and fermionic representations
  for characters of ${M}(p,p')$ minimal models, Lett. Math. Phys. 37 (1996)
  49--66.

\bibitem{BMcS1998}
A.~Berkovich, B.~M. McCoy, A.~Schilling, {R}ogers-{S}chur-{R}amanujan type
  identities for the ${M}(p,p')$ minimal models of conformal field theory,
  Comm. Math. Phys. 191 (1998) 325--395.

\bibitem{Zamolodchikov1989}
A.~B. Zamolodchikov, Integrable field theory from conformal field theory, Adv.
  Stud. Pure Maths 19 (1989) 641--674.

\bibitem{LeClair1989}
A.~LeClair, Restricted sine-{G}ordon theory and the minimal conformal models,
  Phys. Lett. {\rm B} 230 (1989) 103--107.

\bibitem{Smirnov1989}
F.~A. Smirnov, The perturbated $c<1$ conformal field theories as reductions of
  sine-{G}ordon model, Int. J. Mod. Phys. {\rm A} 4 (1989) 4213--4220.

\bibitem{Smirnov1990}
F.~A. Smirnov, Reductions of the sine-{G}ordon model as a perturbation of
  minimal models of conformal field theory, Nucl. Phys. {\rm B} 337 (1990)
  156--180.

\bibitem{Smirnov1991}
F.~A. Smirnov, Exact ${S}$-matrices for $\phi_{1,2}$-perturbated minimal models
  of conformal field theory, Int. J. Mod. Phys. {\rm A} 6 (1991) 1407--1428.

\bibitem{Efthimiou1993}
C.~J. Efthimiou, Quantum group symmetry of the $\phi_{1,2}$-perturbed and
  $\phi_{2,1}$-perturbed minimal models of conformal field theory, Nucl. Phys.
  {\rm B} 398 (1993) 697--740.

\bibitem{IzerginKorepin1981}
A.~G. Izergin, V.~E. Korepin, The inverse scattering method approach to the
  quantum {S}habat-{M}ikhailov model, Comm. Math. Phys. 79 (1981) 303--316.

\bibitem{Kuniba1991}
A.~Kuniba, Exact solution of solid-on-solid models for twisted affine {L}ie
  algebras ${A}^{(2)}_{2n}$ and ${A}^{(2)}_{2n-1}$, Nucl. Phys. {\rm B} 355
  (1991) 801--821.

\bibitem{WNS1993}
S.~O. Warnaar, B.~Nienhuis, K.~A. Seaton, A critical {I}sing model in a
  magnetic field, Int. J. Mod. Phys. {\rm B} 7 (1993) 3727--3736.

\bibitem{KKMMNN1992}
S.-J. Kang, M.~Kashiwara, K.~Misra, T.~Miwa, T.~Nakashima, A.~Nakayashiki,
  Affine crystals and vertex models, Int. J. Mod. Phys. {\rm A} 7, Suppl.~1A
  (1992) 449--484.

\bibitem{Baxter1982}
R.~J. Baxter, Exactly Solved Models in Statistical Mechanics, Academic Press,
  London, 1982, (Reprinted 2007 by Dover, New York).

\bibitem{Martins1992}
M.~J. Martins, Renormalization group trajectories from resonance factorized {S}
  matrices, Phys. Rev. Lett. 69 (1992) 2461--2464.

\bibitem{Martins1993}
M.~J. Martins, Exact resonance {ADE} {S}-matrices and their renormalization
  group trajectories, Nucl. Phys. {\rm B} 394 (1993) 339--355.

\bibitem{RST1996}
F.~Ravanini, M.~Stanishkov, R.~Tateo, Integrable perturbations of {CFT} with
  complex parameter: the ${M}_{3/5}$ model and its generalizations, Int. J.
  Mod. Phys. {\rm A} 11 (1996) 677--698.

\bibitem{DDT2000}
P.~Dorey, C.~Dunning, R.~Tateo, New families of flows between two-dimensional
  conformal field theories, Nucl. Phys. {\rm B} 578 (2000) 699--727.

\bibitem{Zamolodchikov1987}
A.~B. Zamolodchikov, Renormalizaiton group and perturbation theory about fixed
  points in two-dimension field theory, Sov. J. Nucl. Phys. 46 (1987)
  1090--1096.

\bibitem{LudwigCardy1987}
A.~W.~W. Ludwig, J.~L. Cardy, Perturbative evaluation of the conformal anomaly
  at new critical point with applications to random systems, Nucl. Phys. {\rm
  B} 285 (1987) 687--718.

\bibitem{Lassig1992}
M.~Lassig, New hierarchy of multicriticality in two-dimension field theory,
  Phys. Lett. {\rm B} 278 (1992) 439--442.

\bibitem{Ahn1992}
C.~Ahn, {RG} flows of non-unitary {CFT}s, Phys. Lett. {\rm B} 294 (1992)
  204--208.

\bibitem{BfMW2010}
O.~Blondeau-Fournier, P.~Mathieu, T.~A. Welsh, A bijection between paths for
  the ${M}(p,2p+1)$ minimal model {V}irasoro characters, Ann. Henri Poincar\'e
  11 (2010) 101--125.

\bibitem{BfMW2012}
O.~Blondeau-Fournier, P.~Mathieu, T.~A. Welsh, Half-lattice paths and
  {V}irasoro characters, Fundamenta Informaticae 117 (2012) 57--83.

\bibitem{Bressoud1989}
D.~Bressoud, Lattice paths and the {R}ogers-{R}amanujan identities, in:
  K.~Alladi (Ed.), Number Theory, Madras 1987, Vol. 1395 of Lecture Notes in
  Mathematics, Springer Berlin / Heidelberg, 1989, pp. 140--172.

\bibitem{Welsh2006}
T.~A. Welsh, Paths, {V}irasoro characters and fermionic expressions, J. Phys.:
  Conf. Ser. 30 (2006) 119--132.

\bibitem{Andrews1976}
G.~E. Andrews, The Theory of Partitions, Addison-Wesley, Reading, MA, 1976.

\bibitem{AndrewsBaxter1987}
G.~E. Andrews, R.~J. Baxter, Lattice gas generalization of the hard hexagon
  model. {III}. $q$-trinomial coefficients, J. Stat. Phys. 47 (1987) 297--330.

\bibitem{CRS1998}
J.~M. Camino, A.~V. Ramallo, J.~M. {Sanchez de Santos}, Graded parafermions,
  Nucl. Phys. {\rm B} 530 (1998) 715--741.

\bibitem{VJS2014}
{\'E}.~Vernier, J.~L. Jacobsen, H.~Saleur, Non compact conformal field theory
  and the $a_2^{(2)}$ ({I}zergin-{K}orepin) model in regime {III}, J. Phys.
  {\rm A} 47 (2014) 285202.

\bibitem{GasperRahman1990}
G.~Gasper, M.~Rahman, Basic Hypergeometric Series, Encyclopedia of mathematics
  and its applications, Cambridge University Press, Cambridge, 1990.

\bibitem{SeatonScott1997}
K.~A. Seaton, L.~C. Scott, $q$-trinomial coefficients and the dilute {A} model,
  J. Phys. {\rm A} 30 (1997) 7667--7676.

\end{thebibliography}

\end{document}